%% file: ANA-TCAL-2021-01-PAPER.tex
\documentclass[cernpreprint, PAPER, atlasdraft=false, texlive=2023, UKenglish, texmf, orcidlogo]{atlasdoc}
\usepackage{atlaspackage}
\usepackage{atlasbiblatex}

\usepackage{atlasphysics}
\graphicspath{{logos/}{figures/}}

\usepackage{ANA-TCAL-2021-01-PAPER-defs}

\usepackage{multirow}
\usepackage{xspace}
\usepackage{mathrsfs}
\usepackage{rotating}
\usepackage[export]{adjustbox}

\AtlasTitle{Operation and performance \\of the ATLAS tile calorimeter in LHC Run 2}

\AtlasAbstract{The ATLAS tile calorimeter (TileCal) is the hadronic sampling calorimeter covering the central region of the ATLAS detector at the Large Hadron Collider (LHC).
This paper gives an overview of the calorimeter's operation and performance during the years 2015--2018 (Run~2).
In this period, ATLAS collected proton--proton collision data at a centre-of-mass energy of 13~\TeV
and the TileCal was 99.65\% efficient for data-taking. The signal reconstruction, the calibration procedures, and the detector operational status are presented.
The performance of two ATLAS trigger systems making use of TileCal information, the minimum-bias trigger scintillators and the tile muon trigger, is discussed.
Studies of radiation effects allow the degradation of the output signals at the end of the LHC and HL-LHC operations to be estimated.
Finally, the TileCal response to isolated muons, hadrons and jets from proton--proton collisions is presented.
The energy and time calibration methods performed excellently, resulting in good stability and uniformity of the calorimeter response during Run~2.
The setting of the energy scale was performed with an uncertainty of 2\%. The results demonstrate that the performance is in accordance with specifications defined in the Technical Design Report.
}

\author{The ATLAS Collaboration}
\AtlasRefCode{TCAL-2021-01}

\PreprintIdNumber{CERN-EP-2023-266}

\arXivId{2401.16034}

\AtlasJournalRef{\EPJC 84 (2024) 1313}
\AtlasDOI{10.1140/epjc/s10052-024-13151-4}

\hypersetup{pdftitle={ATLAS document},pdfauthor={The ATLAS Collaboration}}

\begin{document}

\maketitle

\tableofcontents
\clearpage

\section{Introduction}
\label{sec:intro}
%

%

%
\newcommand{\AtlasCoordFootnote}{
ATLAS uses a right-handed coordinate system with its origin at the nominal interaction point (IP)
in the centre of the detector and the \(z\)-axis along the beam pipe.
The \(x\)-axis points from the IP to the centre of the LHC ring,
and the \(y\)-axis points upwards.
Cylindrical coordinates \((r,\phi)\) are used in the transverse plane,
\(\phi\) being the azimuthal angle around the \(z\)-axis.
The pseudorapidity is defined in terms of the polar angle \(\theta\) as \(\eta = -\ln \tan(\theta/2)\).
Angular distance is measured in units of \(\Delta R \equiv \sqrt{(\Delta\eta)^{2} + (\Delta\phi)^{2}}\).}

\newcommand{\DataStructureFootnote}{
In ATLAS, data collected over
periods of time spanning an LHC fill or generally stable conditions
are grouped into a `run', while the entire running period under
similar conditions for several years is referred to as a `Run'. Run~1 refers to runs collected in the period 2008--2012. Data
taken within a run are broken down into elementary units called
luminosity blocks, corresponding to about one minute of collision data
for which the detector conditions or software calibrations remain
approximately constant.
}

\newcommand{\EmptybunchFootnote}{
Empty bunch crossings are those for which the bunches have no protons and are separated from any filled bunch by at least five bunch crossings to ensure signals from collision events are cleared from the detector.
}

ATLAS~\cite{PERF-2007-01} is a general-purpose detector at the Large Hadron Collider (LHC)~\cite{Evans:2008zzb} designed to reconstruct events from colliding hadrons. It has a forward--backward symmetric cylindrical geometry and a near \(4\pi\) coverage in solid angle. ATLAS consists of an inner tracking detector surrounded by a thin superconducting solenoid providing a \SI{2}{\tesla} axial magnetic field, electromagnetic and hadronic calorimeters, and a muon spectrometer. The inner tracking detector covers the pseudorapidity range of \(|\eta| < 2.5\).\footnote{\AtlasCoordFootnote} It consists of silicon pixel, silicon microstrip, and transition radiation tracking detectors. Lead/liquid-Argon (LAr) sampling calorimeters provide electromagnetic (EM) energy measurements. A steel/scintillator-tile hadronic calorimeter covers the central pseudorapidity range of \(|\eta| < 1.7\). The endcap and forward regions are instrumented with LAr calorimeters for both the EM and hadronic energy measurements up to \(|\eta| = 4.9\). The muon spectrometer (MS) surrounds the calorimeters in the pseudorapidity range of \(|\eta| < 2.7\) and is based on three large superconducting air-core toroidal magnets with eight coils each. The field integral of the toroids ranges between \num{2} and \SI{6}{Tm} across most of the detector. The muon spectrometer includes a system of precision tracking chambers and fast detectors for triggering.

The tile calorimeter (TileCal) is the hadronic barrel calorimeter sub-detector of ATLAS and provides essential input to the identification of hadronic jets and measurement of their energy and direction. It also provides information for triggers, participates in the measurement of the missing transverse momentum carried by non-interacting or not detected particles, and assists in the identification of muons.
The TileCal design provides a standalone energy resolution for isolated pions of $\sigma/E = 56.4\%/\sqrt{E(\GeV)}\oplus 5.5\%$, measured at test beams~\cite{PERF-2007-01}, and a linear response with a nonlinearity tolerance of 2\% for simulated jets up to 4~\TeV to be sensitive to the full range of energies expected in the LHC lifetime.
A detailed description of the ATLAS TileCal can be found in the dedicated Technical Design Report~\cite{ATLAS-TDR-03}; the construction, optical instrumentation and installation into the ATLAS detector are described in Refs.~\cite{Abdallah:2013pea, Abdallah:2009zza}; the operation and performance of the TileCal during Run~1 at the LHC are reported in Ref.~\cite{TCAL-2017-01}.

This paper presents the operation and performance of the TileCal during the LHC Run 2 data-taking period (2015--2018).\footnote{\DataStructureFootnote}
The full ATLAS detector recorded events from cosmic-ray muons in the period February--July 2015. The first
proton--proton ($pp$) collisions at a centre of mass energy  $\sqrt{\text{s}}~=~13$~\TeV{} were recorded in April 2015 with a bunch spacing of \SI{50}{ns}~\cite{Aad:2019hmz}. Later in 2015 and the following years, the LHC $pp$ collisions continued to be at $\sqrt{\text{s}}=13~\TeV$, but the instantaneous luminosity and the number of proton collisions per bunch crossing increased and the bunch spacing decreased to 25~ns.
The total integrated recorded (delivered) luminosity was 145.5~\ifb{} (157.4~\ifb{})~\cite{Aad:2019hmz,DAPR-2021-01}. Only events recorded during stable beam conditions and with all ATLAS sub-detectors fully operational are considered in the analysis. The corresponding integrated luminosity is 140.1~\ifb{}~\cite{DAPR-2021-01}.
A summary of the LHC beam conditions in Run~2 is shown in Table~\ref{tab:datasets}. Additional special runs with low integrated
luminosity and a low average number of interactions per bunch crossing, $\langle\mu\rangle$, used for commissioning purposes, were taken with a bunch spacing of 25~ns in 2017 and 2018.
During these years ATLAS also recorded data with lower-energy proton collisions ($\sqrt{\text{s}}=~5~\TeV$), and data produced by lead--lead or xenon--xenon ion collisions. These runs had very low integrated luminosity.

\begin{table}[htbp]
\caption{Summary of the proton--proton collision data collected in the four years of Run 2 and used to obtain the results presented in this paper.
The ATLAS analysis integrated luminosity corresponds to the total integrated luminosity approved for analysis, passing all data quality requirements
ensuring the detector and reconstruction software are properly functioning~\cite{DAPR-2021-01,Aad:2019hmz}.}
\centering
\begin{tabular}{l c c c c}
\toprule
& 2015 & 2016 & 2017 &2018 \\
\midrule
Maximum beam energy [\TeV{}]                                            & 6.5 & 6.5 & 6.5 & 6.5 \\
Delivered integrated luminosity [\ifb{}]                                & 4.0 & 39.0 & 50.6 & 63.8 \\
Recorded integrated luminosity [\ifb{}]                                 & 3.6 & 35.5 & 46.4 & 60.0 \\
ATLAS analysis integrated luminosity [\ifb{}]                           & 3.2 & 33.4 & 44.6 & 58.8 \\
Mean number of interactions per bunch crossing                          & 13 & 25 & 38 & 36 \\
Maximum instantaneous luminosity [$10^{33}\text{cm}^{-2}\text{s}^{-1}$] & 5 & 13 & 16 & 19 \\
\bottomrule
\end{tabular}
\label{tab:datasets}
\end{table}

The recorded events are separated into different streams according to the trigger category for which the event is selected. Physics streams are composed of triggers that are used to identify electrons, photons, muons, jets, hadronically decaying $\tau$-leptons, and missing transverse momentum in collision data. There are also calibration streams used by the various sub-detectors for calibration and monitoring purposes.

This paper is organised as follows. Section~\ref{sec:experimental_setup} describes the experimental set-up. The reconstruction and the calibration of physics events are the subjects of Section~\ref{sec:physics_signals}.
The calibration streams used by the TileCal for calibration purposes were taken in dedicated runs when the beams were off. As discussed in the paper, they include caesium, laser and charge injection system (CIS) events. Laser events taken during empty bunches in the LHC abort gap~\cite{Evans:2008zzb}
are used for timing calibration. Physics events are also used for calibration purposes. To minimise systematic uncertainties, as discussed at the end of the section, the procedure used to reconstruct the experimental energy deposited in the cells is used in the case of simulated events. The TileCal data subsets used in the hardware-based Level-1 (L1) ATLAS trigger system are discussed in Section~\ref{sec:trigger_signals}. The deterioration of the detector performance due to the radiation exposure is reported in Section~\ref{sec:detector_ageing}. These studies allow the amplitude reduction of the output signals at the end of the LHC and HL-LHC operations to be estimated. The online and offline data quality (DQ) checks applied to the hardware and data acquisition systems are explained in Section~\ref{sec:data_quality_analysis_operation}. The validation of the full chain of TileCal calibration and reconstruction using physics events is described in Section~\ref{sec:performance_studies}.
The results are obtained by analysing the physics stream containing muons, isolated hadrons, zero-bias events~\cite{TRIG-2019-04} and jets.
The performance of ATLAS trigger systems based on TileCal information is also discussed. All results presented in this document are summarised in Section~\ref{sec:conclusion}.

The data used in the studies were collected by the tile calorimeter sub-detector using the full ATLAS data acquisition chain. An extensive software suite~\cite{ATL-SOFT-PUB-2021-001} is used for real and simulated data reconstruction and analysis, for operation, and in the trigger and data acquisition systems of the experiment. In addition to the TileCal, the information of other ATLAS sub-detectors is used to assist in particle identification, and track, momentum and energy reconstruction.


\FloatBarrier

\section{Experimental setup}
\label{sec:experimental_setup}

\newcommand{\InteractionLengthFootnote}{
The nuclear interaction length $\lambda$ is defined as the mean path length required to reduce the flux of relativistic primary protons and neutrons to a fraction 1/e.
}

\subsection{Tile calorimeter}
\label{subsec:tilecal_calorimeter}

The tile calorimeter is a sampling calorimeter consisting of tiles of plastic scintillator as active material and low-carbon steel absorber plates.
The scintillator tiles are composed of Polystyrene with a low mass concentration of pTP (1.5\%) and POPOP (0.04\%). Further technical details are
described in Ref.~\cite{Abdallah:2009zza}.
As shown in Figure~\ref{fig:calorimeter_d3_a}, the TileCal surrounds the LAr calorimeter and fills the volume from an inner radius of 2.28~m to an outer radius of 4.23~m.
The amount of material in front of the TileCal at $\eta = 0$~ corresponds to 2.3$\lambda$~\footnote{\InteractionLengthFootnote}~\cite{PERF-2007-01}.
The sub-detector thickness at the outer edge of the tile instrumented region is 9.7$\lambda$ at $\eta =0$. Including 1.3$\lambda$ from the outer support, the total detector thickness is 11$\lambda$ at $\eta =0$, sufficient to reduce punch-through well below the irreducible level of prompt decay muons.
The calorimeter is mechanically divided into three segments along the $z$-axis (the beam axis): one central long barrel (LB) section that is 5.8~m in length ($|\eta| < 1.0$), and two extended barrel (EB) sections on either side of the LB that are each 2.6~m long ($0.8 < |\eta| < 1.7$).

\begin{figure}[t]
\centering
\subfloat[]{\includegraphics[width=0.585\linewidth]{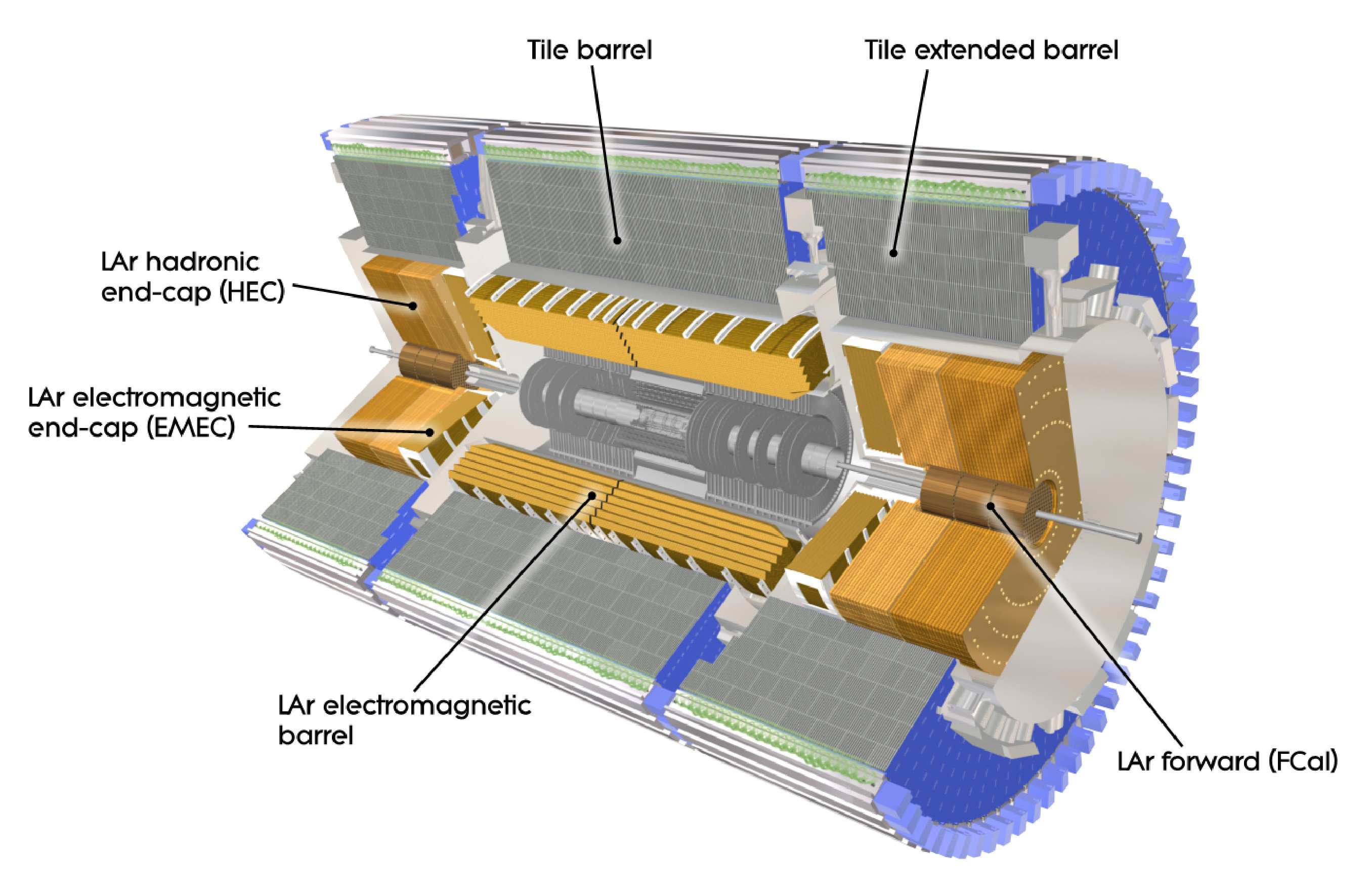}\label{fig:calorimeter_d3_a}}
\subfloat[]{\includegraphics[width=0.315\linewidth]{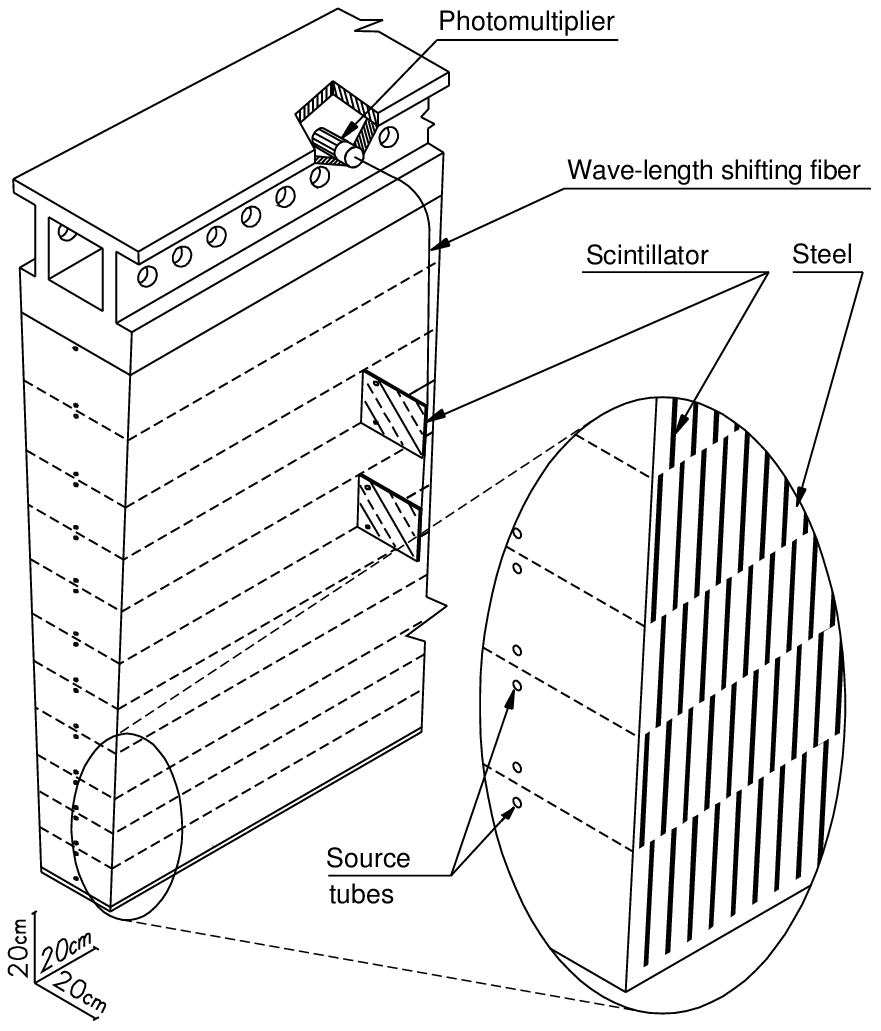}\label{fig:calorimeter_d3_b}}
\caption{(a) Cut-away view of the ATLAS calorimeter system. The TileCal consists of a barrel and two extended barrel sections. The sections of the ATLAS liquid-argon (LAr) calorimeters are also indicated. (b) An illustration of the mechanical assembly and optical readout of a single tile calorimeter module. A total of 256 such modules comprise the full tile calorimeter. Source tubes are used to circulate a $^{137}$Cs radioactive source contained in a capsule for calibration purposes.}
\end{figure}

Each TileCal barrel consists of 64 modules in the angular direction $\phi$ allowing full azimuth coverage. Figure~\ref{fig:calorimeter_d3_b} shows a schematic of a TileCal module. In each module, the steel and scintillator tiles have a thickness of 14~mm and 3~mm respectively. The steel and scintillator tiles are interleaved periodically in the longitudinal direction $z$. The scintillator tiles are organised along the radius
in 11 rows of different sizes, numbered from 1 to 11 starting from the smallest radius. Each row corresponds to a group of tiles at the same distance from the beam axis. Light from the scintillator tiles is collected at their edges by wavelength-shifting (WLS) fibres, arranged in pre-shaped opaque plastic `profiles' attached to both sides of the modules and running radially, as shown in Figure~\ref{fig:calorimeter_d3_b}. The WLS fibres transport the light to photomultiplier tubes (PMTs). The readout cell geometry is defined by grouping the fibres from individual tiles on the given PMTs~\cite{Abdallah:2009zza}. The cell layout for LB and EB modules in a plane containing the beam axis is shown in Figure~\ref{fig:tilecal_cells_and_rows2} for $z>0$. A mirroring of those in the other direction in $z$ defines the three sections of the calorimeter. Three longitudinal layers are obtained. In the long barrels, layers A, BC and D cells are made of  three, six and two rows corresponding to depths of 1.5, 4.1, and 1.8$\lambda$ respectively. In the extended barrels, layers A, B and D cells have three, four and four rows corresponding to depths of 1.5, 2.6, and 3.3$\lambda$ respectively. The layers allow the jet energy depositions to be measured at different depths. The cells have a granularity of $\Delta \eta \times \Delta \phi = 0.1\times 0.1$ in the two innermost layers and $\Delta \eta \times \Delta \phi = 0.2 \times 0.1$ in the outermost one. As shown in the figure, the scheme produces pseudo-projective towers. The total energy in a tower is used for triggering purposes as discussed in Section~\ref{subsec:the_tilecal_readout}.

To correct for energy losses in the gap region between the TileCal LB, and the EB and the LAr sections, where many cables and electronics crates from other ATLAS sub-detectors are located, a special intermediate tile calorimeter (ITC) system is installed. As shown in Figure~\ref{fig:tilecal_cells_and_rows2}, the ITC cells (D4, C10 and E1--E4) are located between the LB and EB, and provide coverage in the range of $0.8 < |\eta| < 1.6$. Some of the C10 and D4 cells have reduced thickness or special geometry to accommodate services and readout electronics for other ATLAS sub-detectors~\cite{ATLAS-TDR-03, TCAL-2010-01}. The gap (E1--E2) and crack (E3--E4) cells are composed of only scintillator media and are read out by one PMT each. The minimum-bias trigger scintillators (MBTS)~\cite{Sidoti_2014}, used to trigger events from colliding particles, are also read out by TileCal EB electronics and provide coverage in the range of $2.08 < |\eta| < 3.86$ (see Section~\ref{subsec:mbts_intro}).
Most TileCal cells are read out by two PMTs, accounting for a total of 9852 readout channels for the 5182 cells.

\begin{figure}[t]
\centering
\includegraphics[width=1.\textwidth]{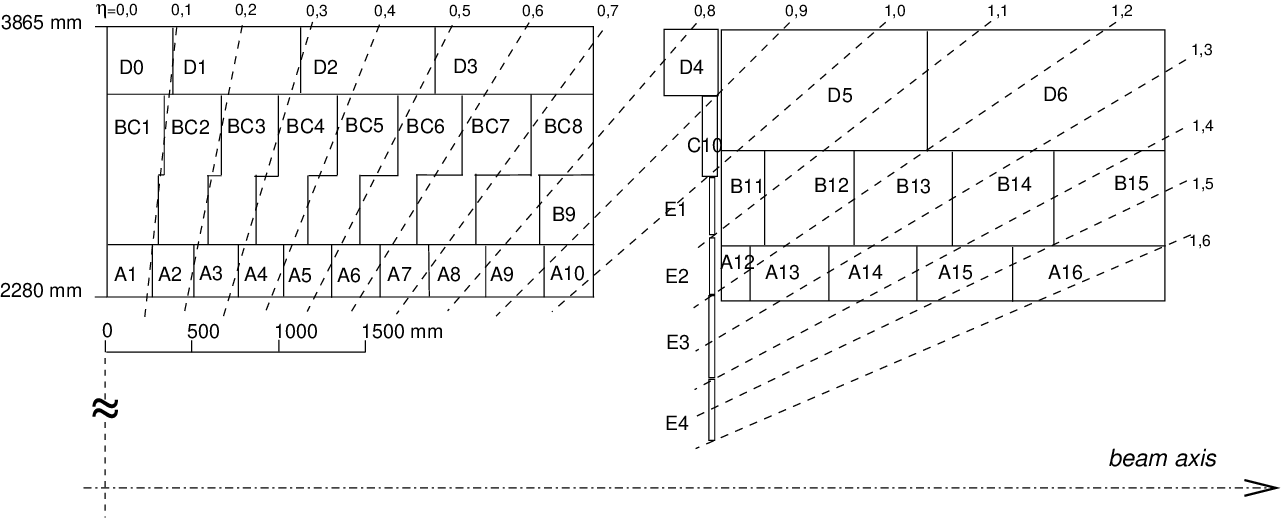}
\caption{The layout of the TileCal cells with $\eta \ge 0$, denoted by one or two letters (A to E) plus an integer number. The layer A is closest to the beamline.The naming convention is repeated for cells with negative $\eta$. The long barrel (extended barrel) cells are shown at the left (right).}
\label{fig:tilecal_cells_and_rows2}
\end{figure}

\subsection{Readout electronics}
\label{subsec:the_tilecal_readout}

The PMTs and front-end electronics are placed in aluminium units housed in a steel girder on the outer radius of each module. The units can be completely extracted and are therefore called electronics drawers.
Two contiguous connected drawers form a super-drawer.
In each module of the LB there are two super-drawers hosting the electronics of the cells with $\eta > 0$ (A-side) and $\eta < 0$ (C-side) respectively. The electronics of the cells of each of the modules of the EBs are located in a super-drawer. In this way, from an electronic point of view, one identifies four barrels, the LBA and EBA with $\eta > 0$ and the LBC and EBC with $\eta < 0$. There are 45 and 32 channels per super-drawer in the LB and EB, respectively.

Each channel consists of a unit called a PMT block, which contains the light-mixer, PMT tube and High-Voltage (HV) divider, and a `3-in-1 card'~\cite{Anderson:2005ym, Adragna:2006ur}. The card receives the PMT signal and provides three output signals as follows:

\begin{itemize}
\item A shaper with a shaping time of 50~ns shapes the fast PMT signal to the requirements of the 10-bit, 40M sample per second ADCs used to digitise the signal at the LHC bunch crossing rate~\cite{Berglund:2008zz}. The amplitude of the shaped signal is proportional to the integrated charge of the PMT pulse. Two linear outputs are produced with a relative gain of 64 and hence an overall 16-bit dynamic range using two 10-bit ADCs. These data are digitised continuously and seven samples centred around the pulse peak, synchronised with the LHC master clock, are obtained. The seven samples are stored in a circular pipeline buffer and transferred to the off-detector electronics when a Level-1 trigger~\cite{TRIG-2019-04} signal is received.
A gain switch is used to determine which gain information is sent to the back-end electronics for event processing. By default the high-gain (HG) signal is used unless any of the seven samples saturates the ADC, at which point the low-gain (LG) signal samples are transmitted. The signals from the digital readout are used to measure physics data as well as laser and charge injection system calibrations (Section~\ref{subsec:the_calibration/monitor_systems}).
\item An integrator that receives less than 1\% of the PMT current digitized by a 12-bit ADC card~\cite{GonzalezParra:2011cla}. The integrator is a low-pass DC amplifier with six switchable gain settings and a calibration input. With one of the available gains, the currents range between 0.01~nA and 1.4~$\mu$A over a time window of 10--20~ms. The integrator readout is used for calibration and monitoring  purposes (Section~\ref{subsec:the_calibration/monitor_systems}) measuring the caesium and physics runs signals. In the case of physics runs the signals are produced essentially by inelastic proton-proton collisions at low momentum transfer, known as minimum-bias (MB) events. Besides response monitoring, the measurements of MB events also allow to determine the instantaneous luminosity of the colliding beams, not discussed in this article.
\item
A differential signal from the LG output of the shaper is sent to the adder boards located in the drawer to determine the total energy in a calorimeter tower used in the ATLAS trigger (Section~\ref{subsec:level-1_calorimeter_trigger_tileCal_information}). The signals of the D5 and D6 cells are used as input to the tile muon trigger system (Section~\ref{subsec:tile_muon_trigger_system}). The channels reading the MBTS counters provide the analogue HG outputs to the corresponding trigger (Section~\ref{subsec:mbts_intro}).
\end{itemize}

In addition to the PMT signals, the 3-in-1 card also contains an input path for the calibration charge injection system described in Section~\ref{subsubsec:the_charge_injection_system}.

In the case of the digital readout, the DMU (data management unit) chips located on the digitiser boards are responsible for organising the digitized samples in  packets of data. There are eight digitiser boards in each super-drawer, and each one has two DMU devices, so there are 16 DMU devices per super-drawer. The interface board collects data from all the digitisers, serialises and transmits them to the back-end electronics. The integrity of the data received by the back-end is checked using a CRC (cyclic redundancy check) algorithm. The back-end electronics are located in the counting room approximately 100~m away from the ATLAS detector. The data acquisition system of the TileCal is split into four logical partitions associated with the super-drawers of the modules in LBA, LBC, EBA, and EBC\@. Optical fibres transmit signals between each super-drawer and the back-end trigger, timing and control (TTC), and readout driver (ROD~\cite{Valero:2007zz}) crates. There are a total of four TTC and ROD crates, one for each logical partition. The ATLAS TTC system distributes the LHC clock, trigger decisions, and configuration commands to the front-end electronics. If the TTC system sends the trigger acceptance command to the front-end electronics, the corresponding digital samples for all channels of the calorimeter are sent to the ROD where the signal is reconstructed.

The integrator output voltage of each PMT of a module, proportional to the input charge, is read out sequentially using a single ADC\@.
The switching between the corresponding 3-in-1 cards takes place about every 0.1 s. This switching mechanism is known to induce noise for a couple of bunch crossings, thus it is done in the LHC abort gap~\cite{Evans:2008zzb}. The `SHAFT board'~\cite{Solovyanov:2018rfj,Koschik:2008zz} that is responsible for handling the different Tile calibration signals during data-taking,
sends the pulse to trigger the sending of the switching command via TTC to be distributed over the full detector.
The digitized signals are sent to an off-detector readout buffer through CAN bus.

The front-end electronics of a single super-drawer are supplied by a low-voltage power supply (LVPS) source, which is positioned in an external steel box mounted just outside the super-drawer. The HV is set and distributed to each PMT using dedicated boards positioned inside the super-drawers next to the front-end electronics.

\subsection{Calibration systems}
\label{subsec:the_calibration/monitor_systems}

\begin{figure}[t]
\centering
\includegraphics[width=1.\textwidth]{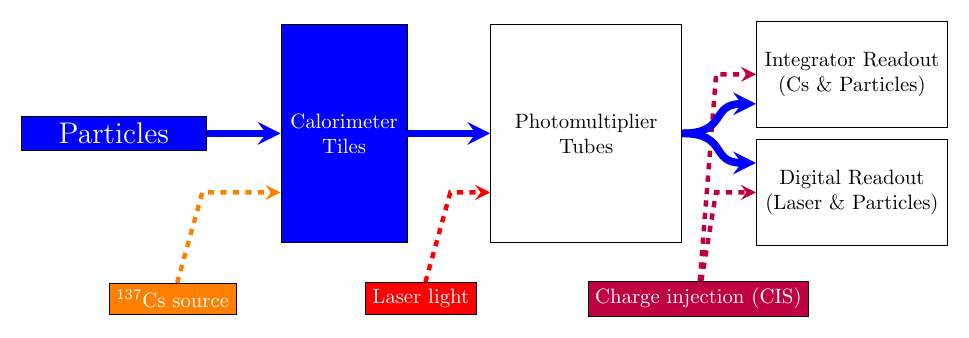}
\caption{The signal paths for each of the three calibration systems used by the TileCal. The signal produced by particles from collisions is denoted by the thick solid line, and the path taken by each of the calibration systems is shown with dashed lines.}
\label{Fig:TileCalCalibrationChain}
\end{figure}

The ATLAS physics performance goals for hadronic jets measurements require a tolerance of 2\% on the non-linearity of the detector response for jets up to 4~\TeV.
Jets are clusters of energy deposits in the calorimeter cells, and jet-level requirements are adopted as conservative requirements for the cell energy measurements. To achieve the aforementioned jet performance goals in the ATLAS environment and over an extended period of time requires calibration of the calorimeter cell response with the same 2\% precision. Four systems, the electronic charge injection system, the integrator readout, the caesium radioactive $\gamma$-source system~\cite{Blanchot_2020} and the laser system~\cite{Agaras:2023vbp} are used, as depicted in Figure~\ref{Fig:TileCalCalibrationChain}. They probe the three steps in the signal reconstruction of TileCal and MBTS cells (see Section~\ref{subsec:mbts_intro}): production and collection of light in scintillator tiles and fibres, conversion of light to electrical signal in PMTs, and calibration of electrical signals in the electronic readout. The aim is to maintain a uniform and stable response of the measurements of the energy deposited in all cells accounting for changes in the optics and electronics.
The calibration systems allow the scale of the cell energy measurement to be set, as described in Section~\ref{sec:physics_signals}, and the determinations obtained using experimental and simulated data to be compared. The caesium source, MB currents and the laser system allow to determine the degradation of the TileCal and MBTS signals due to their exposure to a high radiation level produced during the LHC operations. The latter is discussed in Section~\ref{sec:detector_ageing}. Moreover, the three complementary calibration systems also aid in identifying the source of problematic channels (see Section~\ref{subsec:calibration_run_checks}). Problems originating strictly in the readout electronics are seen by both the laser and CIS, problems related solely to the PMTs are detected exclusively by the laser system, while caesium sources and MB events probe also scintillation light production and transport in the WLS fibres.

\subsubsection{Charge Injection system}
\label{subsubsec:the_charge_injection_system}

The charge injection system calibrates the front-end electronics. The calibration is performed by injecting a full dynamic range of input charge signals, similar to what the detector experiences in physics runs, to the readout electronics of all channels through the respective 3-in-1 cards.
The magnitude of each injected charge is controlled by a 10-bit digital to analogue converter (DAC). The DAC's non-linearity is better than 0.4\%. The conversion factor is 0.801 pC per DAC setting count.
The DAC setting is increased from 0 to 15 with a step size of 1 for the HG readout, and varied from 32 to 992 with a step size of 32 for the LG. The two readouts provide a comprehensive test of the electronics response over the full range of energies expected during regular running of the LHC\@. As discussed in Section~\ref{subsubsec:integrator readout system}, a part of this system is also used to calibrate the gain conversion constant of the integrator readout (see~Figure~\ref{Fig:TileCalCalibrationChain}).

\subsubsection{Integrator readout of minimum-bias collisions}
\label{subsubsec:integrator readout system}

The integrator readout measurements of the PMT current induced by MB $pp$ interactions are used to monitor the variations of the sub-detector response over time. The current, proportional to the LHC luminosity, varies with the position of the cell. To avoid saturation and maintain an adequate resolution, the integrator gain can be selected by choosing one of the six predefined resistors that also define the integration time. The PMT current is obtained from the ADC voltage measurement as

\begin{equation}
I\ \mathrm{[nA]=\frac{ADC\ [mV] - ped\ [mV]}{Int.\ gain\ [M\Omega]}}
\label{eq:mb}
\end{equation}

where ped is the pedestal, measured with circulating beams before collisions to account for beam background sources such as beam halo and beam-gas interactions, and Int. gain is the integrator gain constant determined by an internal calibration circuit with a high precision DAC on the 3-in-1 card.

Dedicated runs are periodically taken to calculate the values for each of the six gain settings by fitting the linear relationship between the injected current and measured voltage response. For an instantaneous luminosity between $1\times 10^{30}$ and $3\times 10^{34}$~cm$^{-2}$s$^{-1}$, the non--linearity of the integrator response is $< 1\%$.
The deviation from stability of the integrator gains is better than 0.05\% for individual channels and better than 0.01\% on average.

\subsubsection{Caesium system}
\label{subsubsec:the_caesium_system}

The caesium system employs three $^{137}$Cs radioactive $\gamma$-sources, one for the LB, one for the EBA and one for the EBC modules~\cite{Blanchot_2020}.
A hydraulic system moves the source through the calorimeter using a network of stainless steel tubes parallel to the beam line running through small holes in each tile scintillator (see Figure~\ref{fig:calorimeter_d3_b}). It produces 0.662~\MeV\ photons that generate scintillation light in each tile. The average activity of the three sources was approximately 323~MBq in March 2015 and dropped to approximately 296~MBq in December 2018 (about $-2.3$\% per year). To collect a sufficient signal, the electrical readout of the caesium calibration is performed using the integrator readout path (see Figure~\ref{Fig:TileCalCalibrationChain}). The readout employs a resistor of 28.81 M$\Omega$ with corresponding integration time of 13.9 ms. The caesium system allows the cells' response to be equalised and the combined variations of the optical components and the PMTs to be monitored.

\subsubsection{Laser system}
\label{subsubsec:the_laser_system}

The laser calibration system consists of a single laser source, located off detector, able to produce controlled short light pulses that are simultaneously distributed by optical fibres to the photocathode of all 9852 PMTs. During the LHC long shutdown before the start of Run~2, a new laser system~\cite{Agaras:2023vbp,Scuri:2016ctn} was developed to correct shortcomings in electronics and light monitoring of the first system used during Run~1, which in turn resulted in an improved long term reliability. The new system has been used since the beginning of Run~2. The intrinsic stability of the laser light was found to be within 3\%, so to measure the PMT gain variations to a better precision using the laser source, the response of the PMTs is normalised to the signal measured by a dedicated photodiode. Its stability is monitored by an $\alpha$-source and, in 2015, the variation over one month was shown to be 0.5\%, and the linearity of the associated electronics response was within 0.2\%. As shown in Figure~\ref{Fig:TileCalCalibrationChain}, the laser system monitors the PMTs and electronic components of the digital readout. Since the optical fibres have the same length, the system is also used to equalise the timing of the digitizers before absolute time calibration with physics events (see Section~\ref{subsubsec:time_calibration}).


\FloatBarrier

\section{Reconstruction and calibration of signals in physics events}
\label{sec:physics_signals}

To reconstruct physics events, the energy deposited in a cell is obtained by summing the energy measurements from the connected PMT channels, with each channel energy $E$ given by

\begin{equation}
E\ [\GeV]=\frac{A\ [\mathrm{ADC}]}{C_{\mathrm{ADC}\to \mathrm{pC}}\times C_{\mathrm{pC}\to \GeV}\times C_{\mathrm{Cs}}\times C_{\mathrm{MB}}\times C_{\mathrm{Las}}} .
\label{eq:calibration}
\end{equation}

The signal amplitude $A$
is determined by using the seven signal samples of the digital readout (Section~\ref{subsec:the_tilecal_readout}) as discussed in Section~\ref{subsec:digital_signal}. The calibration constant $C_{\mathrm{ADC}\to \mathrm{pC}}$ converting the signal from ADC channel units to pC units is estimated using the CIS system as reported in Section~\ref{subsec:monitor_conversion_factor_ADC_pC}. The factor $C_{\mathrm{pC}\to \GeV}$ was determined by measuring the response of the sub-detector to electrons with test beams (TBs) (Section~\ref{subsec:the_electromagnetic_energy_scale}). It allows the signals to be expressed in \GeV\ and the experimental energy determinations to be compared with the ones obtained using simulated events. %
The factors $C_{\mathrm{Cs}}$, $C_{\mathrm{MB}}$ and $C_{\mathrm{Las}}$, %
discussed in Section~\ref{subsec:monitor_PMT_tiles_fibres_variation_response_Cs}, Section~\ref{subsec:monitor_PMT_tiles_fibres_variation_response_MB} and Section~\ref{subsec:monitor_of_PMT_variation_response} respectively, are used to monitor the stability of the responses and provide corrections for each channel during the ATLAS operations. The procedure is described in Section~\ref{subsec:summary_calibration_procedure}. In the case of simulated events, the cell energy deposited is also obtained using Eq.~(\ref{eq:calibration}) with the $C_{\mathrm{Cs}}$, $C_{\mathrm{MB}}$ and $C_{\mathrm{Las}}$ constants being equal to 1. The procedure is discussed in Section~\ref{subsec:signal_reconstruction_simulated_events}.

\subsection{Signal amplitude}
\label{subsec:digital_signal}

The optimal filtering (OF) algorithm is used to reconstruct the amplitude $A$ in units of ADC counts, the time offset $\tau$ (phase) relative to the 40 MHz clock and the pedestal, $ped$, of the
pulse of the shaped PMT signal~\cite{Cleland:2002rya,Fullana:816152}. The method linearly combines the seven signal samples $S_i$, $i=1, ..., 7$ of the digital readout, which are read with 25 ns spacing:

\begin{equation}
A=\sum_{i=1}^{n=7}a_iS_i,\hspace{3em}
A\tau=\sum_{i=1}^{n=7}b_iS_i,\hspace{3em}
ped=\sum_{i=1}^{n=7}c_iS_i.
\label{eq:of}
\end{equation}

The coefficients $a_i$, $b_i$ and $c_i$ are optimised using TB data to minimise the bias on the reconstructed quantities introduced by the electronic noise (see Section~\ref{subsubsec:electronic_noise})~\cite{Fullana:816152}.
The normalised pulse shape function, taken as the average pulse shape from test beam data, is used to determine the coefficients. As shown in  Figure~\ref{fig:ps_lo_hi}, separate functions are defined for HG and LG modes.
The pulse shape and coefficients are stored in a dedicated database. For each sample, gain and phase there are five coefficients: the three ($a_i$, $b_i$ and $c_i$) that appear in Eq.~(\ref{eq:of}) and two additional for the calculation of quality factors. In total there are 70 coefficients per channel and the constants are the same for all the channels with the same timing. Pileup effects are not considered. A total of 2001 phase values in the range -100~ns and +100~ns with steps of 0.1 ns are considered, resulting in 140070 constants stored in the database.

The values of $A$, $\tau$ and $ped$ are obtained by solving the set of equations in Eq.~(\ref{eq:of}). The signal reconstruction is performed twice: i) in real time by the RODs (referred to as online) for use in the trigger, and ii) after the data have been recorded (referred to as offline) for use in the data analysis.

\begin{figure}[t]
\centering
\includegraphics[width=0.5\textwidth]{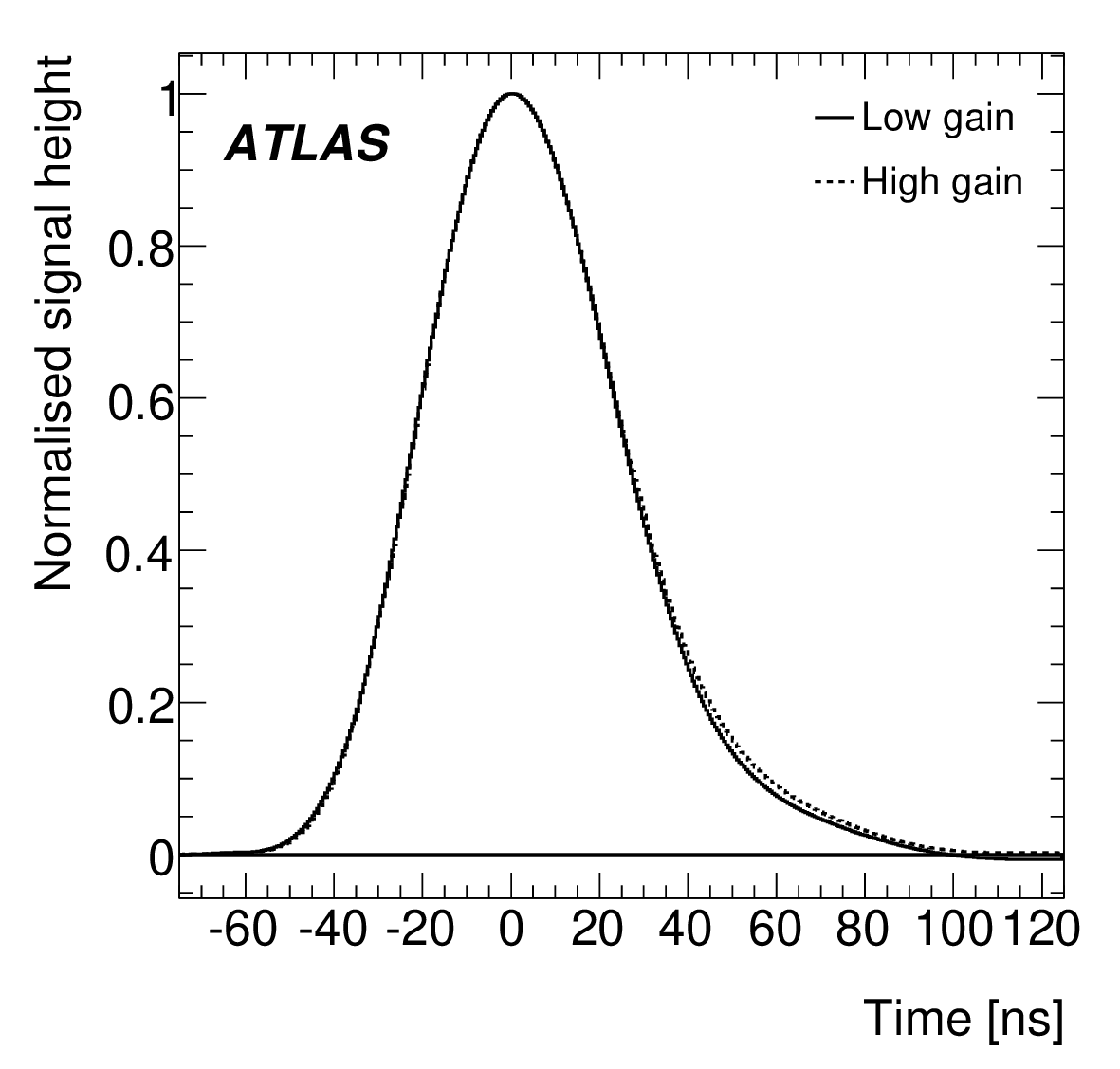}
\caption{The reference pulse shapes for high gain (dotted curve) and low gain (solid curve), shown in arbitrary units~\cite{TCAL-2010-01}.}
\label{fig:ps_lo_hi}
\end{figure}

The expected time of the pulse peak is calibrated such that, for particles originating from collisions at the interaction point, the pulse should peak at the central (fourth) sample, synchronous with the LHC clock. The reconstructed value of $\tau$ represents the time offset in nanoseconds between the expected pulse peak and the time of the actual reconstructed signal peak, arising from fluctuations in particle travel time and uncertainties in the electronics readout (see Section~\ref{subsubsec:time_calibration}).

There are two versions of the optimal filtering algorithm, iterative and non-iterative. The coefficients $a_i$, $b_i$ and $c_i$ are functions of the pulse's true time offset relative to the 40~MHz clock. This time offset is known only approximately a priori before the reconstruction. The iterative OF method takes the time of the maximum sample as an initial value of the time offset. In the subsequent iterations, the input time offset is taken to be equal to $\tau$ calculated in the previous iteration. The algorithm converges to the actual time offset value with an accuracy better than 0.5~ns in the absence of \pileup pulses (see Section~\ref{subsubsec:pileup_noise}). Typically, three iterations are needed to converge. This method is used when reconstructing events occurring asynchronously with the LHC clock, such as cosmic-ray muons.
In the case of LHC physics events, with an increasing number of events per bunch crossing, the non-iterative method is used both online and offline, for being faster and more robust against \pileup.
The out-of-time \pileup, see Section~\ref{sec:resolution_pileup}, can lead to a reconstructed time offset value far from the expected one, biasing the energy measurement when the iterative method is used. The non-iterative OF method performs only a single iteration of the optimal filtering method and uses as input the time offset determined for each TileCal channel from prior timing calibration runs. This is the time offset expected for in-time pulses from the collisions of interest. By forcing the time offset to its predetermined expected value, the non-iterative optimal filtering method better reconstructs the energy of the in-time pulse from a collision of interest in the presence of out-of-time \pileup. This method is also more robust against electronic noise for very low amplitude signals.

In real time, or online, the digital signal processor (DSP) in the ROD performs the signal reconstruction using the OF technique and provides the channel energy and time to the High-Level Trigger~\cite{TRIG-2019-04}. The conversion between signal amplitude in ADC counts and energy units of \GeV\ is done by applying channel-dependent calibration constants that are described in the following sections. The DSP reconstruction is limited by the use of fixed point arithmetic, which has a precision of 0.0625 ADC counts (approximately 0.75~\MeV\ in HG), and imposes precision limitations for the channel-dependent calibration constants. The offline signal is reconstructed using the same iterative or
non-iterative OF technique as online but using floating point arithmetic.

To avoid saturation of the output ROD bandwidth
in high instantaneous luminosity conditions, all LG channels and only HG channels for which the difference between the maximum and minimum $S_i$ is larger than five ADC counts (approximately 60~\MeV) have the raw data transmitted from the ROD for offline signal reconstruction. Otherwise, the ROD signal reconstruction results are used for the offline data processing. The procedure has a negligible impact on the performance of the energy reconstruction.

\subsubsection{Channel timing calibration}
\label{subsubsec:time_calibration}

To allow for optimal energy reconstruction by the non-iterative OF method and to enable precise time-of-flight measurement in certain physics analyses, the time difference between the digitising sampling clock and the peak of the PMT pulses must be minimised and measured with a precision of 1~ns. To achieve this, the clock phases in the DMUs in the front-end hardware (see Section~\ref{subsec:the_tilecal_readout}) are adjusted in multiples of 104~ps. The hardware time offset can be set for groups of six channels.

The initial time calibration for Run~2 was performed using the relative timing differences per channel relative to the end of Run~1 measured by the laser system (see Section~\ref{subsubsec:the_laser_system}). The calibration was later refined using beam-splash events from a single LHC beam utilising the same method as before the start of Run~1~\cite{TCAL-2010-01}. Finally, the time calibration is established with the $pp$ collision data. To avoid possible bias from non-collision beam backgrounds, only channels belonging to reconstructed jets satisfying standard quality criteria~\cite{ATLAS-CONF-2015-029,PERF-2016-04} are considered in each event. Given that the timing is slightly dependent on the energy deposited in a cell, in the case of HG signals, the offset is determined in the range $\SI{2}{\GeV} < E < \SI{4}{\GeV}$. In the case of LG signals the behaviour is smoother and a broader energy range of $\SI{15}{\GeV} < E < \SI{50}{\GeV}$ is chosen. The distributions show that more than 99.5\% of the reconstructed $\tau$ values are found to be between $-10$~ns and $+10$~ns. A non-zero value of the phase $\tau$ causes the reconstructed amplitude to be underestimated.
A correction (parabolic correction) based on the phase is applied when the phase is reconstructed within half the LHC bunch spacing (12.5~ns) and the channel amplitude is larger than 15 ADC counts, to reduce contributions from noise~\cite{TCAL-2017-01}. After applying the correction, the reconstructed amplitudes are found to underestimate the actual signal by less than 1\% in magnitude within time phases of $\pm$10 ns.
\Pileup is included in simulated events, such that the signal amplitude reconstruction proceeds in the same way for simulated and real events (see Section~\ref{subsec:signal_reconstruction_simulated_events}).

\subsubsection{Channel timing monitoring}
\label{subsec:time_minitoring}

As discussed, the time settings in each
channel are adjusted so that the channel signal pulse produced by a collision particle peaks at the central sample of the digital readout.
Two complementary procedures are developed to monitor the time calibration. The first one exploits the laser calibration events recorded during the empty bunch crossings of physics runs with a frequency of about 3~Hz.
This tool provides the reconstructed time as a function of the
luminosity block in each channel. The second tool directly uses physics events. Only HG signals are monitored. Both tools were systematically used in all runs and revealed three main types of problems: the so-called timing jumps, bunch-crossing offsets, and bad or unstable channels.

\begin{figure}[t]
\centering
\subfloat[]{\includegraphics[height=0.415\linewidth]{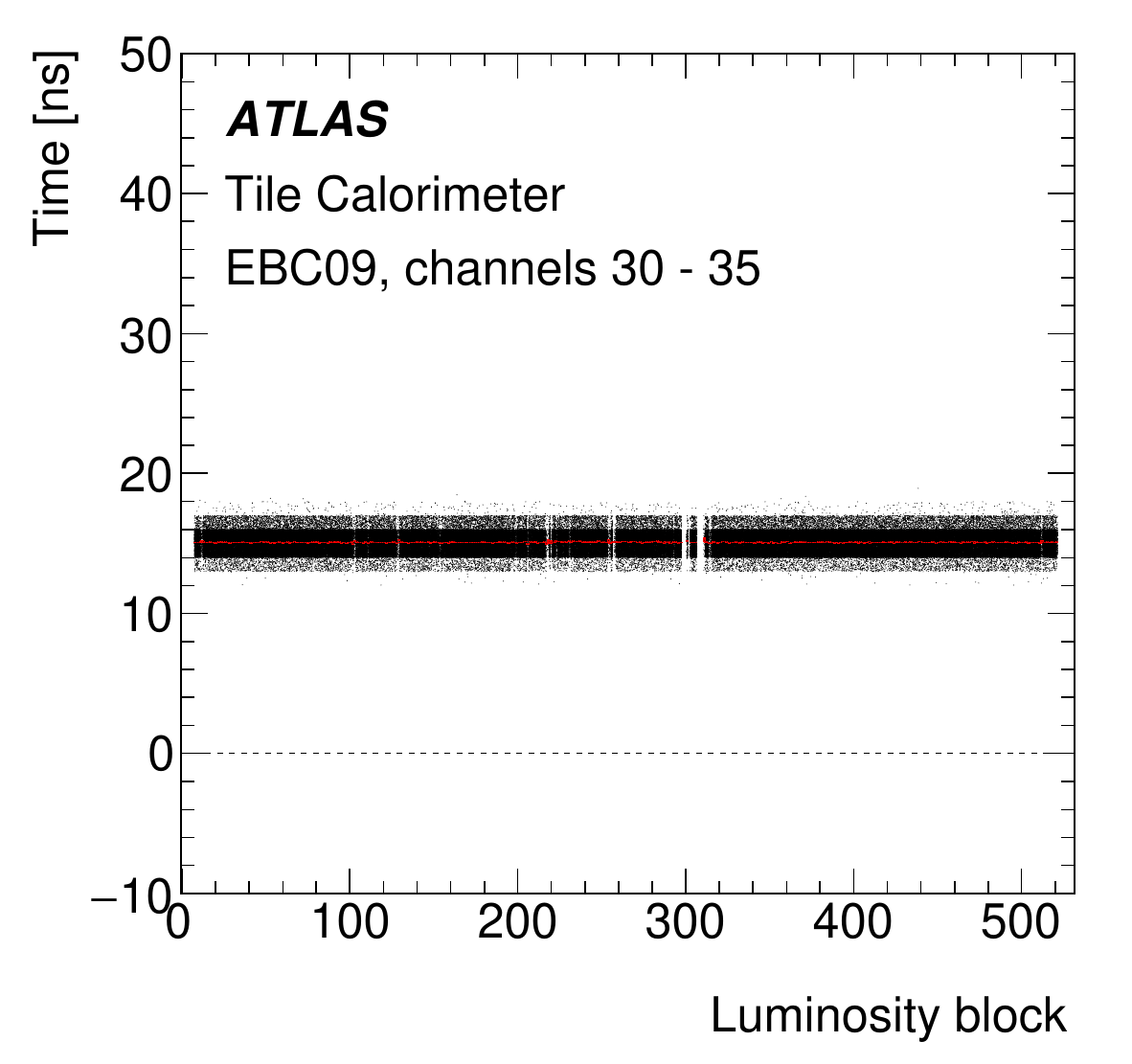}\label{fig:timing_jump_laser_physics_a}}\quad
\subfloat[]{\includegraphics[height=0.415\linewidth]{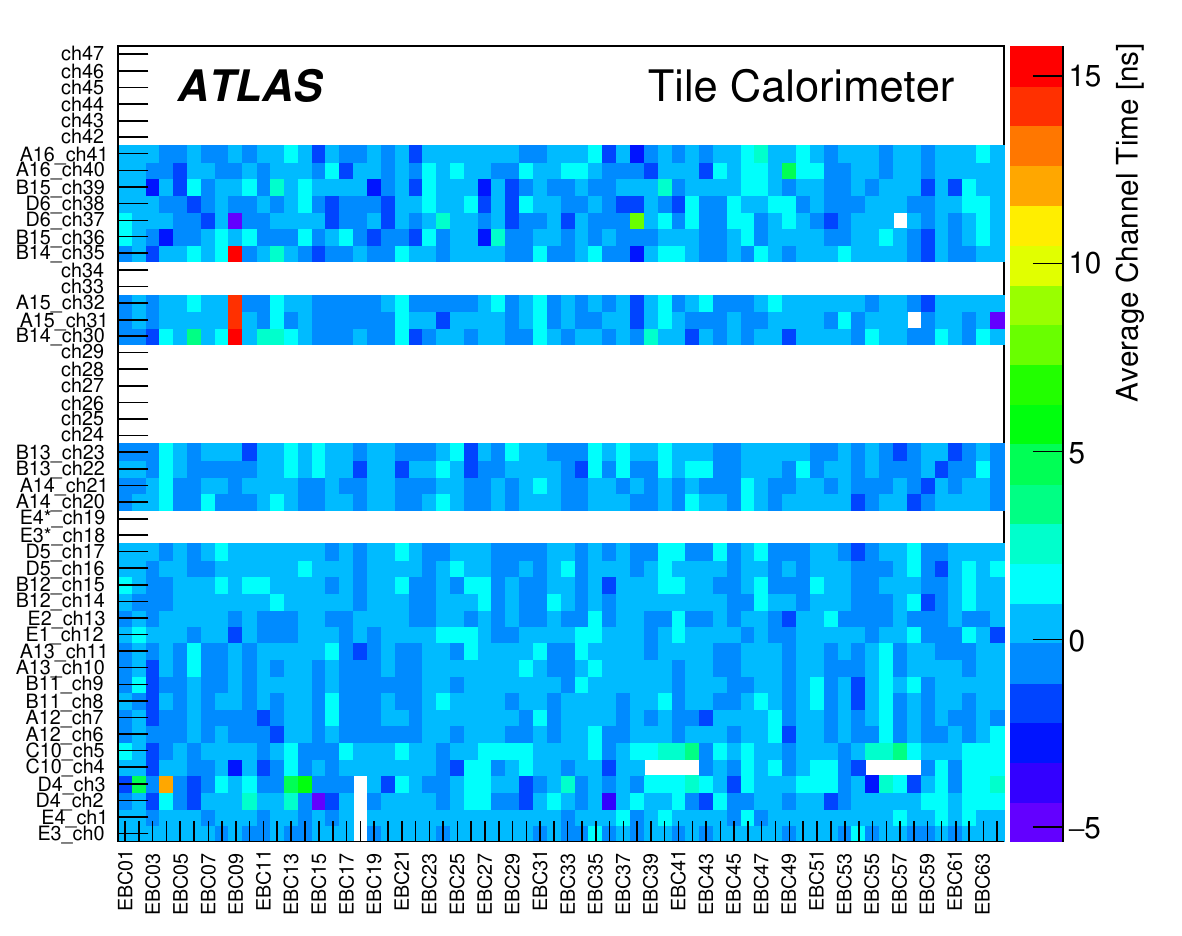}\label{fig:timing_jump_laser_physics_b}}
\caption{(a) The reconstructed time of laser events as a function
of the luminosity block. Data from six channels belonging to the same digitizer are superimposed. The timing jump (from the expected mean time of 0~ns indicated by the dashed line) lasted the entire duration of the run and all events are centered around $+\SI{15}{ns}$. (b) The 2D histogram shows the average channel time in physics events on a colour scale as a function of module number ($x$-axis) and channel number ($y$-axis).
}
\label{fig:timing_jump_laser_physics}
\end{figure}

In a timing jump, the reconstructed time
suddenly changes for a group of six channels belonging to
the same digitiser board. An example of such a case, identified by
both monitoring tools, is shown in
Figure~\ref{fig:timing_jump_laser_physics}.
Timing jumps are corrected by adjusting the
corresponding time constants in the affected period.
The observed frequency of timing jumps was much lower than in
Run~1~\cite{TCAL-2017-01} due to the improved stability of the LVPS, and they have no direct impact on the overall timing performance.

\begin{figure}[t]
\centering
\subfloat[]{\includegraphics[width=0.5\linewidth]{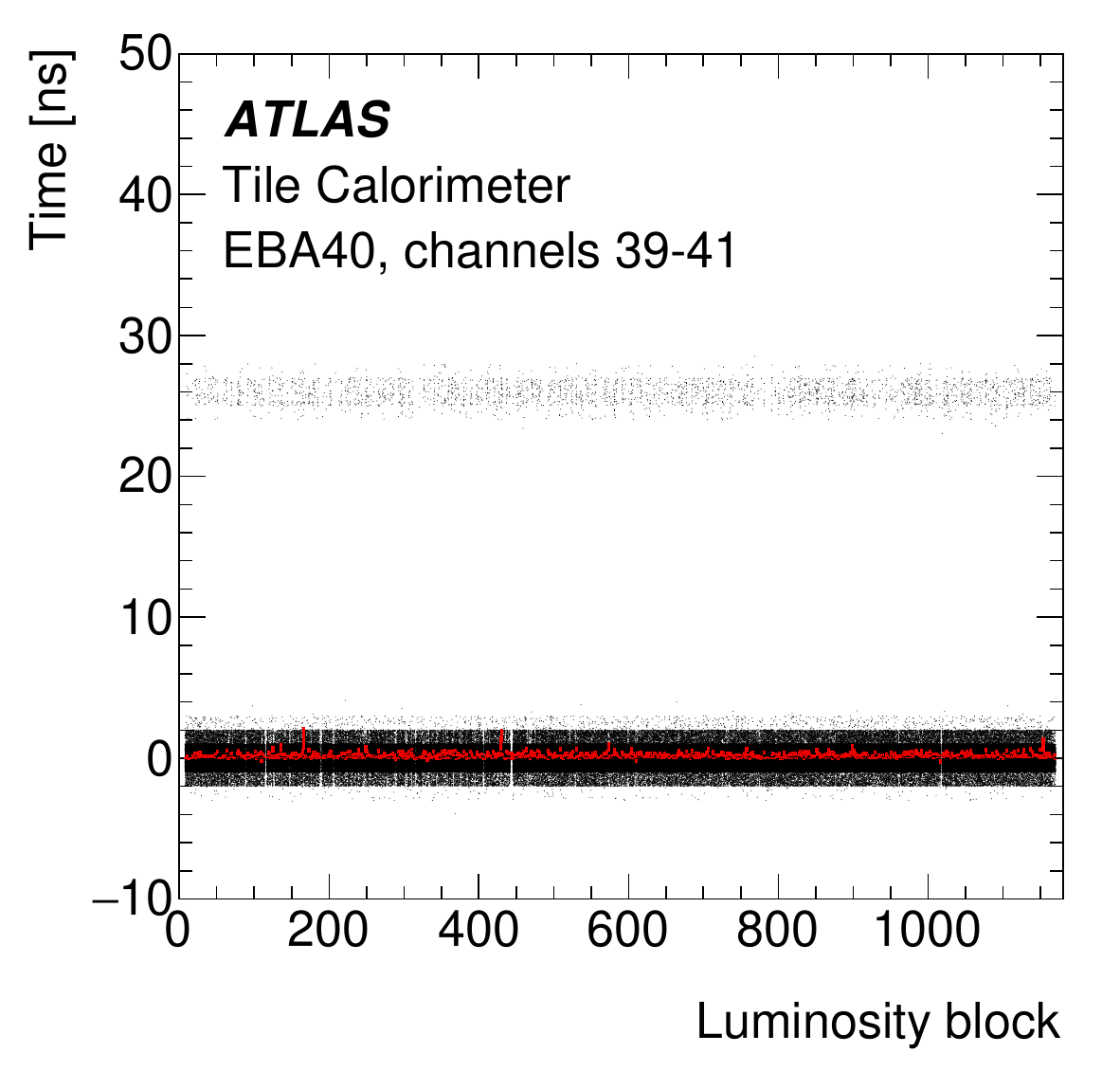}\label{fig:timing_BC_offset_a}}
\subfloat[]{\includegraphics[width=0.5\linewidth]{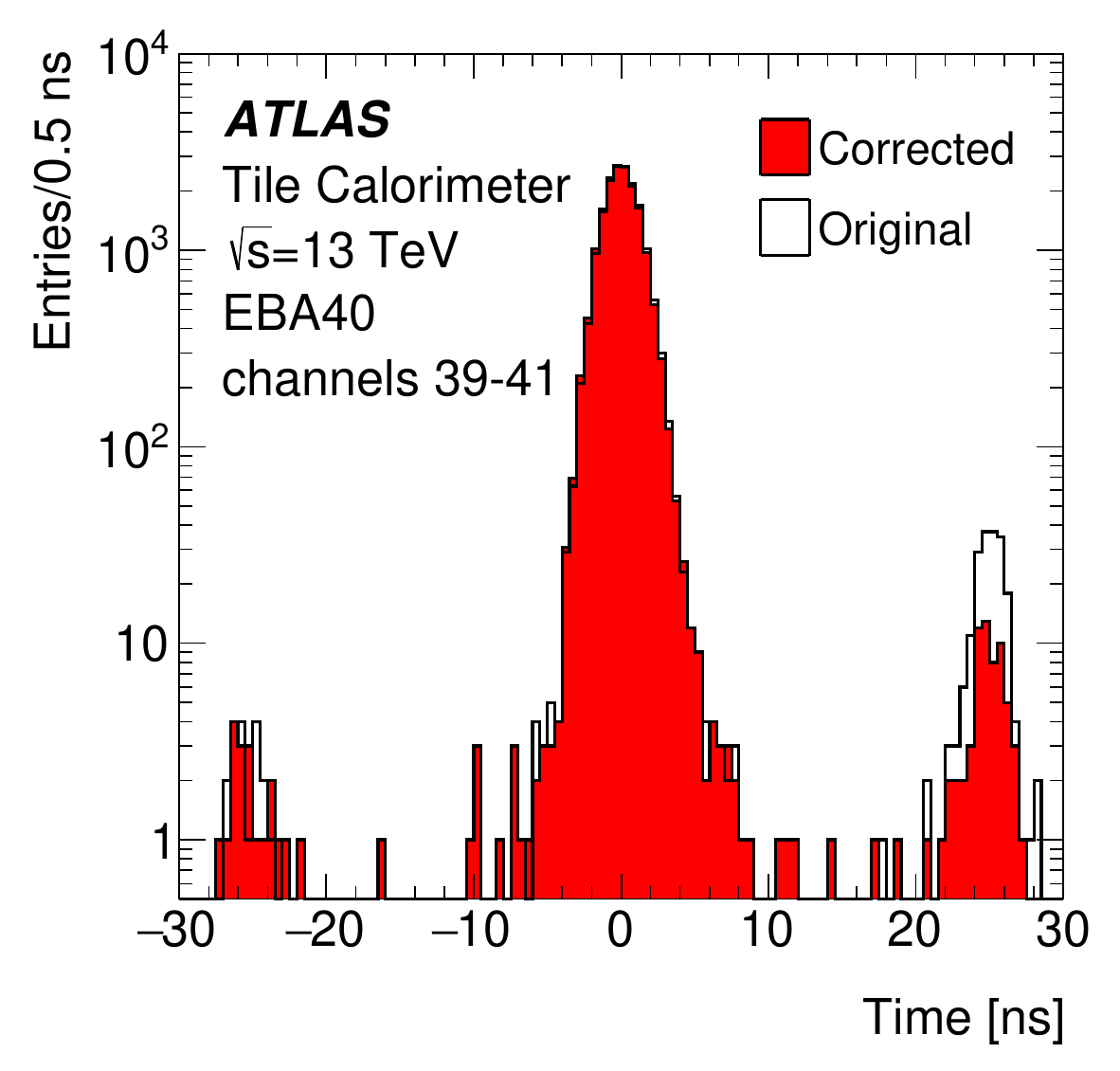}\label{fig:timing_BC_offset_b}}
\caption{(a) The reconstructed time of laser events as a function
of the luminosity block. Data from three channels belonging to the same DMU are superimposed. The majority of events, centred around zero, are
well timed in. The events affected by the the bunch-crossing offset are
centred at $+\SI{25}{ns}$. (b) The reconstructed time in
physics events in the same three channels with (corrected) and
without (original) applying the algorithm mitigating the bunch-crossing
offset events. The algorithm significantly reduces events
centred around $+\SI{25}{ns}$.}
\label{fig:timing_BC_offset}
\end{figure}

The bunch-crossing offsets correspond to the cases where channels have
reconstructed times occasionally deviating by one or two bunch crossings, i.e. $\pm
25$ or $\pm\SI{50}{ns}$. This feature affects
three channels connected to the same DMU~\cite{Berglund:2008zz}, described in Section~\ref{subsec:the_tilecal_readout}, and thus, the magnitude of the observed offsets and
affected events are fully correlated across the three channels.
The problem is intermittent and the number of affected events are typically at the percent-level.
The bunch-crossing offsets are identified with laser and physics events.
An example is shown in Figure~\ref{fig:timing_BC_offset_a}.
In order to mitigate this problem, the affected channels are flagged in the conditions database (Section~\ref{subsec:offline_data_quality_review}) and a dedicated software algorithm
was developed to identify individual problematic events in the affected channels and exclude them from further data processing.
Figure~\ref{fig:timing_BC_offset_b} compares the
reconstructed time in physics events with and without this algorithm
applied. A significant fraction of affected events close to $+\SI{25}{ns}$ is removed.
This algorithm was used during the data reprocessing campaign.

Some channels have a distorted pulse shape resulting in a wrong reconstructed time or they exhibit instabilities in the reconstruction.
This is typically caused by a malfunctioning or damaged hardware component in a given channel. A special flag is
assigned to such channels to prevent the incorrect time from further propagation in the object
reconstruction and subsequent data analysis. In total, 35 channels were
flagged as having bad or unstable timing at the end of Run~2.

\subsection{The conversion factor from ADC counts to pC}
\label{subsec:monitor_conversion_factor_ADC_pC}

The CIS system described in Section~\ref{subsubsec:the_charge_injection_system} is used to determine the factor $C_{\mathrm{ADC}\to \mathrm{pC}}$. Runs are typically taken daily for both gains in the absence of colliding beams.
The procedure of injecting charges, sampling the analogue pulse, and
measuring the fitted amplitude is repeated as the DAC value is increased. The relation between the reconstructed amplitude (in ADC counts) discussed in Section~\ref{subsec:digital_signal} and injected charge (in pC) is obtained by performing a single parameter linear fit for specific ranges of charge (3--10~pC for HG ADCs, 300--700 pC for LG ADCs).\footnote{The lower limit of each fit range is chosen to ensure the precision of each injection to be at least 0.4$\%$, which ensures a measured error value of less than 0.5$\%$. The upper limit on the HG fit range eliminates injections that saturate the ADC, as the amplitude of charge values $>$10~pC are all read out at 1023~ADC.} The slope of the linear fit yields the calibration constant $C_{\mathrm{ADC}\to \mathrm{pC}}$ for the ADC of each channel in units of ADC/pC.
For channels where the calibration constant differs by more than 1.0\% from the previous value, the constant is updated for the energy reconstruction.

Figure~\ref{fig:DTSrun2} shows the stability of the charge injection constants, as a function of time, observed during the entire Run~2 for the HG and LG ADC channels.
There is a 0.7\% systematic uncertainty present in
individual calibrations, represented by the shaded error band.  This
uncertainty comes from the observed peak output amplitudes and is
taken as characteristic of the channel-to-channel variation from this
source, prior to any calibration.
It can be seen that the detector-wide mean CIS constant over all non-problematic
channels falls within the systematic error band of the typical channel
plotted, throughout Run~2.
Figure~\ref{fig:histrun2} shows the change in the average CIS constant in per cent for every channel in the detector
between August 2015 (at the beginning of Run~2) and October 2018 (at the end of Run~2). In all figures
shown, channels that are unresponsive or have CIS constants that
fluctuate run-to-run are not included.
For the duration of Run~2 the CIS constants of only 32 channels in the detector changed by more than $\pm$4$\%$.

\begin{figure}[t]
\centering
\subfloat[]{\includegraphics[width=0.5\linewidth]{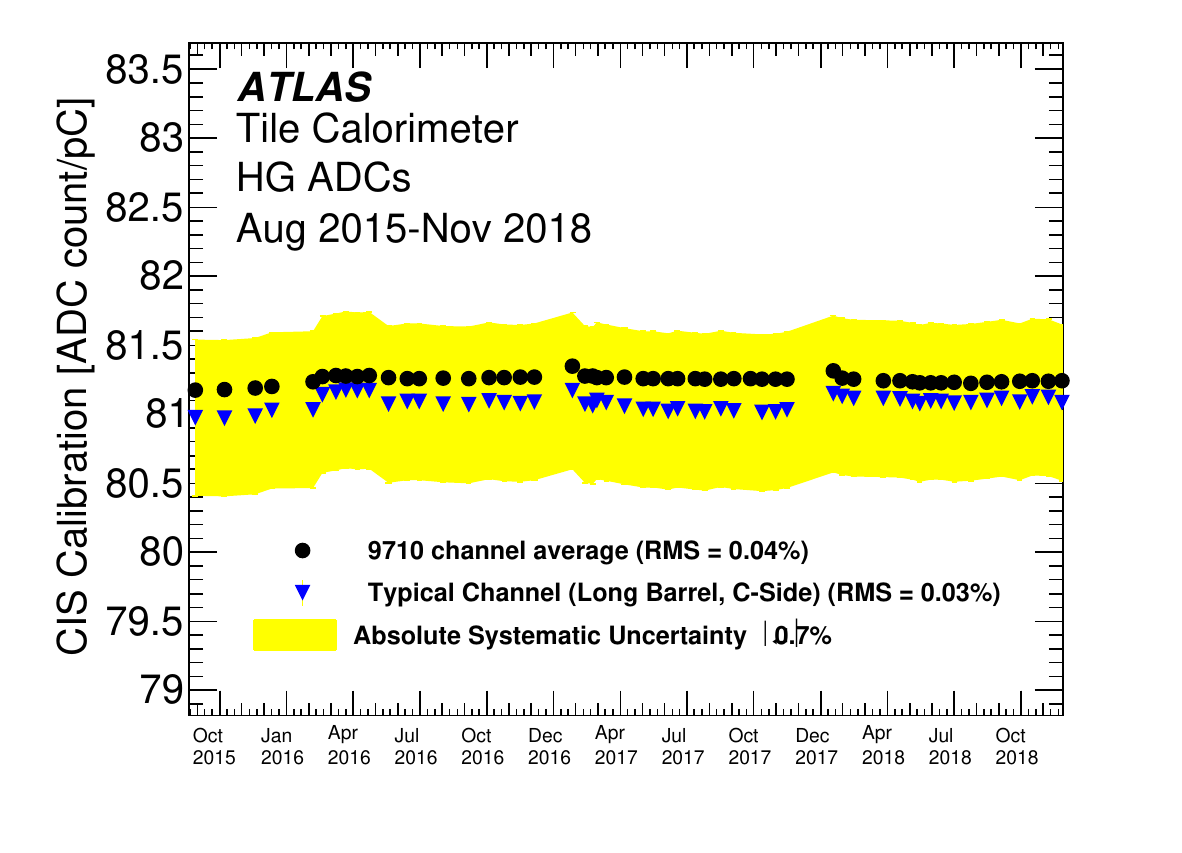}\label{fig:DTSrun2L}}
\subfloat[]{\includegraphics[width=0.5\linewidth]{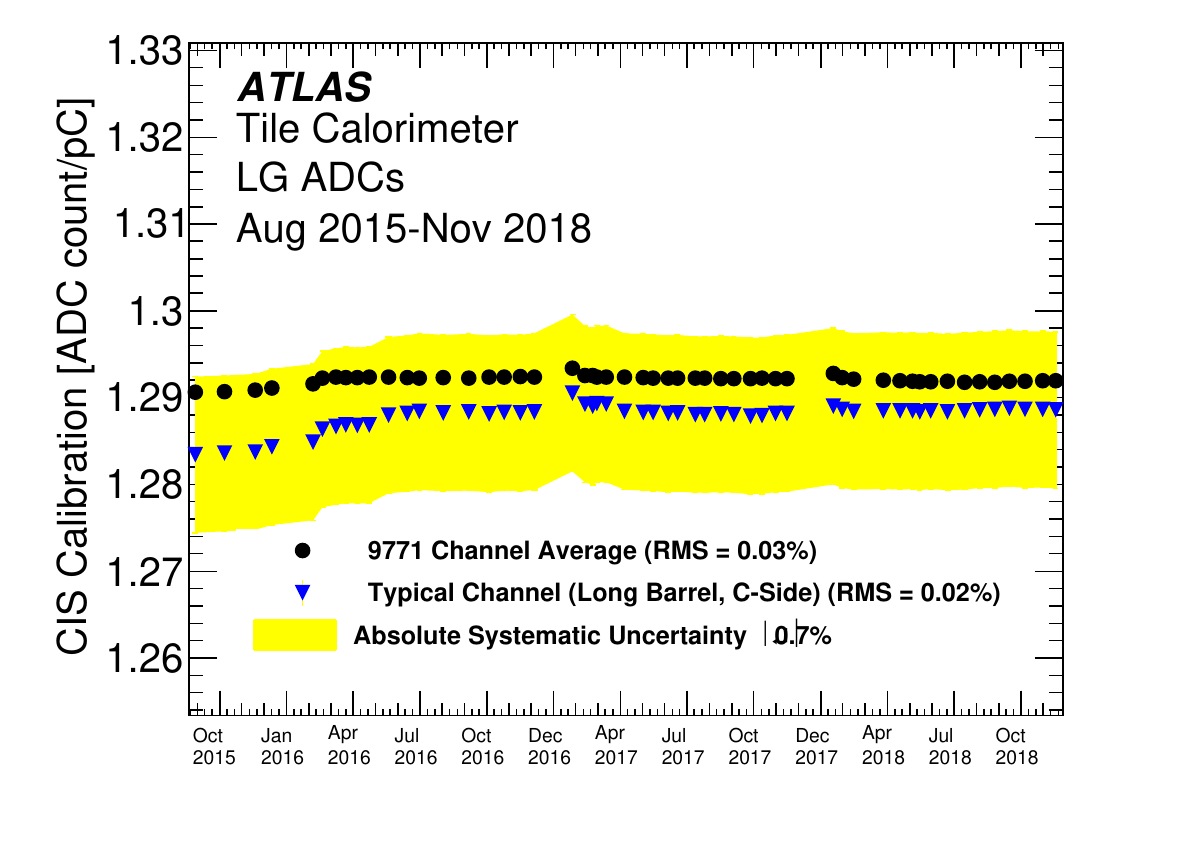}\label{fig:DTSrun2R}}
\caption{
The charge injection system constants ($C_{\mathrm{ADC}\to \mathrm{pC}}$) for the
(a) high-gain and (b) low-gain
ADCs,
as a function of time, observed during the entire Run~2 (between CIS calibration runs taken on August 2015 and November 2018).
Values for the average over all channels and for one typical channel are shown.
The RMS values indicate the fluctuations present in calibrations. In
addition, there is a 0.7\% systematic uncertainty present in
individual calibrations, represented by the shaded error band. Only good channels not suffering from damaged components relevant to the charge injection calibration are included.
\label{fig:DTSrun2}}
\end{figure}

\begin{figure}[t]
\centering
\includegraphics[width=0.7\textwidth]{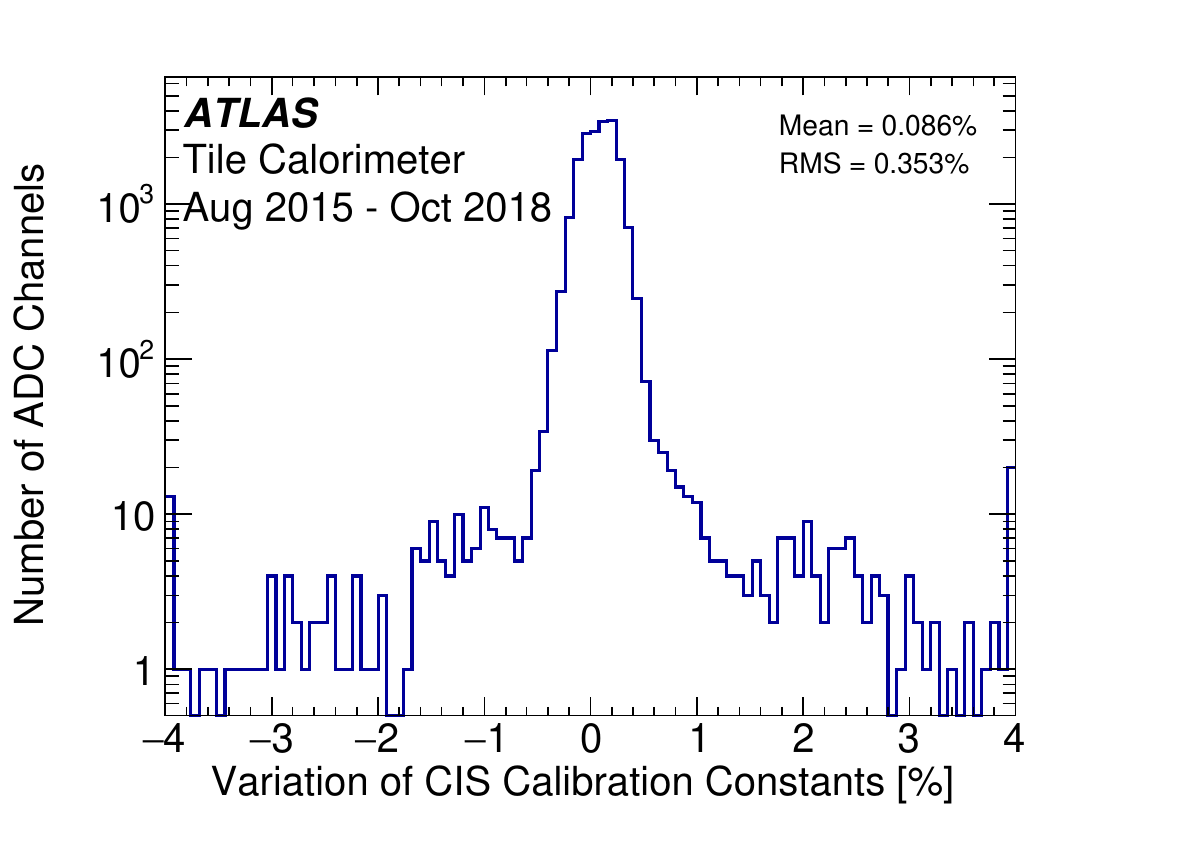}
\caption{Variations in the detector-wide CIS constants (in per cent) between August 2015 and October 2018. The RMS variation is approximately 0.35$\%$. Only good channels not suffering from damaged components relevant to the charge injection calibration are included. The first and last bins contain underflow and overflow, respectively.\label{fig:histrun2}}
\end{figure}

\subsection{Signal measurement in \GeV}
\label{subsec:the_electromagnetic_energy_scale}

The electromagnetic (EM) scale calibration constant $C_{\mathrm{pC}\to \GeV}$, converting the cell signals measured as an electric charge in pC to \GeV, is obtained from the measured amplitude $A\ [\mathrm{ADC}]$ of the signal produced by electrons at dedicated test beam campaigns using 11\% of the production modules~\cite{TCAL-2010-01, Adragna:2009zz}:

\begin{equation}
C_{\mathrm{pC}\to \GeV}=\frac{A\ [\mathrm{ADC}]}{ C_{\mathrm{ADC}\to \mathrm{pC}} \times E_e}
\label{eq:EM_scale}
\end{equation}

At TBs, the electrons hit the centre of the A cells at an angle of $20^\circ$ relative to the cell surface normal. The energy $E_e$ of the incident electron is completely deposited in the hit cell. The calibration constant $C_{\mathrm{ADC}\to \mathrm{pC}}$, obtained using CIS measurements, allows the measured amplitude in ADC counts to be converted to pC (Section~\ref{subsec:monitor_conversion_factor_ADC_pC}). The value of $C_{\mathrm{pC}\to \GeV}$ amounts to $1.050\pm 0.003$~pC/\GeV\ with an RMS spread of ($2.4\pm 0.1$)\%.

To transport to ATLAS the value of the EM scale determined at the TB, at the start of Run~2, the response of all TileCal cells was equalised using the same procedure as before the electron runs at the TB and the start of Run~1~\cite{TCAL-2017-01}. In February 2015, ahead of the start of collisions, the HV of each PMT was adjusted so that the integrator response to the $^{137}$Cs source in the PMTs was equal to that observed before the start of Run~1 and also equal to
the response measured during TB campaigns. Corrections were applied taking into account the activity decrease of the $^{137}$Cs source, which is about 2.3\% per year. After the equalisation, the HV applied to the PMTs was kept unchanged during the entirety of Run~2.
Due to light attenuation, the scintillator tile response depends on the impact point position of the particle on the tile, as well as on the tile size. Correction factors are applied for each layer of the calorimeter to account for both effects.
Those values are determined from TB data, measuring the response to muons impinging on the calorimeter with a direction parallel to the $z$-axis (beam axis in Figure~\ref{fig:tilecal_cells_and_rows2}), and from the measurements obtained using a $^{90}$Sr source~\cite{Adragna:2009zz}.

\subsection{Monitoring of the PMT, tile and fibre response with the caesium system}
\label{subsec:monitor_PMT_tiles_fibres_variation_response_Cs}

Since a caesium scan~\cite{Blanchot_2020} needs a pause in the $pp$ collisions of at least six hours, this calibration cannot be performed very often. Moreover, during the LHC technical stop at the beginning of the data-taking period in 2016, a few traces of liquid coming from the caesium hydraulic system were found in the detector cavern. Since then until the end of Run~2, caesium scans were restricted to the end of year technical stops due to risk of the liquid leak and were performed only a few times per year in Run~2.

Figure~\ref{fig:cesium_driftL} shows the response deviation from the expected value, $\Delta R_{\mathrm{Cs}}$ in per cent, as a function of time, averaged over all cells in a given radial layer. These drifts are a combination of different sources (scintillator tiles, WLS fibres and PMTs), as detailed in Section~\ref{subsec:combination_calibration} discussing the radiation effects to the detector. Due to higher radiation exposure, the most affected cells are located at the inner radius in layer~A. %
Figure~\ref{fig:cesium_driftR} shows the difference in cell responses recorded over the period of Run~2 (between February 2015 and October 2018), for cells located at different $\eta$.
It can be seen that the degradation is not uniform across $\eta$, an effect of the different radiation doses received.
At the end of Run 2, the response of the most irradiated cells in layer A had drifted downward by 18\%, while central cells in outer layer D drifted up by 2\%.

\begin{figure}[t]
\centering
\subfloat[]{\includegraphics[width=0.503\linewidth]{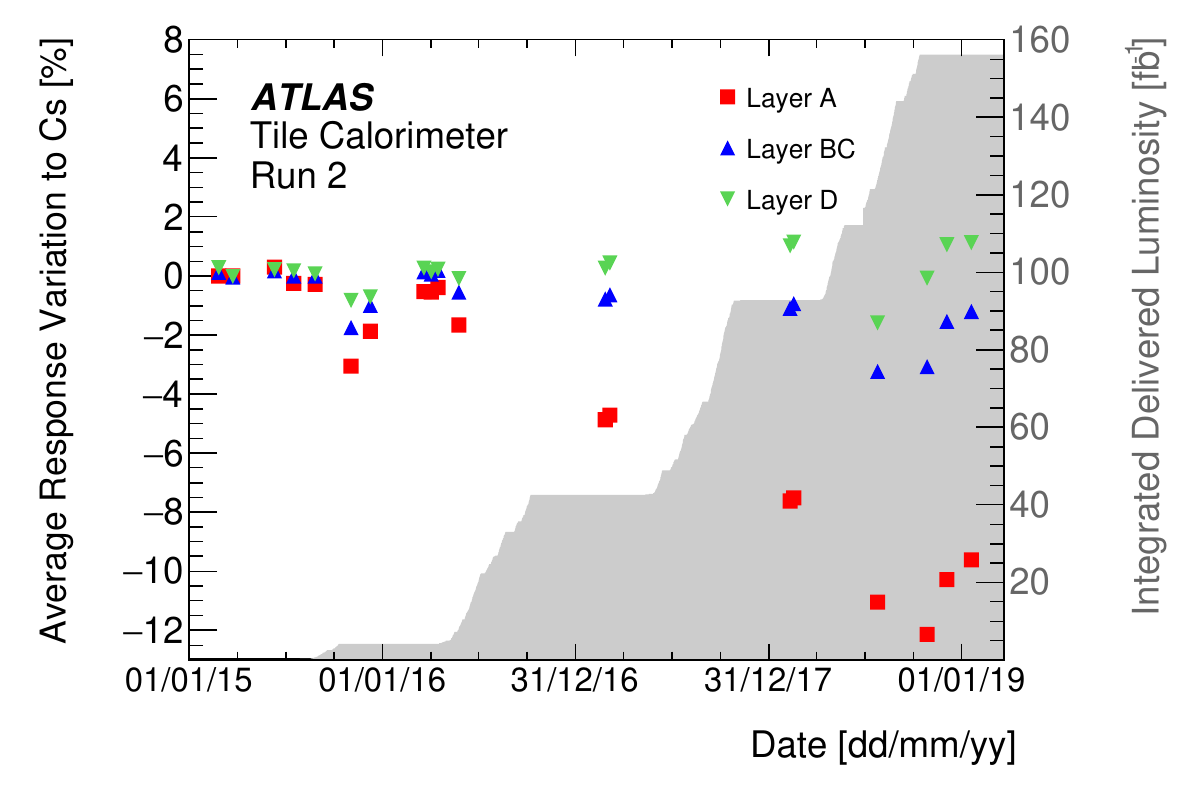}\label{fig:cesium_driftL}}\quad\quad
\subfloat[]{\includegraphics[width=0.437\linewidth]{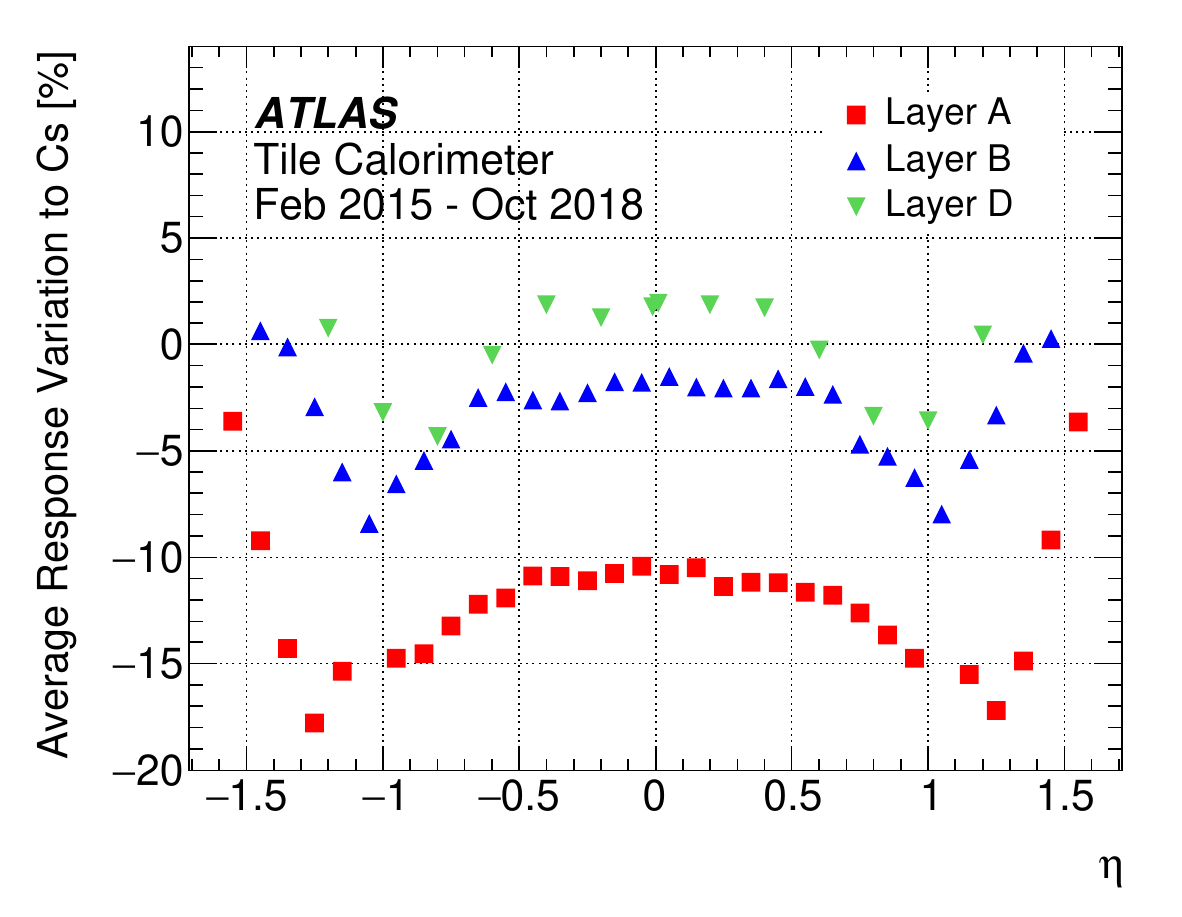}\label{fig:cesium_driftR}}
\caption{
(a) The average response variation of the TileCal cells to the $^{137}$Cs source relative to the expected value, $\Delta R_{\mathrm{Cs}}$, as a function of time. The average runs over all cells in three radial layers. The increasing response corresponds to the periods without collisions. The LHC delivered integrated luminosity is shown by the shaded area. (b) The average response variation of the TileCal cells to the $^{137}$Cs source relative to the expected value, $\Delta R_{\mathrm{Cs}}$, after Run~2 data-taking, as a function of the cell position in $\eta$, for three different layers.
}
\label{fig:cesium_drift}
\end{figure}

The data from Figure~\ref{fig:cesium_driftR} is one of the inputs for the combined channel calibration, as detailed in Section~\ref{subsec:summary_calibration_procedure}. In each channel, the calibration constant, $C_{\mathrm{Cs}}$ in Eq.~(\ref{eq:calibration}) is related to $\Delta R_{\mathrm{Cs}}$ according to

\begin{equation}
C_{\mathrm{Cs}} = 1 + \Delta R_{\mathrm{Cs}} .
\label{eq:C_Cs}
\end{equation}

The precision of the caesium calibration in a typical cell is approximately 0.3\%. For cells on the extreme sides of a partition, the precision is 0.5\% due to larger uncertainties associated with the source position. The precision for the narrow C10 and D4 ITC cells is 3\% and 1\%, respectively.

\subsection{Monitoring of the PMT, tile and fibre response with minimum-bias collisions}
\label{subsec:monitor_PMT_tiles_fibres_variation_response_MB}

Corrections based on MB measurements are applied during the reprocessing of the data. Since the MB response is proportional to the instantaneous luminosity, the cell response to MB events is normalised to luminosity measurements obtained using the inner tracker~\cite{DAPR-2021-01}. The response deviation in each cell type, $\Delta R_{\mathrm{MB}}$, is determined relative to a nominal value and then turned into a calibration constant, $C_{\mathrm{MB}}$ in Eq.~(\ref{eq:calibration}), as described in Section~\ref{subsec:summary_calibration_procedure}.
The precision of the measurements is approximately 1.2\%. %
As an example, Figure~\ref{fig:E_minimum_bias} shows the variation of the average response to MB events for the cells in the gap/crack region of the Extended Barrel as a function of time. These cells are exposed to high radiation doses.

\begin{figure}[t]
\centering
\includegraphics[width=0.8\linewidth]{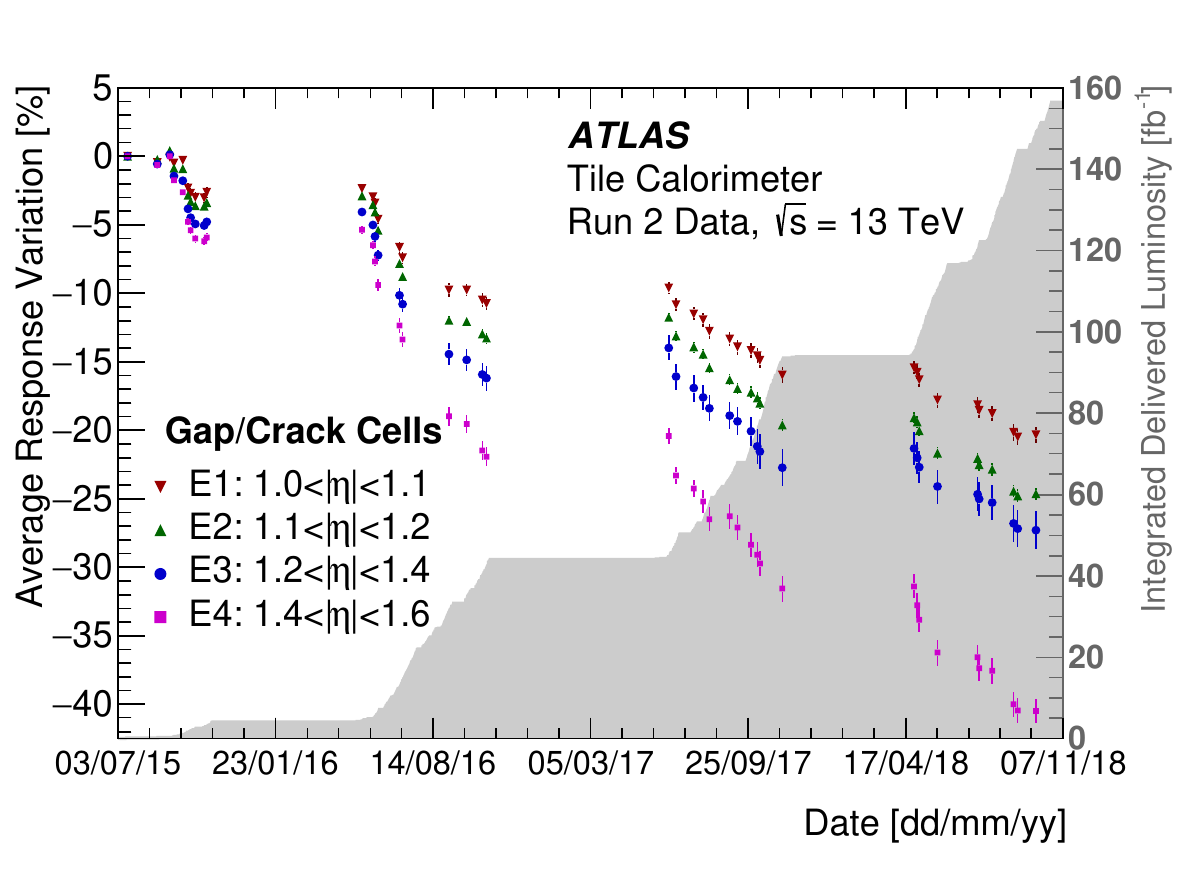}
\caption{The variation of the average response to MB events $\Delta R_{\mathrm{MB}}$ for the cells in the gap/crack region of the extended barrel as a function of time. This quantity is derived relative to the luminosity measured using the inner tracker. The error bars correspond to the RMS of all the response distributions. The results are normalised to the values measured in the first run of Run 2 (16th July 2015). The integrated luminosity delivered by the LHC is shown by the shaded area.}
\label{fig:E_minimum_bias}
\end{figure}

\subsection{Monitoring of the PMT response with the laser system}
\label{subsec:monitor_of_PMT_variation_response}

The laser system is used to monitor the variation of the PMT response~\cite{Agaras:2023vbp}.
Deviation in each channel response relative to its nominal value, $\Delta R_{\mathrm{Las}}$, is translated into a calibration constant, $C_{\mathrm{Las}}$ in Eq.~(\ref{eq:calibration}) using the equation

\begin{equation}
C_{\mathrm{Las}} = 1 + \Delta R_{\mathrm{Las}} .
\label{eq:C_Las}
\end{equation}

The laser calibration runs are usually taken daily for both gains, in the absence of colliding beams (standalone runs).
To address the fast drift of PMT response caused by the large instantaneous luminosity, the laser calibration constants were updated every 1--2 weeks.

Figure~\ref{fig:laser_map_2018} shows the mean of the $\Delta R_{\mathrm{Las}}$ values for each cell type (average over $\phi$) measured with the laser system during the entire $pp$ collisions period in 2018. The most affected cells are those located at the inner radius and in the gap and crack region with down-drift up to 4.5\% and 6\%, respectively. Those cells are the most irradiated and their readout PMTs experience the largest anode current for a given luminosity. The EB partitions experience larger current draws than the LBs due to higher exposure. Figure~\ref{fig:laser_evolution_run2} shows the time evolution of the mean response variation in the PMTs of each layer observed during the entire Run~2. The PMT response exhibits a correlation with the LHC operation, as detailed in Section~\ref{subsec:combination_calibration}. For data taken in 2015 and 2016, the laser calibration constants were calculated and applied for channels with PMT response variations larger than 1.5\% (2\%) in the LB (EB). In 2017 and 2018, the thresholds on PMT response variations were removed and all channels were corrected weekly.
The total statistical and systematic errors in the laser calibration constants depend on the integrated luminosity $\mathscr{L}$ and are found to be $\sigma/C_{\mathrm{Las}} = 0.16\%(0.32\%) \cdot \mathscr{L}(\mathrm{fb^{-1}})\oplus 0.5\%$ for the LB (EBs).

\begin{figure}[t]
\centering
\includegraphics[width=1.\linewidth]{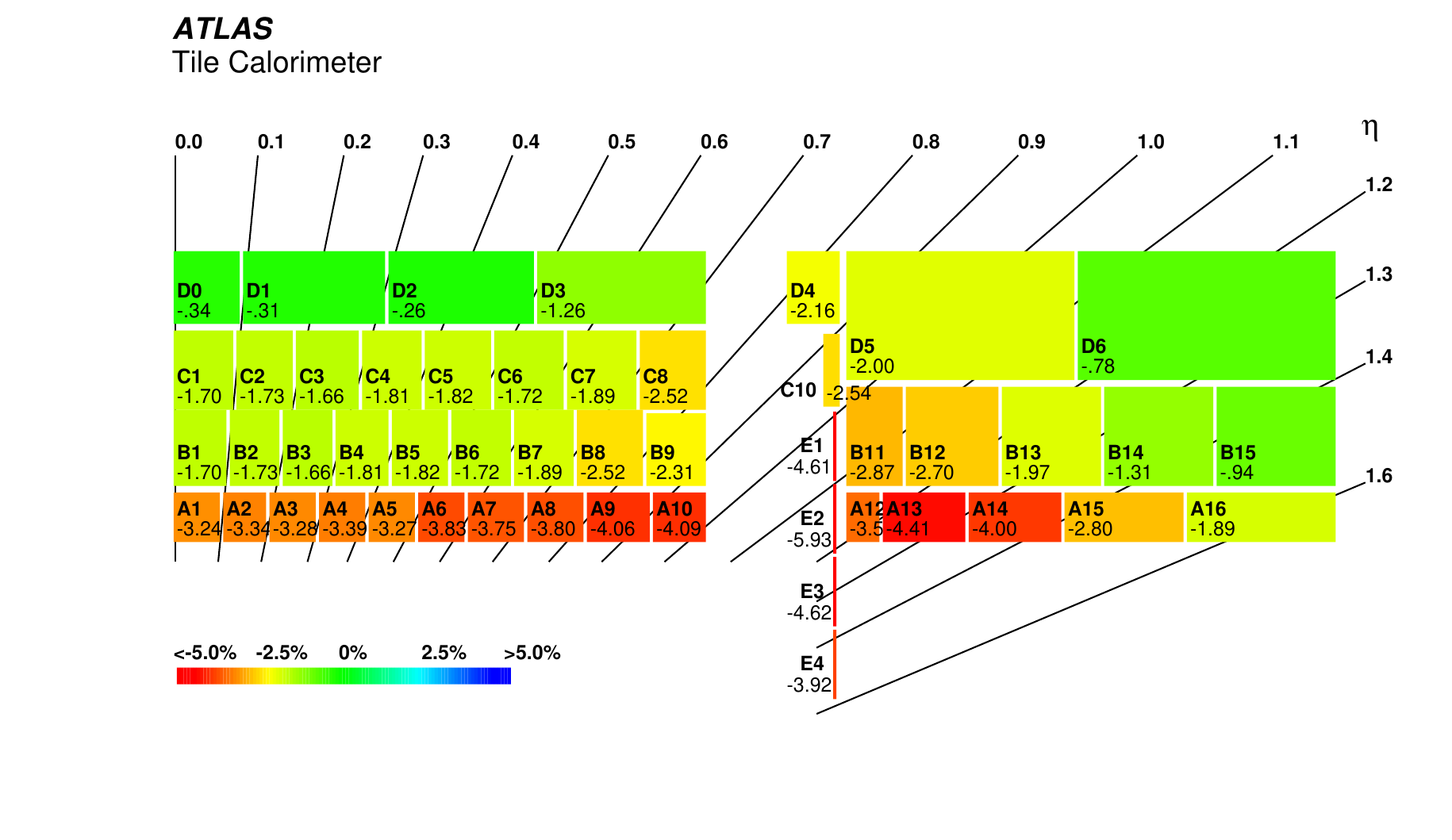}
\caption{The mean response variation $\Delta R_{\mathrm{Las}}$ in the PMTs for each cell type, in percent, averaged over $\phi$, observed during the entire $pp$ collisions data-taking period in 2018 (between standalone laser calibration runs taken on 18 April 2018 and 22 October 2018). For each cell type, the response variation is defined as the mean of a Gaussian function fit to the response variations in the channels associated with given cell type. A total of 64 modules in $\phi$ are used for each cell type, with the exclusion of known bad channels.}
\label{fig:laser_map_2018}
\end{figure}

\begin{figure}[t]
\centering
\includegraphics[width=0.7\linewidth]{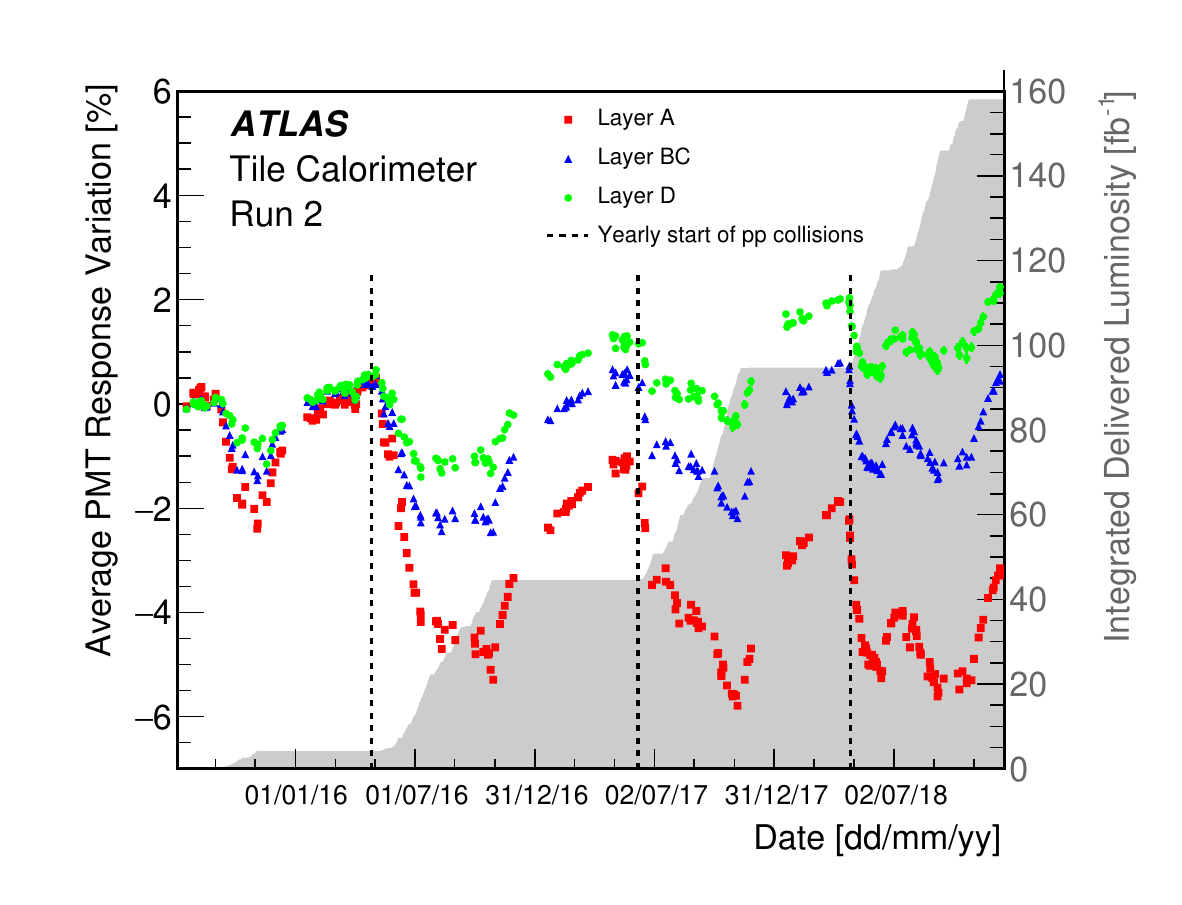}
\caption{The mean response variation $\Delta R_{\mathrm{Las}}$ in the PMTs for each layer, as a function of time, observed during the entire Run~2 (between standalone laser calibration runs taken on 17 July 2015 and 22 October 2018). For each layer, the response variation is defined as the mean of a Gaussian function fit to the response variations in the channels associated with given layer. Known bad channels are excluded. The LHC integrated delivered luminosity is shown by the shaded area.}
\label{fig:laser_evolution_run2}
\end{figure}

\subsection{The calibration procedure}
\label{subsec:summary_calibration_procedure}

During Run 2, data from the four sub-detector calibration systems CIS, caesium, integrator readout and laser, were used to correct for the response variation of the different calorimeter components. The detector calibration constants entering in Eq.~(\ref{eq:calibration}) $C_{\mathrm{ADC}\to \mathrm{pC}}$, $C_{\mathrm{Cs}}$, $C_{\mathrm{MB}}$  and $C_{\mathrm{Las}}$ are stored in a database consulted during online and offine energy reconstruction and are defined within
a time interval or a range of run numbers or a range of luminosity blocks where they are applicable.

The calibration activities started in 2015 with the equalisation of the detector calibration through HV adjustments to all PMTs and the channel timing calibration, as discussed in Sections~\ref{subsec:the_electromagnetic_energy_scale} and~\ref{subsubsec:time_calibration}, respectively. Afterwards, regular updates of the caesium, laser and CIS calibration constants in Eq.~(\ref{eq:calibration}) were performed during the data-taking period for prompt data processing. The laser system was used to calibrate the response of the PMTs and readout electronics by frequently updating the $C_{\mathrm{Las}}$ factors calculated relative to the previous caesium scan. When the rare caesium scans were performed, the values of the $C_{\mathrm{Cs}}$ constants were updated, the laser reference signals for each PMT were written to the database and the laser constants were reset to 1. This procedure permitted the overall response of the detector to be calibrated with a negative offset smaller than 1\% in absolute value
for most of the detector cells, which have not been harmed by significant radiation damage. Due to the accumulated radiation exposure, since 2016 it is not possible to correct for the response reduction of the most exposed cells with an accuracy down to 2\% by relying exclusively on the sparse caesium scans and laser measurements.

Every year, the data recorded by the ATLAS detector are reprocessed to apply finer calibrations and improved algorithms in the ATLAS reconstruction chain, as well as to rectify any missed or previously unattainable corrections during prompt data processing. In particular, in the LHC Long Shutdown 2 at the end of Run 2, the full Run 2 data set was reprocessed. For the TileCal calibration, the reprocessing campaigns also included the determination of the calibration constant $C_{\mathrm{MB}}$ appearing in Eq.~(\ref{eq:calibration}) from the analysis of the acquired collision data. The correction was introduced to address the residual miscalibration effects due to scintillator and WLS fibre degradation beyond the PMT response variation, more precisely calibrated with the laser system, as follows:

\begin{equation}
C_{\mathrm{MB}} = 1 + \langle \Delta R_{\mathrm{MB}} - \Delta R_{\mathrm{Las}} \rangle.
\label{eq:C_MB}
\end{equation}

The cell response variations to MB events, $\Delta R_{\mathrm{MB}}$, are calculated yearly relative to their nominal values at the beginning of the data-taking campaign and close to the Cs scan, when $C_{\mathrm{MB}}$ is reset to 1. These are computed as an average of the values per cell type since the optics response variations are found to be similar accross the different detector modules. They were introduced for the E-cells and a few cells mostly in layer A (A13 in 2016, A12 and A13 in 2017, and A12, A13, A14 and especially narrow C10 in 2018), which show larger response variations during LHC operation.

\subsection{Signal reconstruction of simulated events}
\label{subsec:signal_reconstruction_simulated_events}

In physics analysis, the measured energy is compared with that obtained by analysing simulated events. The ATLAS Monte Carlo (MC) simulation~\cite{SOFT-2010-01} relies on
the \textsc{Geant4} toolkit~\cite{Agostinelli:2002hh} to model the detector and the interactions of
particles with the detector material.
The MC process is divided into four steps: event generation, simulation, digitisation, and reconstruction.
Various event generators are used for the ATLAS physics analyses and performance studies, see Section~\ref{sec:performance_studies}.
During Run 2, ATLAS used the `FTFP\_BERT\_ATL'
physics model to describe the hadronic interactions with matter, where at high energies the hadron showers are modelled using the Fritiof string model.
The Bertini intra-nuclear cascade model is used for lower-energy hadrons~\cite{Bertini:1971xb}. The transition between those two models takes place in the energy region 9--12~\GeV~\cite{Geant4PhysicsLists}.

The input to the digitisation is a collection of hits in the active scintillator
material, characterised by energy, time and position. The amount
of energy deposited in the scintillator is divided by the calorimeter
sampling fraction to correct for energy deposited in the inactive material.
To obtain the channel energy at the electromagnetic scale~\cite{Budagov:962065} the sampling fraction is obtained by simulating electrons with the same kinematics features as the ones used at the TBs (Section~\ref{subsec:the_electromagnetic_energy_scale}).
In the case of a cell read out by two PMTs, the energy is shared between the two PMTs according to the distance in $r\phi$ of each hit from the edges of the cell.
Moreover, due to cell geometry and the presence of the caesium system pipes, the cell energy response has an azimuthal dependence, measured using $W\to\mu\nu$ events in the 2012 $pp$ collision data~\cite{Alberich_2018}. This dependence is implemented in the MC simulations.

To follow the same procedure used in the reconstruction of experimental events, in the digitisation step, the channel energy in \GeV{} is converted into its equivalent charge using the electromagnetic scale constant measured at TBs. The charge is subsequently translated
into the signal amplitude in ADC counts using the corresponding
calibration constant as explained in Section~\ref{subsec:monitor_conversion_factor_ADC_pC}. The amplitude is convoluted with the pulse shape and digitized every
25~ns as in real data. The effects of the electronics are emulated and added to the digitized samples as described in Section~\ref{subsubsec:electronic_noise}.
The generation of the simulated event samples includes the effect of multiple $pp$ interactions per bunch crossing, and the effect on the detector response as described in Section~\ref{subsubsec:pileup_noise}, due to bunch crossings close to the one containing the analysed interaction.


\FloatBarrier

\section{Trigger signals}
\label{sec:trigger_signals}

A two-level trigger system~\cite{TRIG-2019-04} was used by ATLAS in Run~2 to reduce the event rate from a maximum raw rate of 40~MHz to 1~kHz, which is written to disk. The Level-1 trigger is implemented in hardware and uses a subset of calorimeter and muon sub-detector information to accept events at a rate below \SI{100}{\kHz}. The data are searched for signatures such as large energy deposits (Level-1 calorimeter trigger) or high-pT muon tracks (Level-1 muon trigger). This is followed by a software-based high-level trigger (HLT), implemented in a large dedicated computer farm adjacent to the cavern, which reduces the accepted event rate to \SI{1}{\kHz} on average depending on the data-taking conditions. The TileCal data subsets used in the L1 trigger are discussed in the next sections. As discussed in Section~\ref{subsec:mbts_intro}, information collected by the TileCal is also used to select events during low luminosity LHC runs.

\subsection{Level-1 calorimeter trigger}
\label{subsec:level-1_calorimeter_trigger_tileCal_information}

The Level-1 calorimeter trigger exploits the tower structure in the $\eta-\phi$ plane of the TileCal (see Figure~\ref{fig:calorimeter_d3_b}) and the LAr calorimeter. The TileCal information used in L1 comes from the adder boards, referred to in Section~\ref{subsec:the_tilecal_readout}. These boards sum the 3-in-1 card low-gain analogue signals of the six channels connected to cells with the same $\eta$ and $\phi$ in the layers A, BC and D of the LB and in the layers A, B and D of the EB. %

\subsection{Tile muon trigger}
\label{subsec:tile_muon_trigger_system}

The tile muon trigger uses information from the TileCal outermost layer cells (D cells) of the EB
to reduce the rate of the ATLAS L1 muon trigger~\cite{PERF-2007-01} background events due to low momentum protons emerging from the endcap toroid and beam shielding. Figure~\ref{detector_cut} shows that muons arising from $pp$ collisions with $1.0<|\eta|<1.3$, cross the D5 and D6 cells of the TileCal extended barrel before reaching the endcap muon chambers (TGC)~\cite{PERF-2007-01}. This is not the case for background events. The requirement of a coincidence between the TileCal and the muon sub-detector signals reduces the background rate.

\begin{figure}[t]
\centering
\includegraphics[width=0.8\textwidth]{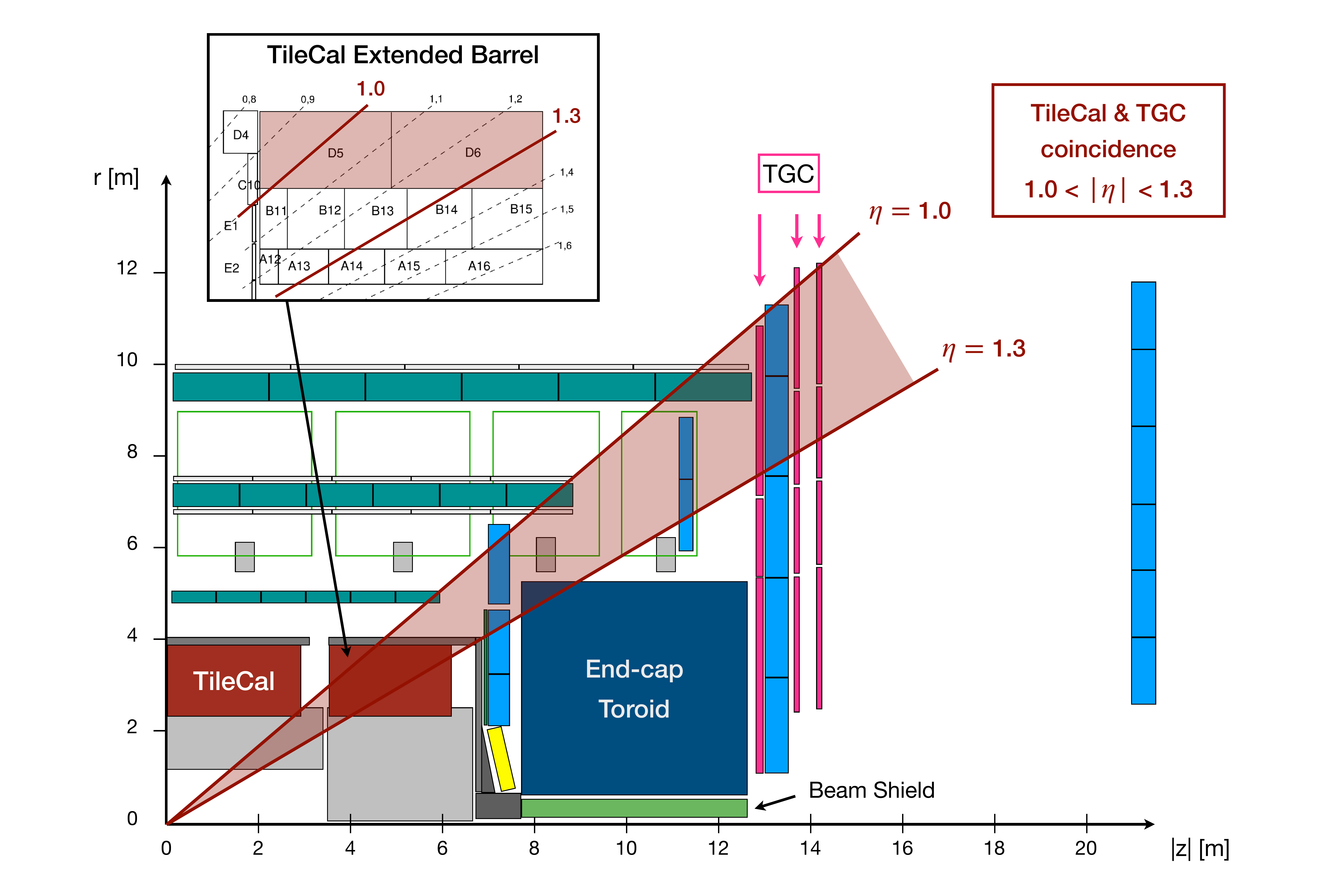}
\caption{Schematic of a partial longitudinal cut of the ATLAS detector showing the $\eta$ region covered by the tile muon trigger system. The azimuth angle coverage is $0 \le \phi <2\pi$. The sub-systems not used in the tile muon trigger are sketched for completeness and their drawings do not represent the accurate geometry.}
\label{detector_cut}
\end{figure}

The tile muon trigger is based on the information of $2\times 64$ TileCal azimuth sectors, each corresponding to one EB module, and $2\times 24$ TGC chamber sectors.  To provide the TileCal and TGC sector matching, a tile muon digitiser board (TMDB) is required to process the D5 and D6 signals from eight TileCal modules and interface with three Level-1 TGC sector logic blocks~\cite{Ryzhov:2016oib,Ciodaro:2014swa} (Figure~\ref{tileMuon_overview}).
To cover the entire detector, 16 TMDBs housed in a VME 9U crate in the ATLAS service cavern are required.

For each TileCal sector, the TMDB provides four energy values, one for each of the two PMT’s of the D5 and  D6 cells.
The analogue signals are digitized at a frequency of 40~MHz and a window of seven samples in time is used to represent the readout pulses. The method adopted for energy estimation is based on the Matched Filter (MF)~\cite{Peralva:2021jsy}, which has successfully been tested for the TileCal signals~\cite{BPeralva}. Unlike the OF method discussed in Section~\ref{subsec:digital_signal}, the MF approach aims at maximizing the signal-to-noise ratio for muon signals while increasing the detection efficiency. For the TMDB, the MF employed for each channel is implemented by a 7-coefficient digital filter where the signal energies are estimated by performing an inner product between the MF coefficients and the incoming time samples in ADC counts. The presence of a signal produced by a muon in the D5, in the D6, and in the D5 and D6 cells is obtained by comparing the four energy determinations with the corresponding threshold values. The MF coefficients and threshold values are loaded in the TMDB\@. The board transmits the $\eta$ and $\phi$ coordinates of the identified cells to the three TGC logic blocks through three Gigabit links (GLink). The information is also transmitted to neighbour receiver boards to accommodate the non-perfect matching between the eight TileCal modules and the three muon sector logic blocks. Each TGC sector Logic board performs the coincidence. The performance of the tile muon trigger systems is discussed in Section~\ref{subsubsec:performance_of_the_tilemuon_trigger}.

\begin{figure}[t]
\centering
\includegraphics[width=0.5\textwidth]{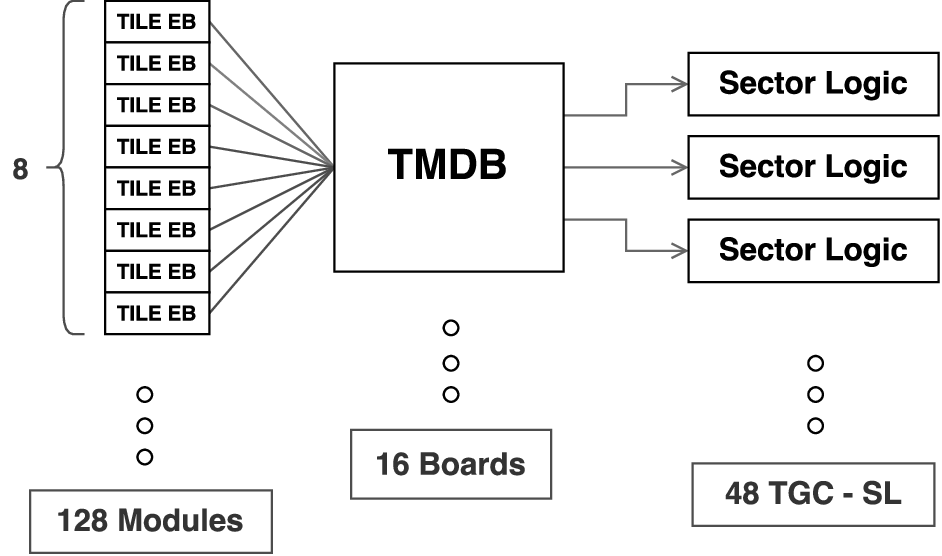}
\caption{The basic architecture of the tile muon trigger system.
}
\label{tileMuon_overview}
\end{figure}

\subsection{Minimum-bias trigger scintillators}
\label{subsec:mbts_intro}

During low luminosity LHC runs, the MBTS system allowed the online selection of events with the highest efficiency and the lowest possible bias~\cite{Sidoti_2014}. %
The system consists of 2~cm thick polystyrene scintillator counters made by PS-PTP-POPOP polyethylene located on both A- and C-sides of the ATLAS detector at a distance of 3.6~m from the interaction point.
Each side is composed of eight inner counters ($2.76 \leq |\eta|< 3.86$) and eight outer counters ($2.08\leq |\eta|< 2.76$) instrumented with WLS fibres embedded in grooves at the edges of the counters. The light going through the fibres is directed to PMTs~\cite{Abdallah:2009zza} where the signal readout is done using the same electronics as the TileCal cells (Section~\ref{subsec:the_tilecal_readout}).
The signals from the shaper output (Section~\ref{subsec:the_tilecal_readout}) are fed into leading edge discriminators and sent as 25~ns NIM pulses to the Central Trigger Processor~\cite{Ask:2008zz}.
As shown in Figure~\ref{fig:mbts_layout}, in the period 2015--2017 there are twelve readout channels in total in each side. Each inner counter is connected to one PMT and adjacent pairs of outer counters are read by a single PMT grouping the corresponding WLS fibres.

Due to the high irradiation dose, the performance of the scintillators, fibres and PMTs of the MBTS degraded throughout Run~2 and a large signal reduction was observed. This is discussed in detail in Section~\ref{subsec:degradation_mbts}. The MBTS efficiencies, observed at the beginning of Run 2 and after two years of LHC operations, are reported in Section~\ref{subsec:efficiency_mbts}. Due to the very large degradation of the response, the MBTS counters, installed before the start of Run 2,  were replaced at the beginning of Run 3.

\begin{figure}[t]
\begin{center}
\includegraphics[width=0.5\linewidth]{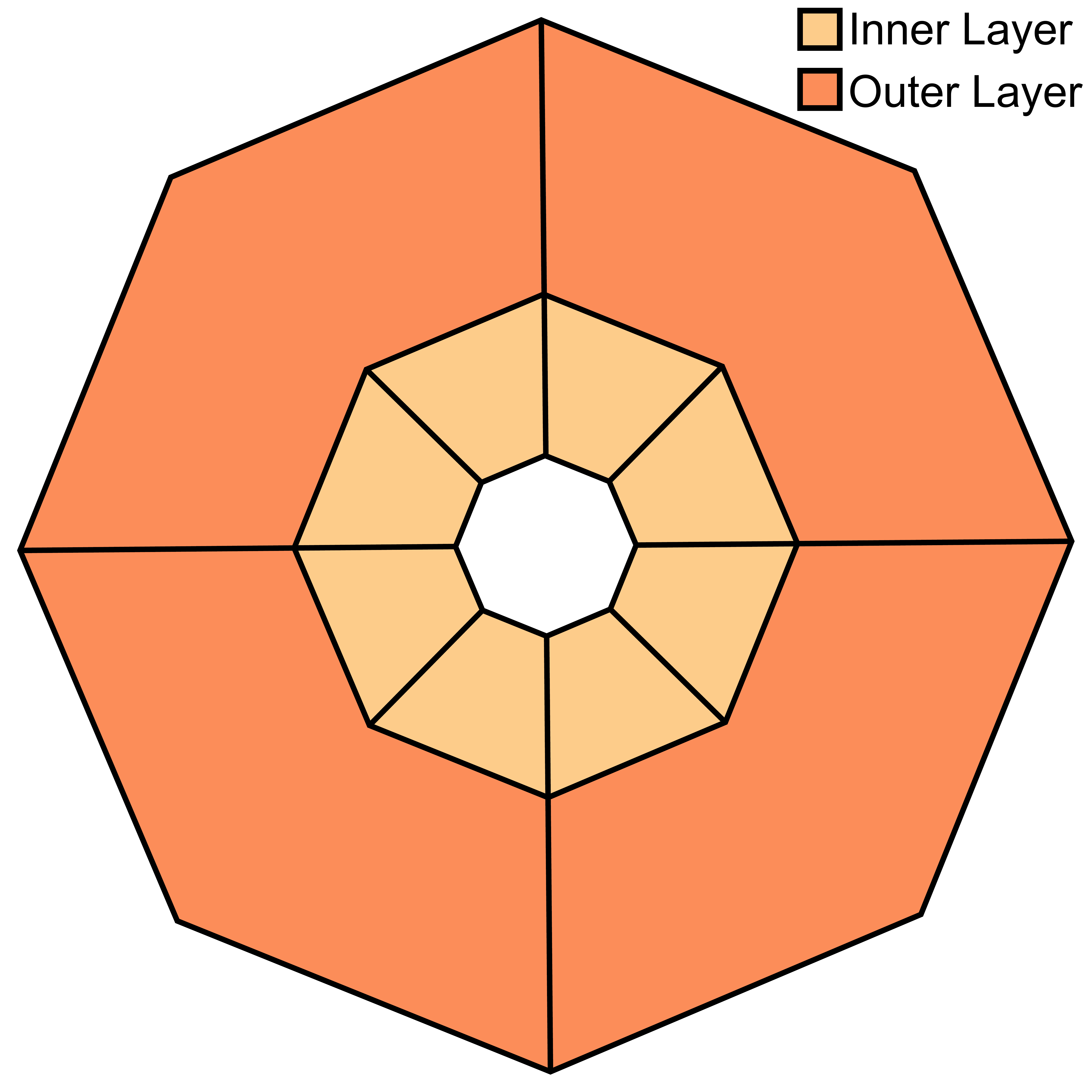}
\end{center}
\caption{The layout of one side of the MBTS sub-detector in the period 2015--2017. The energy deposited in each sensitive region is read by one readout channel. Due to the large reduction of the scintillator response, the number of sensitive regions in the outer layer was doubled in 2018.}
\label{fig:mbts_layout}
\end{figure}


\FloatBarrier

\section{Radiation exposure effects}
\label{sec:detector_ageing}

The LHC operation affects the performance of the TileCal and of the MBTS\@. Prolonged exposure to intense radiation causes a decrease in the optical transmission of the scintillator tiles and of the WLS fibres, inducing a decrease in pulse height which may result in a deterioration of the energy resolution of the detector. Simulation studies show that a 50\% reduction of the cell light yield (from the typical 70 photoelectrons/\GeV) can be acommodated without significant impact on the energy resolution for 20 and 160~\GeV muons~\cite{ATLAS-TDR-28}. On the other hand, the accumulated charge affects the gain of the PMTs (see Section~\ref{subsec:combination_calibration}). The laser measurements allow the effects of the PMTs on the calorimeter response to be determined while the caesium system and the MB events probe the effects of PMTs, tiles and fibres, as described in Section~\ref{sec:physics_signals}. Combining their information allows to study the degradation of the optical system (tiles and fibres) of the TileCal, presented in Sections~\ref{subsec:ageing_scintillating_tiles_and_wls_fibres}~and~\ref{subsec:degradation_mbts}.

\subsection{Comparison of laser, caesium and minimum-bias measurements}
\label{subsec:combination_calibration}

The effects of the LHC conditions on the cell response during Run 2 are discussed in this section. Figure~\ref{fig:combined_calibration} shows
the relative response variations of the A13 calorimeter cells to caesium, MB events and laser pulses, respectively $\Delta R_{\mathrm{Cs}}$, $\Delta R_{\mathrm{MB}}$ and $\Delta R_{\mathrm{Las}}$, as a function of time in Run~2. The assessment of these quantities is detailed in Sections~\ref{subsec:monitor_PMT_tiles_fibres_variation_response_Cs} and \ref{subsec:monitor_PMT_tiles_fibres_variation_response_MB} and Section~\ref{subsec:monitor_of_PMT_variation_response}. The cell A13 is located in the EB, and due to the smaller amount of upstream material, it is exposed to one of the highest radiation doses as shown in Figure~\ref{fig:tilecal_cells_and_rows2}.

\begin{figure}[t]
\centering
\includegraphics[width=0.7\textwidth]{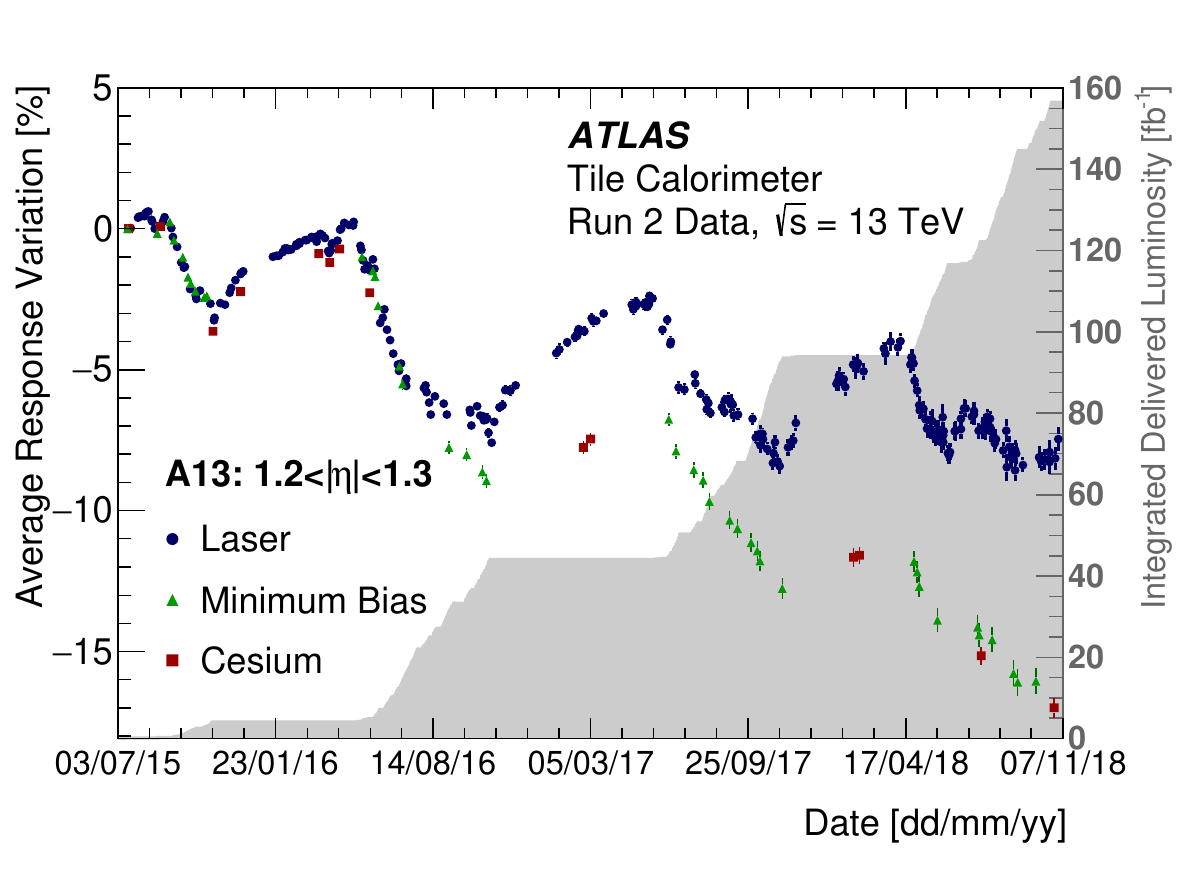}
\caption{
The average response variation of A13 cells measured by the laser, caesium and minimum-bias integrator systems as a function of time during the entire Run~2.
Known bad channels are excluded. As discussed in the text the response behaviours are connected to the LHC delivered luminosity shown by the shaded area.
}
\label{fig:combined_calibration}
\end{figure}

The PMT gain, monitored with the laser system, is known to decrease with increasing light exposure due to lower secondary emissions from the dynode surfaces~\cite{GUPTA1967352}. The decrease in gain depends on several factors, including temperature, intensity and duration of light exposure, as well as the previous history of the PMT\@. When a PMT is initially exposed to light after a long `no light' period, its gain decreases rapidly and then slow stabilisation occurs~\cite{Hillert_1951}. This behaviour is demonstrated in Figures~\ref{fig:laser_evolution_run2} and~\ref{fig:combined_calibration}. The data-taking in each year started after periods of inactivity. The recovery periods, in which the PMT response to the laser tends towards the initial conditions, coincide with the periods in which the LHC was not colliding protons. This is consistent with the known behaviour of fatigued PMTs that gradually return to their original operating condition after the exposure is removed~\cite{WEITKAMP1968122}. A global PMT gain increase of 0.9\% per year is observed without any exposure (e.g. between data-taking periods).

As already discussed, the responses to the caesium system and MB events are sensitive to both the PMT gain changes and the scintillator/fibre degradation.
The transparency of these systems is reduced after radiation exposure~\cite{ZORN1993377}. In the TileCal this is evident in the continued downward response to caesium (or MB events) with increasing integrated luminosity of the collisions, despite the eventual slow recovery of the PMTs as described above. In the absence of the radiation source the annealing process is believed to slowly restore the scintillator material, hence improving the collected light yield. This can be seen in Figures~\ref{fig:cesium_driftL},~\ref{fig:E_minimum_bias} and~\ref{fig:combined_calibration}. The rate and amount of scintillator damage and recovery are complex combinations of factors, such as particle energies, temperatures, exposure rates and duration, and are difficult to quantify.

To disentangle the effects of PMT and scintillator changes one can study the laser compared to MB or caesium responses. It can be seen in Figure~\ref{fig:combined_calibration}, that the responses to caesium, laser and MB integrator systems show similar behaviour in 2015. This indicates no effect of scintillator/fibre irradiation. In 2016--2018, when the LHC delivered significantly larger luminosity, the response to MB (and caesium) events compared to the laser system diverges, indicating the
degradation of the scintillators and the fibers. At the end of Run~2, the maximum response loss in A13 is approximately 16\%, where around 8\% is associated with PMT response loss and the remaining 8\% with scintillator degradation.

\subsection{Degradation of scintillator tiles and WLS fibres}
\label{subsec:ageing_scintillating_tiles_and_wls_fibres}

Scintillator tile and WLS fibre ageing effects were determined during Run~2 using the signals produced by the $^{137}$Cs source, MB collision events and the laser pulses. The relative light yield $I/I_0$ of the cells due to scintillator and fibre degradation is obtained by factoring out the PMT response fluctuations through the ratio between Cs/MB and Laser relative responses $R_{\mathrm{Cs/MB/Las}}$ as:

\begin{equation} \label{eq:I/I_0}
\centering
I/I_0 = \frac{R_{\mathrm{Cs/MB}}}{R_{\mathrm{Las}}}.
\end{equation}

\begin{figure}[t]
\centering
\includegraphics[width=1.\textwidth]{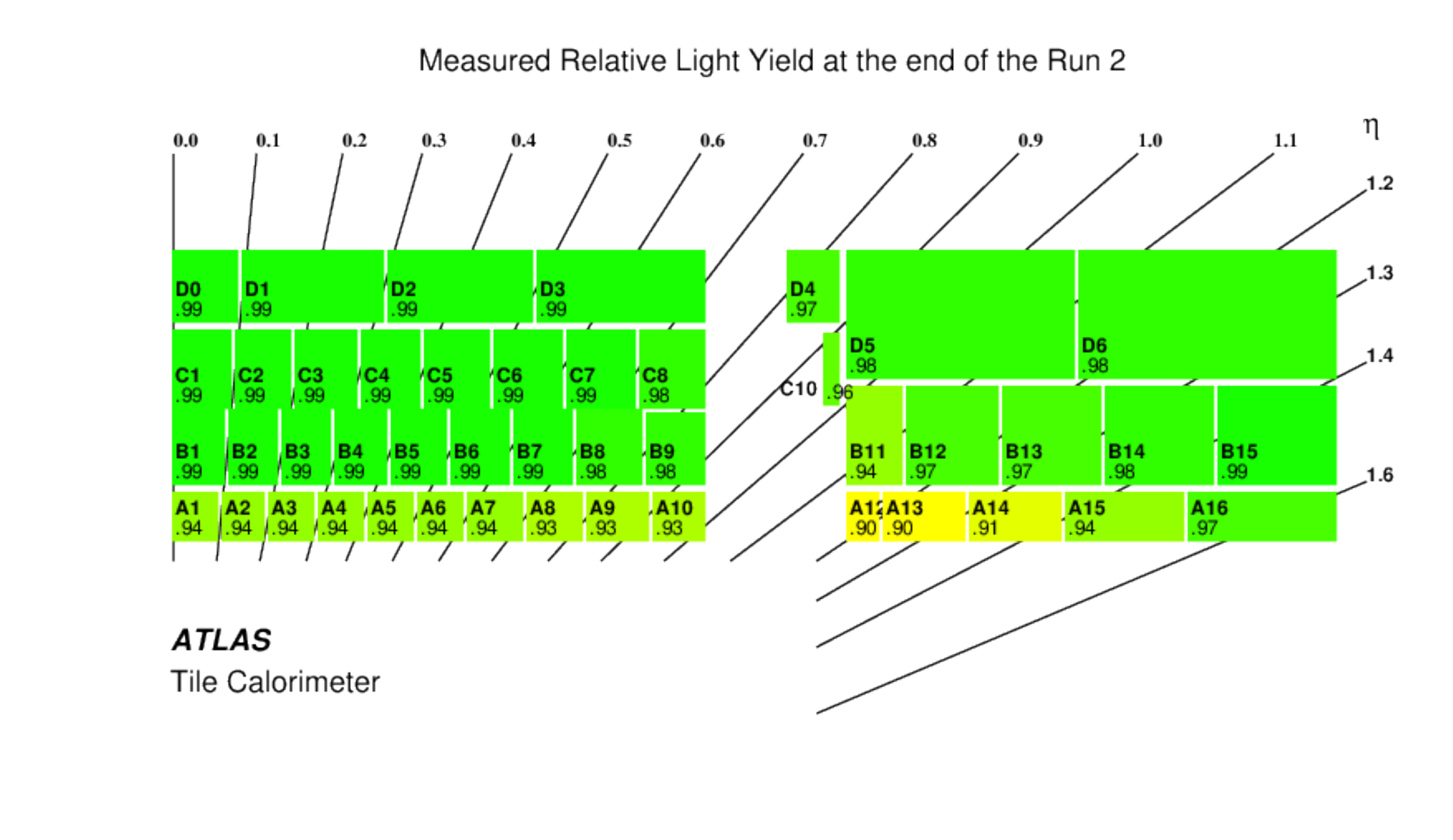}
\caption{\label{fig:II0_Run2} The measured relative light yield $I/I_0$ (Eq.~(\ref{eq:I/I_0})) of the TileCal cells at the end of Run~2. The uncertainty is of the order of 1\%.}
\end{figure}

Figure~\ref{fig:II0_Run2} shows the relative light yield measured at the end of Run~2 averaged over all TileCal barrel cells~\cite{BPereira}. The innermost layer A and the B11 and C10 cells have a light yield loss between 4 and 10\%. The uncertainty is around 1\% and the measurements are not yet sensitive to light yield degradation of B/BC and D cells. The observed light response degradation is more important in the innermost cells, subjected to larger particle fluence and thus more ionising dose. The ionisation doses deposited in the different calorimeter cells are obtained from simulation of $pp$ collisions at a centre-of-mass energy of $\sqrt{s}=13$~\TeV\ using the \textsc{Pythia~8} event generator and \textsc{Geant4}~\cite{Pedro:2019lgd,DoseSimulation}. Figure~\ref{fig:Tile_TID_prelim} shows the simulated ionising dose corresponding to an integrated luminosity of 1~fb$^{-1}$. The largest doses, up to 20 Gy/fb$^{-1}$, occur in the E cells. The most irradiated cells in the calorimeter barrel are the A12 and A13 cells, with doses of 0.5 and 0.4~Gy/fb$^{-1}$, respectively.

\begin{figure}[t]
\centering
\includegraphics[width=1.\textwidth]{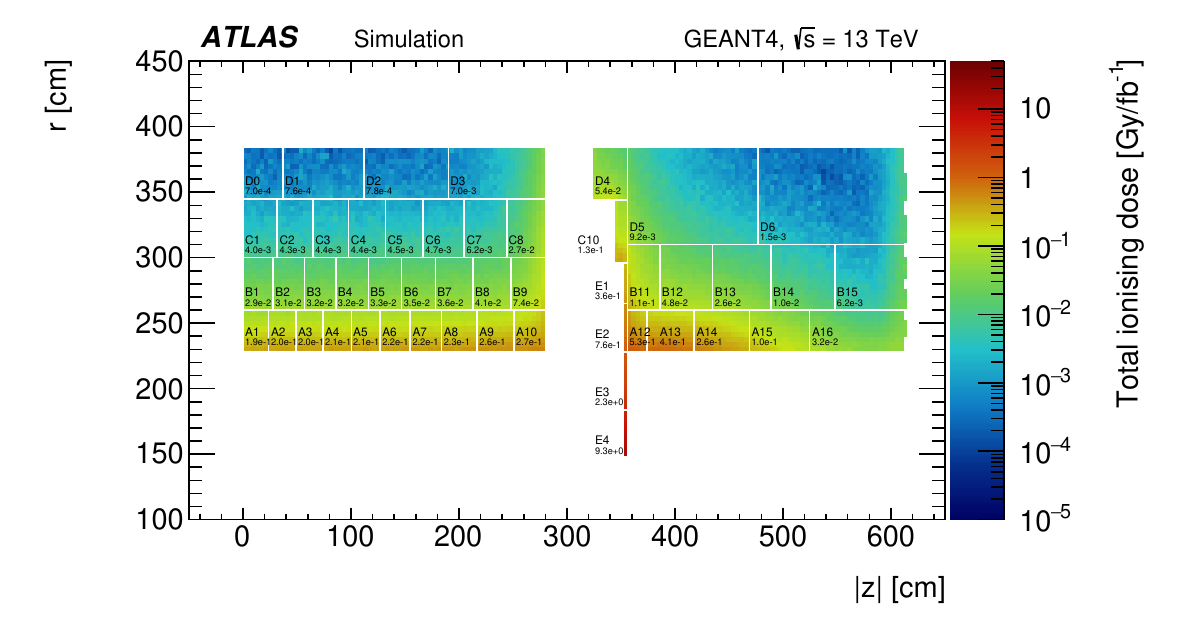}
\caption{\label{fig:Tile_TID_prelim}
Simulated ionisation dose deposited in the scintillator tiles of the cells and in the gap/crack scintillators in $4\times 4$~cm$^2$ bins in $r\times z$. The study was performed using 50~000 inelastic $pp$ collisions at $\sqrt{s}=13$~\TeV\ generated with \textsc{Pythia~8}. The response of the detector was obtained using the simulation program \textsc{Geant4}. The results are normalised to a cross-section of $\sigma_{\mathrm{inel}}$ = 78.42~mb and an integrated luminosity of 1~fb$^{-1}$~\cite{Pedro:2019lgd,DoseSimulation}.}
\end{figure}

To further investigate the effects of radiation exposure on the performance of the TileCal active material, $I/I_0$ was monitored during Run~2. The study allows the light yield till the end of LHC operations and during the HL-LHC data-taking period to be estimated. Figure~\ref{fig:II0_A13_log} shows these measurements as a function of the integrated delivered luminosity and the average simulated dose deposited for A13 cells. The Run~2 integrated dose is obtained by normalising the cell averaged dose coefficients presented in Figure~\ref{fig:Tile_TID_prelim} to the measured integrated luminosity delivered by the LHC~\cite{DAPR-2021-01}. The loss of light yield is quite smooth and increases with exposure to radiation. In the study, the damage as a function of dose $d$ is modelled by an exponential function:

\begin{equation}
I/I_0=p_0e^{-d/p_1}
\label{eq: degradation}
\end{equation}

The parameters $p_0$ and $p_1$ are obtained by minimising the $\chi^2$ function considering only the statistical uncertainty in $I/I_0$. The systematic uncertainties in the caesium, MB and laser calibration systems (around 0.5\%, 1.2\% and 0.5\%) are propagated to the $I/I_0$ uncertainty. This uncertainty is represented by the opaque band, which additionally covers the spread of the dose within the large cell volume (around 50\%). Based on the fit result, the extrapolated light output of the A13 cells is
$75^{+9}_{-25}\%$
at the end of the LHC operation (dose around 180~Gy).

\begin{figure}[t]
\centering
\includegraphics[width=0.7\textwidth]{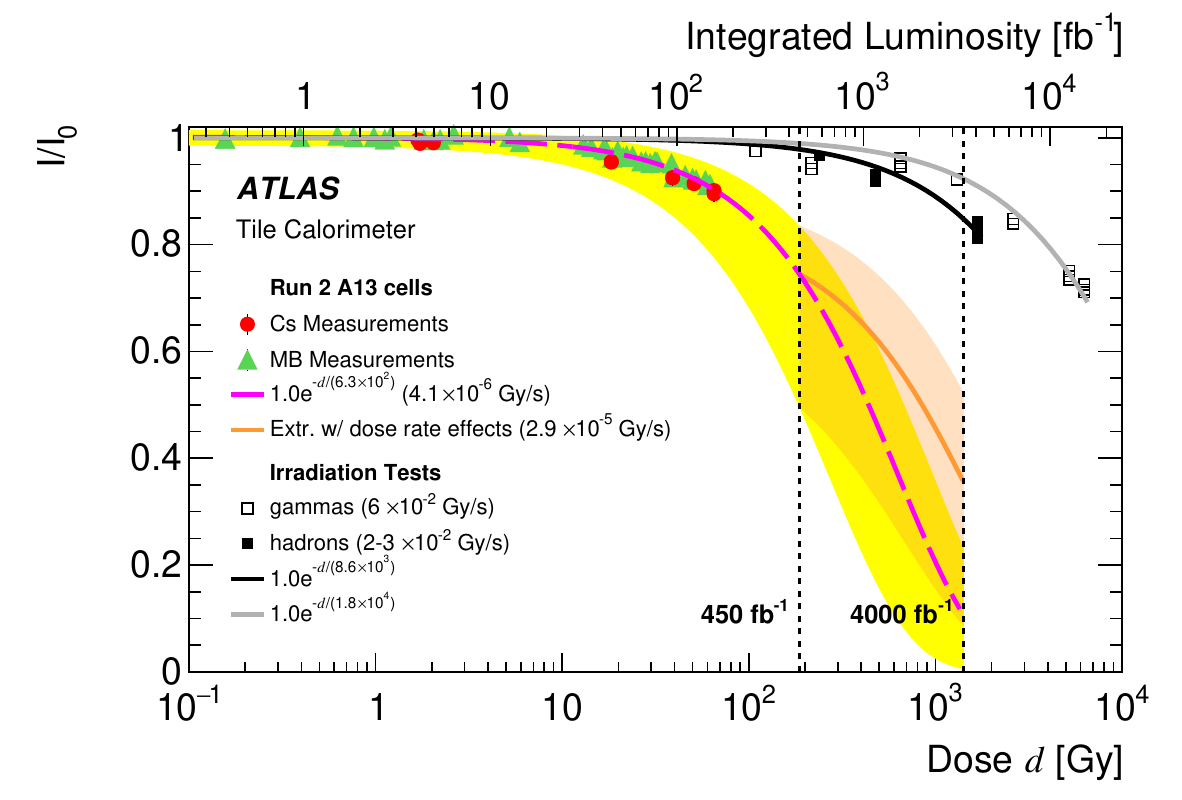}
\caption{\label{fig:II0_A13_log}
Average relative light yield ($I/I_0$) measurements based on the caesium system (dots) and integrated minimum-bias currents (triangles) for A13 cells as a function of average simulated dose $d$ and LHC integrated luminosity. The dashed curve corresponds to the fit to the function in Eq.~(\ref{eq: degradation}) to the data. The surrounding opaque band represents the total uncertainty in the fit including the RMS of the dose distribution within the cell and systematic uncertainties in $I/I_0$ due to the intrinsic precision of the caesium, MB and laser measurements. The solid curve represents the expected average $I/I_0$ of the A13 cells in the HL-LHC phase including dose rate effects (see the text). The surrounding semi-transparent band is the total uncertainty on this extrapolation, obtained by propagating the uncertainty sources of the study. Results from measurements of bare scintillators performed one month after irradiations with much higher dose rate made in the laboratory before the detector construction are also shown~\cite{Abdallah:2007cca}. An exponential function is fitted to the data obtained from irradiations with $\gamma$s (open squares) and hadrons (full squares). Dashed vertical lines represent the expected dose by the end of the LHC (450~fb$^{-1}$) and HL-LHC (4000~fb$^{-1}$)~\cite{ZurbanoFernandez:2020cco}.
}
\end{figure}

\begin{figure}[t]
\centering
\includegraphics[width=0.7\textwidth]{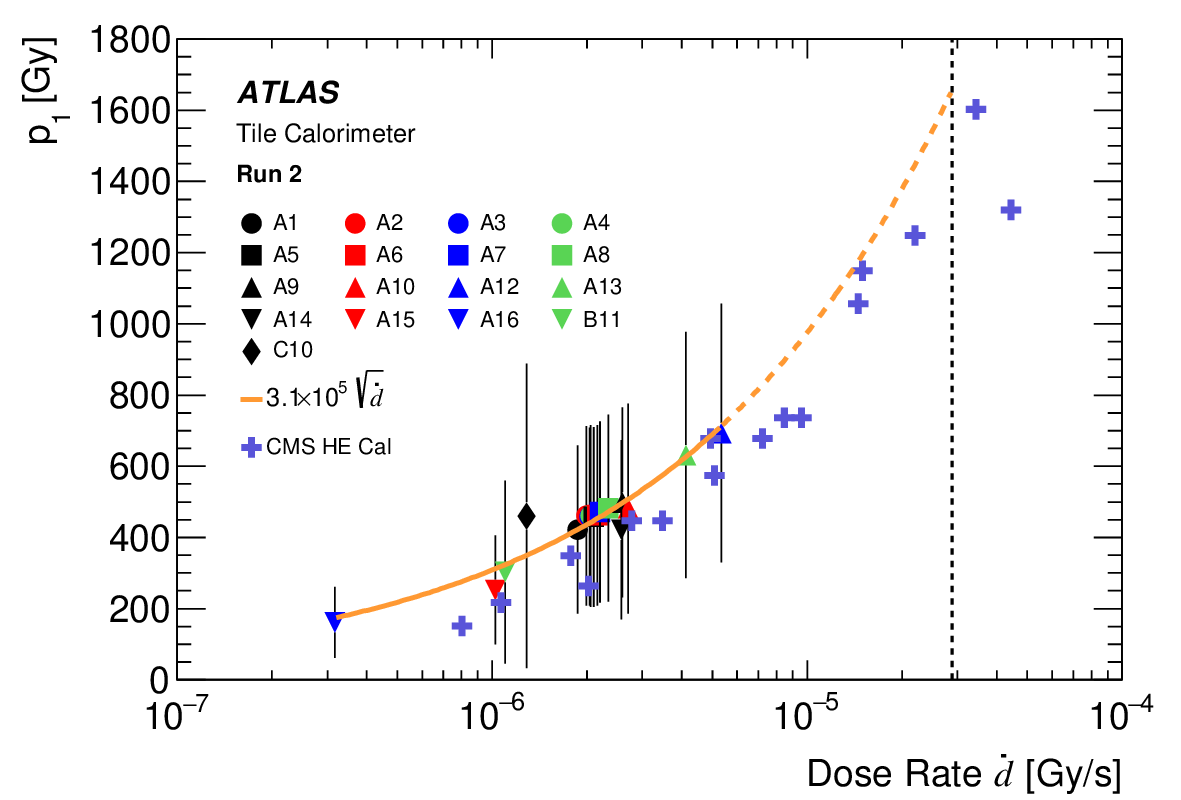}
\caption{\label{fig:dose_rate}
The degradation rate parameter $p_1$ obtained from the simple exponential model ($I/I_0=p_0e^{-d/p_1}$) as a function of the average simulated dose rate $\dot d$ for the most exposed cells. Results from a similar study performed using the CMS Hadron Endcap Calorimeter measurements are also displayed (crosses)~\cite{CMSHCAL:2016dvd}. The vertical error bars on the TileCal data points represent the total uncertainty including the $I/I_0$ measurement uncertainty and the dose spread within the cell volume. The nominal points are fitted with a power law function (continuous curve). This function is extrapolated to the higher dose rate region (dashed curve) expected at the HL-LHC phase and populated by the CMS data. The dashed vertical line indicates the expected dose rate of the A13 cells in the HL-LHC.}
\end{figure}

While it is important to carefully monitor the performance of the TileCal active material and make predictions well ahead of time, doing such an extrapolation with current data is affected by uncertainties not taken into account in this study. For instance, no explicit recovery is incorporated in the degradation model for long shutdown times. Furthermore, dose rate conditions may affect significantly the degradation rate. Measurements from the CMS-HCAL Collaboration~\cite{CMSHCAL:2016dvd} conclude that smaller dose rates produce larger degradation rates. In this scenario, the degradation rate, with respect to dose, of TileCal cells would substantially decrease in the HL-LHC phase given the increase of collision (dose) rate. To estimate the decrease of the degradation effects, fits of the function in Eq.~(\ref{eq: degradation}) are performed to the relative light yield determinations as a function of the simulated dose for the TileCal cells exhibiting the largest degradation, i.e. those in the A layer, B11 and C10, see Figure~\ref{fig:II0_Run2}. The obtained fitted parameters, driving the degradation rate, $p_1$, are represented as a function of the average dose rates of the cells in Figure~\ref{fig:dose_rate}. The average dose rates correspond to the cell averaged dose coefficients presented in Figure~\ref{fig:Tile_TID_prelim} multiplied by the LHC average instantaneous luminosity in Run 2. The results obtained by CMS for the first and seventh HCAL layer featuring Polystyrene-based scintillators from Kuraray (SCSN-81)~\cite{CMSHCAL:2016dvd} are also shown in the figure and, despite differences in calorimeter arrangement and scintillator material (TileCal uses Polystyrene+(1.5\%)pTP+(0.04\%)POPOP~\cite{Abdallah:2009zza}), the agreement with TileCal is good. The TileCal data are fitted with a power law function to model the dose rate dependency of the cell ageing allowing an extrapolation to the HL-LHC that takes into account an expected dose rate being seven times larger (dashed vertical line). The obtained $p_1$ value for the A13 cells in the HL-LHC phase is 1724~Gy. This extrapolation is also drawn in Figure~\ref{fig:II0_A13_log} (solid curve). According to this study, one expects for the A13 cells a normalised light response of
$35^{+17}_{-27}\%$
at the end of the HL-LHC runs, approximately two times larger than the value obtained without taking into account the different dose rate expected at HL-LHC (dashed curve).
In the figure, the extrapolated curve can also be compared with the results of laboratory irradiations~\cite{Abdallah:2007cca} where the dose rates were enormously higher and the scintillators were measured one month after irradiations, having recovered most of the suffered damage. Also the results obtained by analysing the MBTS data discussed in Section~\ref{subsec:degradation_mbts} highlight the reduction of the response degradation by increasing the dose rate.

\subsection{Degradation of the MBTS system}
\label{subsec:degradation_mbts}

The counters of the MBTS system are read out using the same electronics as the TileCal. Due to high irradiation dose, scintillator, fibre and PMT performances degraded throughout Run~2 and large MBTS signal reductions were observed. The average total ionising doses, estimated by using the method described in Section~\ref{subsec:ageing_scintillating_tiles_and_wls_fibres}, are $0.62\times 10^3$~Gy/fb$^{-1}$ for inner counters and $0.83\times 10^2$~Gy/fb$^{-1}$ for outer counters~\cite{Pedro:2019lgd, DoseSimulation}.
The degradation of the MBTS system is determined using laser and MB data collected during $pp$ collision runs. The PMTs were operated at 500~V. The MBTS response is obtained by normalising the measured MB currents to the ATLAS instantaneous luminosity~\cite{DAPR-2021-01}.

Figure~\ref{fig:mbts-lumi} shows the response variation of the inner and outer counters relative to the first run of Run~2, as a function of time.
The relative variations of the PMT responses as measured by the laser system are also shown.

\begin{figure}[t]
\centering
\subfloat[]{\includegraphics[width=0.5\textwidth]{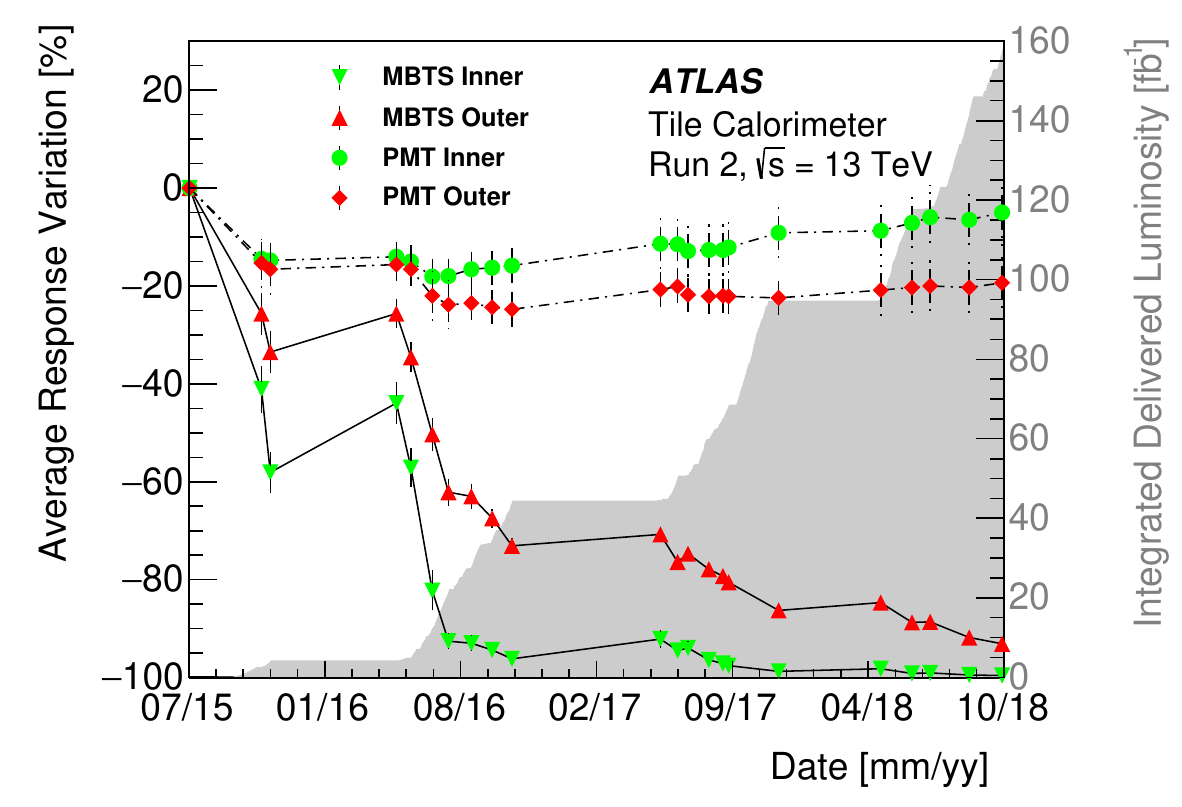}\label{fig:mbts-lumi}}
\subfloat[]{\includegraphics[width=0.5\textwidth]{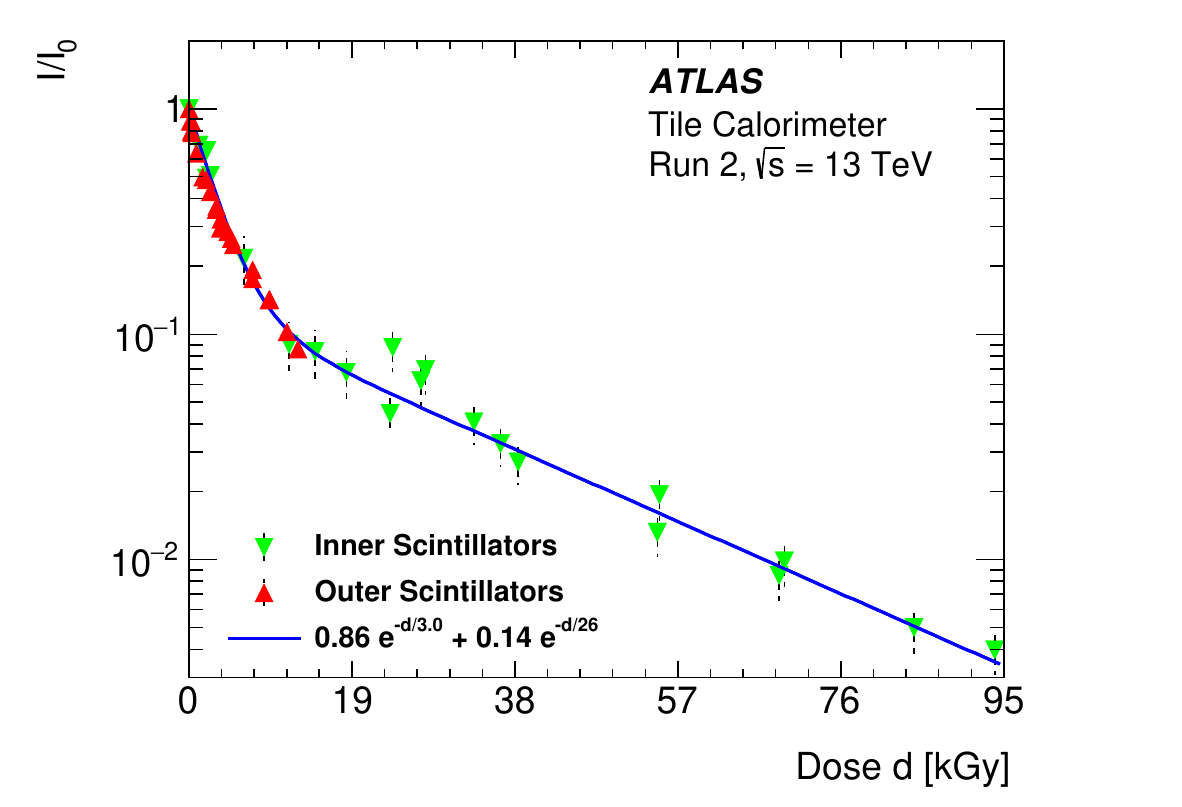}\label{fig:mbts-dose}}
\caption{(a) The average response variation of the MBTS inner (down triangles) and outer (up triangles) counters as a function of time during Run~2. The circle (diamond) markers show the relative response of the PMTs of the inner (outer) counters. The LHC delivered luminosity is shown by the shaded area. (b) The average relative light yield ($I/I_0$) of scintillators and fibres of inner (down triangles) and outer (up triangles) counters as a function of total ionising dose during Run~2. The values are the averages of the corresponding determinations obtained for the inner and outer counters. The uncertainties correspond to the RMS of the counter response distributions. The function obtained by fitting the inner MBTS data points is also shown (solid curve).}
\label{fig:mbts-lumi-dose}
\end{figure}

At the beginning of Run~2, the MB PMT currents were at the level of 15 and 10~nA/10$^{30}$s$^{-1}$cm$^{-2}$ for the inner and outer counters respectively. By end of 2015 (3.2~\ifb{} of integrated luminosity), the inner (outer) counters have lost almost 55\% (35\%) of their response. This difference is due to the larger irradiation density of the inner counters, being closer to the beam line. At the beginning of each new year's data-taking period, a relative recovery of the response of about 30\% is observed. This is due to the technical stop periods where the scintillators and fibres are not irradiated further and partially recover some of their efficiency.

Laser data allowed the PMT variations over the Run~2 period to be evaluated. The inner and outer counters show a rapid decrease of the average PMT response by about 20\% at the beginning of Run~2. This degradation is expected as the PMTs are receiving high light output from the scintillators at that time, inducing a large anode current and a significant amount of charge being integrated. After this rapid decrease, the responses of the PMTs become more stable and a general up-drift can be noticed. This behaviour is understood by the rapid degradation of the MBTS scintillators/fibres that results in much less light received by the PMTs. Consequently the integrated PMT anode currents decrease causing the down-drift to cease and eventually to reverse its course.

To measure the light yield degradation of the MBTS scintillators and fibres, PMT variations are factored out from MB data as expressed in Eq.~(\ref{eq:I/I_0}). The results as a function of the total ionising dose are shown in Figure~\ref{fig:mbts-dose}. The decline appears to follow an exponential decay curve up to 15~kGy with similar degradation rate for inner and outer counters. With the increase of the doses, the inner counters continue to decrease exponentially with a slowing rate caused presumably by saturation effects in the scintillating material.


\FloatBarrier

\section{Data quality analysis and operation}
\label{sec:data_quality_analysis_operation}

A collection of tools continuously monitors the detector hardware and
data acquisition systems during their operation. Some are fully automated
to effectively address problems in real time (online) to protect
hardware and scrutinise the quality of the output data. In the following, the Detector Control System (DCS), Section~\ref{subsec:atlas_detector_control_system}, and the use of calibration runs to check for faulty hardware components, Section~\ref{subsec:calibration_run_checks}, are discussed. Online data quality assessment and monitoring are presented in Section~\ref{subsec:online_data_quality_assessment_monitoring}. After data are recorded, offline monitoring tools provide key information for
analysers to improve data quality and to address more challenging data
corruption issues. For cases of (semi-) permanent problems,
data quality flags are assigned to portions of the affected data,
indicating whether those data are usable for physics analyses with
care (on an analysis-dependent basis) or must be discarded entirely.

\subsection{Detector control system}
\label{subsec:atlas_detector_control_system}

The ATLAS detector control system (DCS) ensures the coherent and safe operation of the whole experiment~\cite{ATLASDCS}.
The TileCal DCS~\cite{Martins:2016hrt} is part of the ATLAS DCS being responsible for the control of the low- and high-voltage systems and the detector's infrastructure.
The DCS executes actions initialised by the operator or automatic actions based on conditions for the monitored data, such as automatic recovery of power supplies from an abnormal shutdown or shutdown electronics due to high temperature or over-current. Alarms and other notifications (email, SMS) are triggered to alert the experts about any abnormality. The relevant monitored data and all alarms are stored in
databases, allowing easy access to the data.
For example, the DCS is able to adjust the various high-voltage levels for each of the TileCal's PMTs, in the range of $[\mathrm{HV_{input}+1~V}$, $\mathrm{HV_{input}+360~V}]$~\cite{Calvet_2018}, while monitoring it with 0.1~V accuracy. Keeping a stable applied voltage is the main goal of the high-voltage system since the gain of the PMT is a function of the applied voltage, HV:

\begin{equation}
G=\alpha \times \mathrm{HV}^{\beta}
\end{equation}

where $\alpha$ and $\beta$ are characteristic parameters of each PMT. In the TileCal LB and EB, the average HV is 640~V and 700~V, respectively, while the average $\beta$ is 7.07 and 6.95. Therefore, a variation in voltage of 1~V causes a gain variation of 1.1\% (1\%) in the LB (EB).

The archived DCS data allow the stability of the HV applied to the TileCal's PMTs to be studied. The analysis of the data collected in 2015 and 2016 indicated very few problematic channels. Only 0.2\% of the PMTs operated with an HV instability larger than 0.5~V~\cite{Calvet_2018}. Consistent results were obtained by analysing data collected over the entire Run~2.
A portion of the faulty channels may be due to monitoring problems. However, hardware failures are the most common cause.

\begin{figure}[t]
\begin{center}
\includegraphics[width=0.7\textwidth]{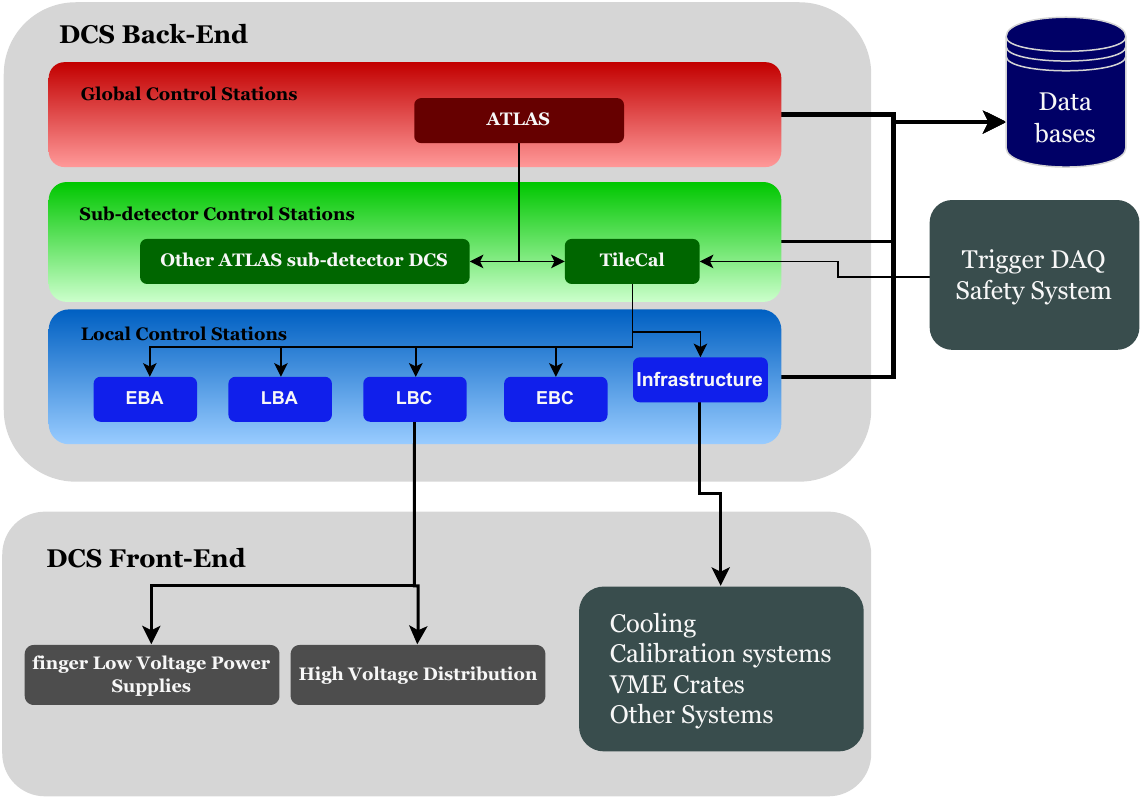}
\caption{\label{fig:Tile:tiledcsherar}The hierarchy of the TileCal DCS within the ATLAS DCS~\cite{Martins:2016hrt}.}
\end{center}
\end{figure}

The ATLAS DCS is based on a supervisory control and data acquisition (SCADA)~\cite{Daneels:1999gj} commercial software. The DCS uses a set of guidelines and software framework components that allow an easy and coherent integration of the ATLAS detectors as a single control system. The TileCal DCS is distributed among two of the three functional levels of the DCS back-end (BE) hierarchy, as can be seen in Figure~\ref{fig:Tile:tiledcsherar}.
It is distributed among six SCADA systems. Four of them are identical and independent in functionality, each being responsible for the low and high-voltage systems of one TileCal partition.
The infrastructure supervises the detector's common components (detector cooling, calibration systems, etc) while the sub-detector control station acts as the top layer of the TileCal hierarchy combining all the SCADA systems.
A finite state machine (FSM) toolkit~\cite{gaspar2006tools} is used to model the DCS BE hierarchy. This allows for full control of the detector and, at the same time, visualisation of either the global detector state and the status at the top level or the individual voltages at the lower level. Due to its simplicity of usage, the FSM is the most commonly used tool by detector experts and users.
In preparation for Run~2, the DCS went through important updates and upgrades~\cite{Martins:2016hrt}.
The main improvement was the migration of the non-SCADA software to a fully Linux based system.
For example, the OLE (object linking and embedding) for process control data access was migrated to Open Platform Communications Unified Architecture.
Apart from the changes required by software migration, it was also necessary to replace control blocks due to the replacement of hardware (a new 200~V DC system and new laser calibration system).

\subsection{Calibration run validation}
\label{subsec:calibration_run_checks}

The analysis of calibration runs mainly allows checks for faulty hardware components. During the long shutdowns, the faulty components are included in the repair list of the maintenance team. The following types of calibration runs are monitored:
\begin{itemize}
\item Laser calibration runs. There are two types of laser runs: `Laser LG' and `Laser HG', which use two different intensities of the laser signal.
\item Charge injection runs. Three types of runs are used: `CIS' calibration runs where a scan with the variable injected charge and phase is performed, and `MonoCIS LG' and `Mono-CIS HG' runs with a constant value of injected charge.
\item Pedestal runs with no beams circulating in the LHC and thus with no signal in the detector.
\end{itemize}

Raw data from calibration runs are reconstructed offline using the dedicated software based on the ATLAS data processing framework Athena~\cite{ATL-SOFT-PUB-2021-001}. It produces plots and histograms
that undergo futher tests to identify potential
problems.
Two types of tests are employed. The first checks the consistency of data (e.g. parity bits, data header, etc) sent by individual DMU chips. The second type of tests monitors the quality of the signal read from individual channels (e.g. digital errors, RMS of signal amplitude, timing
shifts, etc). The first type of tests provides a result for each DMU on the module while the second type provides a result for each channel. Some tests are run for all types of calibration runs while other tests are run-type specific.
Each test takes some specific value obtained for each channel or DMU and compares it to a predefined threshold or a range of values. Based on this comparison it assigns a status (called data quality monitor or DQM status) to each channel (or DMU). The status is Green if the expected value is observed. If a problem is found, the status can be Yellow or Red depending on its
severity. Statuses of individual channels or DMUs are then propagated into the DQM module test status:

\begin{itemize}
\item Green status: all channels/DMUs test results are within expected range.
\item Yellow status: some problems, usually no more than one channel affected.
\item Orange status: more serious problems, usually two or more channels affected.
\item Red status: serious problems.
\item Undefined status: there is not enough data to evaluate the test.
\end{itemize}

The module's overall DQM status is then treated as the worst status of all module tests.

\subsection{Online data quality assessment and monitoring}
\label{subsec:online_data_quality_assessment_monitoring}

Data quality monitoring begins in real time in the ATLAS control room. Online shifters on duty serve as a first line of defence to identify serious detector-related issues.
During periods of physics collisions, the TileCal has experts in the ATLAS control room 24 hours per day and a handful of remote experts available on call to assist in advanced interventions. The primary goal is to quickly identify and possibly correct any problem that cannot be fixed later in software, and that can result in overall data loss.

The ATLAS data quality framework is designed to perform automatic checks of the data and to alert experts to potential problems that warrant further investigation~\cite{Ilchenko_2010}. This framework allows data quality monitoring at various levels of the ATLAS data flow. The subset of data reconstructed online is quickly made available to the online shifters via display tools. The automatic tests performed on the reconstruction output include compatibility checks between the observed distributions from the monitoring data and reference histograms taken from a past run that is both free of data quality issues and taken with similar machine operating conditions. Other tests might involve checks on the number of bins in a histogram above a predefined threshold, or checks on the gradient of a distribution. For example, histograms that monitor readout errors should always be empty under `ideal' conditions. If a bin in such a histogram has a non-zero number of entries, a flag would be raised to alert the shifter to the problem. Online event reconstruction also allows control room experts to monitor reconstructed physics objects, such as jets or missing transverse momentum, permitting real-time monitoring of combined performance in addition to detector status. The data quality monitoring framework  takes the results of tests on individual histograms and propagates them upwards through a tree, resulting in a set of top-level status flags, which can be viewed on the data quality monitoring display. Monitoring histograms are updated to include additional data every few minutes as newly available data are reconstructed. In this way, online monitoring allows hardware- or software-related issues to be caught in real time and rectified to minimise their impact on collected data.

In addition to the global ATLAS data quality monitoring, TileCal specific data are reconstructed and validated.
This allows more events to be reconstructed and more detailed monitoring histograms to be produced. Online reconstruction is done in the same way as it is done offline (see Section~\ref{subsec:offline_data_quality_review}).
TileCal specific automatic checks are focused mainly on monitoring histograms that accumulate entries per luminosity block (one luminosity block nominally spans one minute). These allow detection of problems and to address them within that period during the run. For example, if a test detects that the fraction of corrupted data in a single TileCal module goes above a certain threshold, a request to power-cycle the corresponding module is issued. Then the DCS decides if a power-cycle should be performed immediately or be ignored because the module is in the exclusion list or was recently power-cycled.
During physics runs, in addition to collisions data, the TileCal also collects special laser data when every TileCal PMT receives laser light. These laser data are reconstructed online to monitor timing and address timing jumps, as described in Section~\ref{subsec:time_minitoring}.

Common problems identified by TileCal experts during the online shifts include hardware failures that do not automatically recover, or software configuration problems that might present themselves as data corruption flags from the ROD data integrity checks. The trigger efficiency and data acquisition, as well as higher-level reconstruction data quality, might be influenced by such problems.

\subsection{Offline data quality review}
\label{subsec:offline_data_quality_review}

Once the data are recorded, a two-stage offline DQ assessment ensues~\cite{Aad:2019hmz}.
Soon after the data are taken, a small fraction is quickly reconstructed using the ATLAS Athena software framework~\cite{ATL-SOFT-PUB-2021-001}. Reconstructed data are then used by the offline data quality experts with more complex tools to evaluate the quality of the data. The experts are given 48 hours to identify, and, where possible, to correct problems, before the bulk reconstruction of the entire run is made. The TileCal offline experts can update the conditions database, where information such as the calibration constants and status of each channel is stored. Channels that suffer from high levels of noise have calibration constants in the database updated accordingly. For channels that suffer from intermittent data corruption problems, data quality flags are assigned to the affected data to exclude the channels in the full reconstruction during the problematic period. This 48-hour period is also used to identify cases of digitiser timing jumps and to add the additional time phases to the time constants of the digitiser affected to account for the magnitude of the time jump.

Luminosity blocks can be flagged as defective to identify periods of time when the TileCal is not operating in its nominal configuration. These defects can either be tolerable whereby corrections are applied but additional caution should be taken while analysing these data, or intolerable in which case the data are not deemed suitable for physics analyses. Defects are entered into the ATLAS data quality defect database~\cite{Golling:2011zy} with the information propagating to analyses and to integrated luminosity calculations. Removing all data within that time can accumulate to a significant data loss. For rare situations where only a few events are
affected by the data corruption, an additional error-state flag is introduced into the reconstruction data. This flag is used to remove such events from the analysis.
Once all offline teams perform the review, the entire run is reconstructed using the most up-to-date conditions database. Subsequently the data can be re-reconstructed when reconstruction algorithms are improved and the conditions database is further refined to improve the description of the detector. These data reprocessing campaigns typically occur several months after the data are taken.

\subsection{Summary of TileCal operations in Run~2}
\label{subsec:summary_TileCal_operation_during_Run2}

The TileCal operated very smoothly during Run~2. In 2015, there were no data
lost due to DQ issues in the TileCal. In 2016 most of the inefficiency attributed to the TileCal was due to data lost in two cases when four consecutive modules were
disabled due to a failure in the ROD-ROS links. These events induced a loss of
61~\invpb and 42~\invpb (0.29\% of the total integrated luminosity of 35.5~fb$^{-1}$ in
2016).
Most of the data loss from the TileCal during 2017 corresponded to a single run
when four consecutive modules were disabled
for eight hours due to an auxiliary board~\cite{ATLAS-TDR-03} issue. During the attempt to fix the
auxiliary board the entire LBA partition went off for two hours. This
event induced a loss of 265~\invpb (0.57\% of the total integrated
luminosity of 46.4~fb$^{-1}$ in 2017).
In 2018, there was no data lost due to DQ issues in the TileCal.
The DQ efficiency for each year of Run 2 and for the entire Run 2 are summarised in Table~\ref{tab:DQ_efficiency}.

\begin{table}[t]
\caption{The TileCal DQ efficiency during stable-beams $pp$ collision physics runs at $\sqrt{s}=13$~\TeV\ for each year of Run~2 and for the entire Run~2. For completeness the corresponding ATLAS efficiencies are reported~\cite{Aad:2019hmz}.}
\centering
\begin{tabular}{r c c c c c}
\toprule
& 2015 & 2016 & 2017 &2018 & Entire Run2\\
\midrule
TileCal DQ efficiency [\%]  & 100.00 & 99.31 & 99.41 & 100.00 & 99.65\\
ATLAS DQ efficiency [\%]    &  88.79 & 93.07 & 95.67 &  97.46 & 95.60\\
\bottomrule
\end{tabular}
\label{tab:DQ_efficiency}
\end{table}

\begin{figure}[t]
\centering
\includegraphics[width=0.7\textwidth]{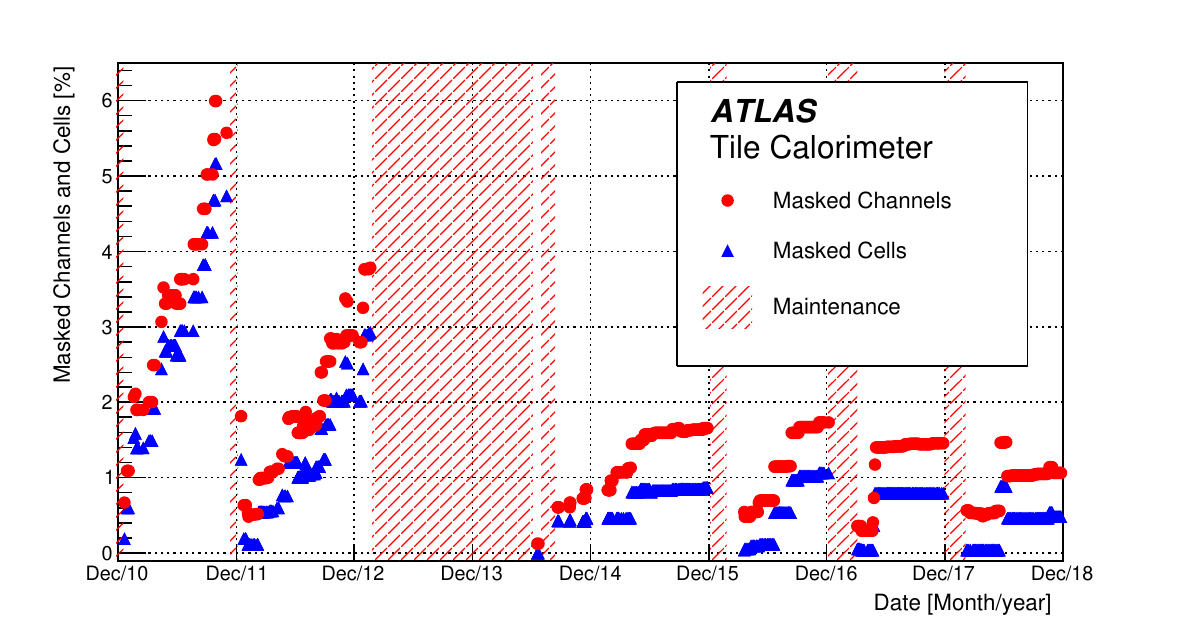}
\caption{The fraction of channels and cells removed from the reconstruction (masked) as a function of time during Run~1 and Run~2.
The number of masked cells (channels) at the end of Run~2, 3 December 2018, is about 0.5\% (1\%). The hatched area represents the maintenance periods of the detector.
}
\label{fig:masked_evolution}
\end{figure}

Failure of some components, like cooling or an
interface card, causes the exclusion of all channels (thus all cells) of a module from the reconstruction. Failure of
other components affects only a part of a module. For example, a failure of a HV
card renders every second channel of a half-module non-operational, while keeping all readout cells operational thanks to the readout redundancy.
The next example is a failure of a
digitiser~\cite{ATLAS-TDR-03}. In this case, six corresponding channels and two or three cells are excluded.
Figure~\ref{fig:masked_evolution} shows the fraction of channels and cells removed (masked) as a function of time in Run~1 and Run~2.
The shaded regions correspond to maintenance periods. The most common issues, which are addressed
and repaired during this time, correspond to cooling, HV, front-end
electronics or the trigger. Regular maintenance helped to keep the fraction of
inefficient cells below 1.1\%. Only a residual number of channels remain inoperational after long maintenance periods. These correspond to cases where there was lack of time for the full repair, or where accessibility conditions and risk management prevented intervention.


\FloatBarrier

\section{Performance studies}
\label{sec:performance_studies}

\newcommand{\transverseMass}{The transverse mass $m_\text{T}$ of a single heavy particle in a semi-invisible decay into particles 1 and 2 is defined by $m_\text{T}=\sqrt{2p_{\text{T},1}p_{\text{T},2}(1-\text{cos}\phi_{12})}$ in the approximation of massless decay products ($m_1=m_2=0$), with $p_{\text{T},1(2)}$ denoting the transverse momentum magnitude of the particle 1 (2) and $\phi_{12}$ the azimuthal distance between the decay products~\cite{ParticleDataGroup:2022pth}.}

The studies reported in this section allow the performance of the reconstruction and calibration methods described in the previous sections to be verified. The results obtained by analysing isolated muons, single hadrons and jets are discussed. The cell noise levels obtained by analysing experimental and simulated data are compared. The performance of the minimum-bias trigger system and the improvements made to the ATLAS muon trigger system by implementing the information provided by the TileCal are also presented.

\subsection{Energy response to single isolated muons}
\label{subsec:energy_response_single_isolated_muons}

The interaction of muons with matter is a well-understood process~\cite{1930AnP...397..325B, ParticleDataGroup:2022pth}, and events with muons allow comparison of the equalisation of the cell response, its stability and the EM scale setting.
In this Section, results of the studies of the calorimeter response to muons produced by $W\to \mu \nu$ decays from $pp$ collisions are reported.
These muons have momenta in the range 20 to 80~\GeV\ and lose energy in the TileCal primarily due to ionisation. The energy loss per unit path is close to the minimum ionising level for the momenta considered and hence small compared to the muon energy over the full path traversed in the calorimeter. This makes the muons from $W\to \mu \nu$ ideal test particles to scan the entire TileCal volume.

In the analysis, the muon tracks measured by the pixel and SCT detectors are extrapolated through the calorimeter volume taking into account detector material and the magnetic field~\cite{Lund:2009zzc}. To compute the path, $\Delta x$, travelled by the muon in a cell, a linear interpolation is
performed to determine the entry and exit points of the muon in the crossed cell. The path and the energy deposited in the cell, $\Delta E$, are used to compute the muon energy loss per unit distance, $\Delta E/\Delta x$. The  distributions of this quantity can be described by a Landau function convoluted with a Gaussian distribution, where the Landau part describes the actual energy loss and the Gaussian part accounts for resolution effects. However, due to long tails from rare energy loss mechanisms, such as
energetic delta electrons and bremsstrahlung, the fits show small $\chi^2$ probability values. For this reason
the truncated mean $\langle \Delta E / \Delta x \rangle_{F = 1\%}$, calculated without considering 1\% of the events in the high tail of the distributions, is used to define the muon response. Typical values are about 1.5~\MeV/mm.
The truncated mean shows slight non-linear behaviour with $\Delta x$. This effect and other non-uniformities, such as \pileup, the differences in momentum and incident angle spectra are, to a large extent, reproduced by the MC simulation. To compensate for these effects, the ratio of the $\langle \Delta E / \Delta x \rangle_{F = 1\%}$ obtained using experimental and simulated data:
\begin{equation} \label{eq:R_trunc}
R \equiv \frac{\langle \Delta E/\Delta x \rangle ^{\mathrm{data}}_{F = 1\%}}{\langle \Delta E/\Delta x \rangle ^{\mathrm{MC}}_{F = 1\%}}
\end{equation}
is then used in the analysis. Differences between the $R$ values obtained for different cells may indicate cell miscalibration. Deviations of the values from unity may indicate an inaccurate setting of the EM energy scale.

\subsubsection{Selection of isolated muons}
\label{subsubsec:isolated_collision_muons}

The results presented here were obtained by analysing proton--proton collision data collected in 2015--2016, 2017 and 2018. The 2015 and 2016 samples were merged together due to low numbers of events in each of them. Events were collected using the L1 muon trigger that accepts events with muons originating from the interaction point~\cite{TRIG-2018-01}. Three further kinematic cuts are used to select $W \to \mu\nu$ events and to suppress background from multijet processes: number of muon tracks $N_{\mu} = 1$, transverse mass~\footnote{\transverseMass}
$40~\GeV \le \mt \le 140~\GeV$
and missing transverse momentum~\cite{PERF-2016-07}
$30~\GeV \le \met \le 120~\GeV$.
The muons with momentum
$20~\GeV \le p \le 80~\GeV$
are retained. The lower limit reduces multiple scattering effects while the upper limit ensures that muons lose their energy primarily via ionisation.
The contribution from nearby particles in the energy deposited in a cell is suppressed by only considering well isolated muon tracks~\cite{TCAL-2017-01}. The selection requirements $\Delta x > 100$~mm and $\Delta E>60$~\MeV\ are applied to reduce noise contributions. The cells with $|\eta|< 0.1$ are not considered in the analysis since in the TileCal the scintillator tiles are oriented radially and the energy deposited by muons with a trajectory close to this direction is measured with poor accuracy due to large variations of sampling fraction as function of the impact point. The number of retained tracks  is about 300~million.

The MC simulation uses the \textsc{Geant4} toolkit that provides the physics models of particle interactions with material, the ATLAS geometry description and the tracking tools as described in Section~\ref{subsec:signal_reconstruction_simulated_events}. The simulated $W\to \mu \nu$ events were generated using \textsc{Sherpa}~\cite{Bothmann:2019yzt} at next-to-leading order interfaced with \textsc{Pythia~8}~\cite{Sjostrand:2007gs} for the parton showering. The number of simulated events is approximately 200 million. The energy deposited in the cells was digitized and reconstructed using the same procedure as applied in the case of experimental data. The retained events pass the same criteria used in the selection of experimental data. The relevant kinematic distributions show good agreement between data and MC with observed discrepancies only in bins whose contents are a negligible proportion of the total yield. Using MC events, the purity of the selected track sample is estimated to be 98\%.

\subsubsection{Cell response uniformity}
\label{subsubsec:cell_response_uniformity}

As shown in Figure~\ref{fig:mu_R_17}, in the calorimeter one can distinguish 74 rings of cells, 46 in the LB and 28 in EB, each consisting of $N_r=64$ cells with the same values of the coordinates $r$ and $z$ and different value of $\phi$. Due to the symmetry of the calorimeter and of the particles produced in $pp$ collisions, one expects that the ratios in Eq.~(\ref{eq:R_trunc}) computed using muons crossing any cell $c$ of the same ring $r$, $R_{r,c}$, are equal.

The estimated value for the ratio defined by Eq.~(\ref{eq:R_trunc}) for a given ring, $R_r$, is determined by maximising the Gaussian likelihood function:

\begin{equation} \label{eq:likelihood_ phi}
\mathcal{L} = \prod_{c=1}^{N_r} \frac{1}{\sqrt{2 \pi} \sqrt{\sigma_{r, c}^{2}+s_{r}^{2}}}  \exp \left[-\frac{1}{2} \frac{\left(R_{r, c}-R_{r}\right)^{2}}{\sigma_{r, c}^{2}+s_{r}^{2}}\right]
\end{equation}

\begin{sidewaysfigure}
\centering
\includegraphics[width=\textwidth]{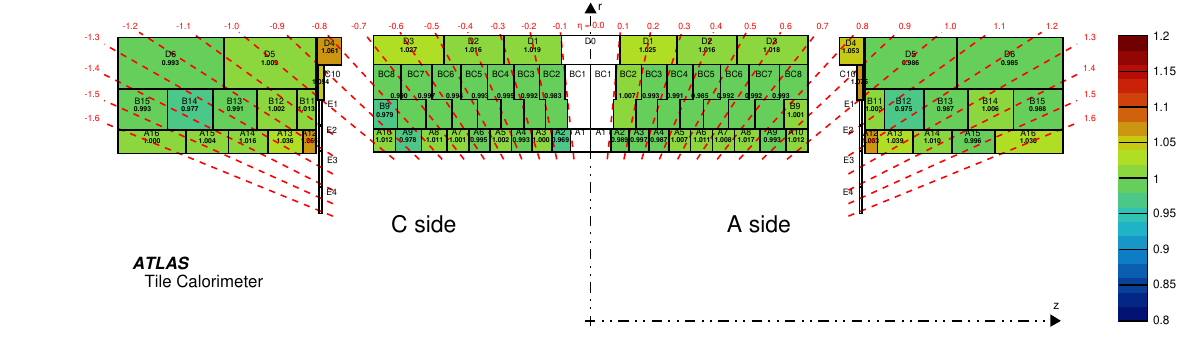}
\caption{Visualisation of the TileCal in the ($z$, $r$) plane showing the 2017 estimated values of the ratios $R_r$ of the cells of a given ring $r$ (see the text). The values are obtained by maximising the likelihood function in Eq.~(\ref{eq:likelihood_ phi}). The statistical uncertainty in each determination is about 0.5\%.}
\label{fig:mu_R_17}
\end{sidewaysfigure}

where $\sigma_{r,c}$ is the statistical uncertainty in the determination of $R_{r,c}$. The other parameter determined in the likelihood function maximisation, $s_{r}$, describes the response non-uniformity of the cells due to differences in the optical and electronic read-out systems not described in the simulation and potential mis-calibration of cells of the ring. The parameter $s_r$ determined from the fit is 2.4\%. The cell non-uniformity was previously measured in ATLAS using cosmic rays~\cite{TCAL-2017-01} and at test beams~\cite{TCAL-2010-01} using electrons incident at the centre of A cells with an angle of 20$^{\circ}$. Similar results are obtained. The analysis shows no problematic cells with a value of $R_{r,c}$ that deviates significantly from $R_r$.

Figure~\ref{fig:mu_R_17} shows the estimates of $R_r$ obtained for the different rings by analysing 2017 data. For a given radial layer, the values are uniform within about 1\%. Similar results are found by analysing the 2015--2016 and 2018 data.

\subsubsection{Radial layer calibration}
\label{subsubsec:radial_layer_intercalibration}

The estimated response $R_l$ for the six radial layers $l$: LB-A, LB-BC, LB-D, EB-A, EB-B, and EB-D are obtained from the truncated mean ratios (Eq.~(\ref{eq:R_trunc})) of the experimental and simulated $\Delta E/\Delta x$ distributions of all the cells belonging to the layer. The results obtained by analysing the data collected in the three periods are reported in Table~\ref{tab:layer_R}. The statistical and systematic uncertainties are shown separately. To estimate systematic uncertainties, the selection parameters were varied around their nominal values (see Section~\ref{subsubsec:isolated_collision_muons}). The systematic uncertainties dominate. In each year the standard deviation of the distributions of the measurements listed in the Table is $1.5\%$. The maximum deviation of the values of $R_l$ from 1 is 0.029.

\begin{table}[t]
\caption{The response ratio $R_l$ of the cells of different radial layers in the LB and EB obtained by analysing 2015--2016, 2017 and 2018 data. Statistical (first value) and systematic (second value) uncertainties are shown.}
\centering
\begin{tabular}{lccc}
\toprule
Layer &               $R_l(2015-2016)$ &                    $R_l(2017)$ &                    $R_l(2018)$ \\
\midrule
LB-A & 0.988 $\pm$ 0.001 $\pm$ 0.003 & 0.996 $\pm$ 0.002 $\pm$ 0.007 & 0.996 $\pm$ 0.001 $\pm$ 0.004 \\
LB-BC & 0.984 $\pm$ 0.001 $\pm$ 0.001 & 0.993 $\pm$ 0.001 $\pm$ 0.002 & 0.992 $\pm$ 0.001 $\pm$ 0.003 \\
LB-D & 1.014 $\pm$ 0.001 $\pm$ 0.004 & 1.019 $\pm$ 0.001 $\pm$ 0.003 & 1.024 $\pm$ 0.001 $\pm$ 0.002 \\
EB-A & 1.006 $\pm$ 0.003 $\pm$ 0.006 & 1.029 $\pm$ 0.002 $\pm$ 0.006 & 1.016 $\pm$ 0.002 $\pm$ 0.006 \\
EB-B & 0.978 $\pm$ 0.002 $\pm$ 0.002 & 0.990 $\pm$ 0.001 $\pm$ 0.006 & 0.989 $\pm$ 0.001 $\pm$ 0.004 \\
EB-D & 0.982 $\pm$ 0.001 $\pm$ 0.004 & 0.990 $\pm$ 0.001 $\pm$ 0.007 & 0.997 $\pm$ 0.001 $\pm$ 0.004 \\
\bottomrule
\end{tabular}
\label{tab:layer_R}
\end{table}

\subsubsection{Time stability}
\label{subsubsec:time_stability}

The determination of the stability of the $R_r$ values discussed in Section~\ref{subsubsec:cell_response_uniformity}  %
allows quantification of how well the calibration compensates for time-dependent variations of the calorimeter response. For the cells of each ring $r$, a determination is obtained by maximising the likelihood function:

\begin{equation} \label{eq:likelihood_time_stab}
\mathcal{L} = \exp(-\frac{1}{2} (\vec{R}_r - \vec{\hat{R}}_r)^T [C_r]^{-1} (\vec{R}_r - \vec{\hat{R}}_r))
\end{equation}
The components of the vector $\vec{R}_r$ = ($R^1_r, R^2_r, R^3_r$) are
the responses of the cells of a ring $r$ obtained by maximising the likelihood function~(Eq.~(\ref{eq:likelihood_ phi})) using data collected in 2015--2016 ($R^1_t$), 2017 ($R^2_t$) and 2018 ($R^3_t$) respectively.
The components of the vector $\vec{\hat{R}}_r$ = ($\hat{R}^1_r, \hat{R}^2_r, \hat{R}^3_r$) are the corresponding parameter estimates. The covariance matrix $C_r$ takes into account correlations between the determinations performed in the different periods.

\begin{sidewaysfigure}
\centering
\subfloat[]{\includegraphics[width=\textwidth]{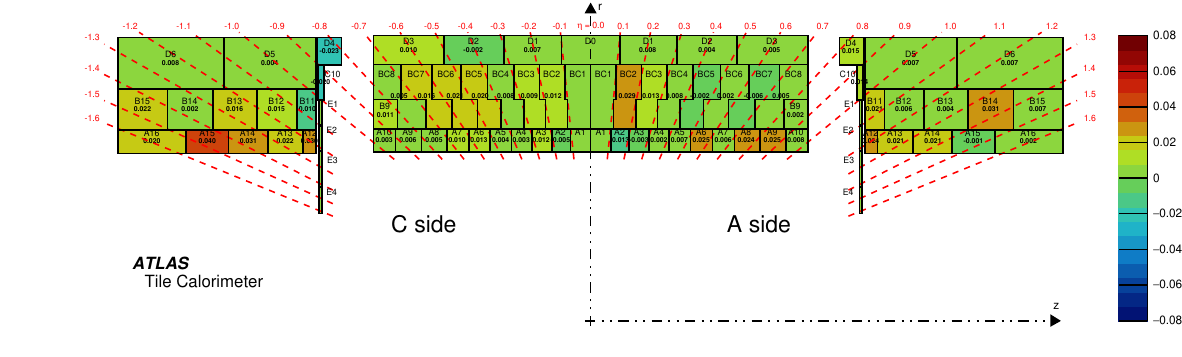}\label{fig:delta_1516_17}}\\
\subfloat[]{\includegraphics[width=\textwidth]{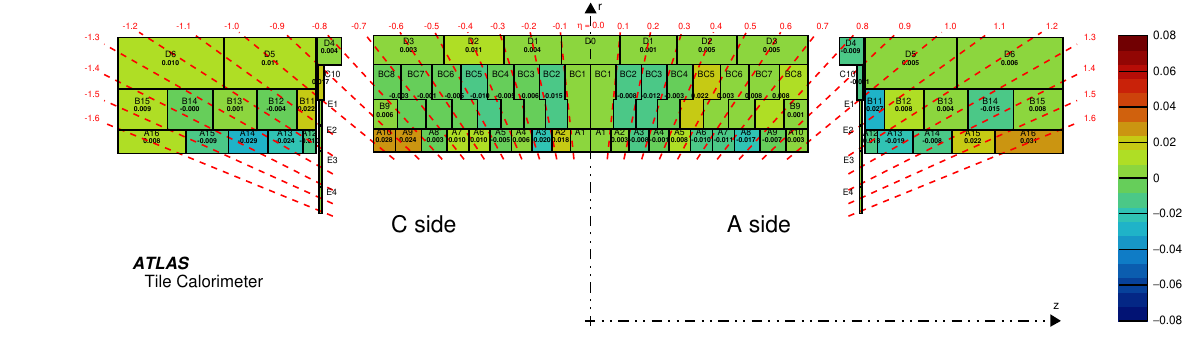}\label{fig:delta_17_18}}
\caption{Visualisation of the TileCal in the ($z$, $r$) plane showing the relative difference of the fitted response ratios (a) $\Delta_r(2015-2016 \to 2017)$, obtained in 2015--2016 and 2017, and $\Delta_r(2017 \to 2018)$, obtained in 2017 and 2018.
The average statistical error in the determinations is about 0.8\%.
}
\end{sidewaysfigure}

Figures~\ref{fig:delta_1516_17} and \ref{fig:delta_17_18} show the relative variations of the ring responses between 2015--2016 and 2017:
\begin{equation} \label{eq:relative_time_variation_2015-2016_and_2017}
\Delta_r(\textrm{2015--2016} \to 2017) \equiv \frac{\hat{R}_r(2017) - \hat{R}_r(\textrm{2015--2016})}{\hat{R}_r(\textrm{2015--2016})}
\end{equation}
and between 2017 and 2018:
\begin{equation} \label{eq:relative_time_variation_2017_and_2018}
\Delta_r(2017 \to 2018) \equiv \frac{\hat{R}_r(2018) - \hat{R}_r(2017)}{\hat{R}_r(2017)}
\end{equation}
respectively. The uncertainties are obtained by combining statistical and systematic effects in quadrature and taking into account correlation effects. In both cases the most probable value is +0.5\%. The 95\% confidence intervals of the variations of the cell response between 2015--2016 and 2017 and between 2017 and 2018 are found to be $[-1.3\%, 3.3\%]$ and $[-2.4\%, +2.5\%]$ respectively.

\subsection{Energy response to single isolated hadrons}
\label{subsec:energy_response_hadrons}

The calorimeter energy response is probed by measuring the energy deposited in the calorimeter by isolated hadrons produced in the $pp$ collisions. The determination is obtained using the ratio:
\begin{equation}
R=E/p
\label{eq:E_over_p_ratio}
\end{equation}
where $E$ is the energy measured by the calorimeters and $p$ the momentum measured by the inner detector. The particles have a momentum below 20~\GeV\ and the precision of the measurement is dominated by the energy resolution~\cite{PERF-2007-01}.
The events analysed were produced in $pp$ interactions in 2017 at a centre-of-mass energy $\sqrt{s} = 13$~\TeV\ with an average of the mean number of interactions per bunch crossing within a luminosity block, $\langle\mu\rangle$, of about 2.

At the trigger level, the events were selected requiring at least one counter with a signal above the threshold value in each side of the MBTS system (see Section~\ref{subsec:mbts_intro}). Each event is required to have a well reconstructed vertex with at least four well reconstructed associated tracks with $\pt > 400~\MeV$. Each track selected for this study is required to have $\pt>500$~\MeV\ and $|\eta| < 2.5$. The reconstruction of the momentum $p$ is described in~\cite{PERF-2007-01}. To select isolated single charged hadrons, no other track is allowed within a cone of $\Delta R < 0.4$ centred on the considered track. The energy measurement $E$ is obtained using ensembles of cells (topological clusters)~\cite{PERF-2014-07}. A cluster of cells $i$ is associated with the track $k$ if:

\begin{equation}
\Delta R = \sqrt{(\Delta \eta (\textrm{track}_{kj}, \textrm{cluster}_{i}) )^2 + (\Delta \phi (\textrm{track}_{kj}, \textrm{cluster}_{i}) )^2} < 0.2
\label{eq:matching-clusters}
\end{equation}
where $\Delta \phi (\textrm{track}_{kj}, \textrm{cluster}_{i})$ is the smallest angle in the transverse plane between track $k$ and cluster $i$. The $\eta(\textrm{track}_{kj})$ and $\phi(\textrm{track}_{kj})$ are the values of the extrapolated track in the layer $j$ of the cluster with the largest deposited energy.
The energy of a track corresponds to the sum of the energy deposited in all the cells of the clusters associated with the track~\cite{PERF-2011-05}. The cell energies are reconstructed and calibrated at EM scale using the procedures described in Sections~\ref{sec:experimental_setup} and \ref{sec:physics_signals}.

Additional requirements are applied to ensure that tracks reach and deposit their energy in the TileCal layers, and to reduce the contamination from neutral particles and muons. These are:
\begin{itemize}
\item Momentum of the track, $p_\mathrm{track}$, larger than $2~\GeV$ and $|\eta_\mathrm{track}| < 1.7$.
\item The energy deposited by the hadron in the LAr EM calorimeter~\cite{ATLAS-TDR-02}, $E_\mathrm{LAr} < 1~\GeV$, compatible
with that of a minimum ionising particle (MIP), so that the hadron deposits almost all its energy in the TileCal.
\item The energy deposited in TileCal is required to be at least 70\% of the total energy deposited in the calorimeters, $E_\mathrm{Tile}/(E_\mathrm{LAr}+E_\mathrm{Tile}) > 0.7$.
\end{itemize}

The last two criteria effectively reduce the contribution from neutral particles decaying electromagnetically since these events deposit most of their energy in the EM calorimeter~\cite{PERF-2011-05}. The experimental results are compared with those obtained analysing simulated data. The simulation program \textsc{Pythia~8.186}~\cite{Sjostrand:2007gs} was used to generate multijet events with the A14 set of tuned parameters (tune)~\cite{ATL-PHYS-PUB-2014-021} and the NNPDF23 leading-order parton distribution function set. The detector response was simulated using the \textsc{Geant4} simulation program. The transverse momentum of the tracks was re-weighted to match the observed spectrum in data.
The same selection criteria and signal reconstruction procedures used in the analysis of the experimental data are used in the analysis of the simulated data.

\begin{figure}[t]
\centering
\includegraphics[width=0.5\linewidth]{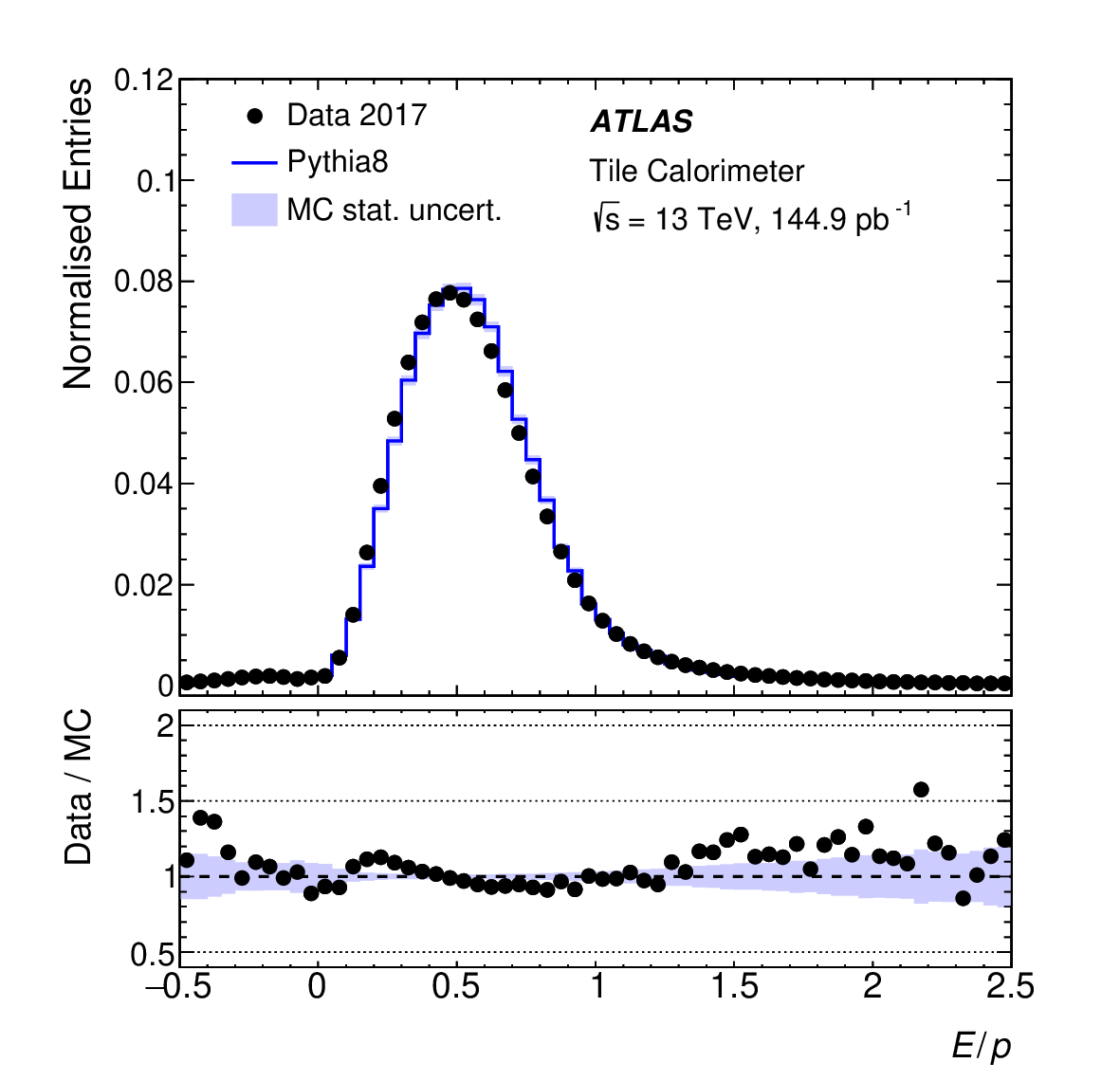}
\caption{The distribution of the ratio of the energy of isolated hadrons measured by the calorimeters divided by the momentum of the track measured by the inner detector ($E/p$).
The distribution obtained by analysing simulated data is also shown. The distributions are normalised to an integrated area of one. The ratios of the experimental values to the simulated ones are plotted in the lower panel. The MC statistical uncertainties are shown.}
\label{fig:eop}
\end{figure}

Good agreement is found between the experimental and simulated results. The $E/p$ distribution of the selected events is shown in Figure~\ref{fig:eop}.
The mean of the distribution obtained by analysing experimental (simulated) data is $\langle E/p\rangle$ is 0.5896 $\pm$ 0.0001 (0.593 $\pm$ 0.001). Only statistical uncertainties are quoted. The values are smaller than one because of the non-compensating nature of the calorimeter. Also energy leakage affects the result.
Figure~\ref{fig:avgeop_vs_eta_p_phi} shows the values of $\langle E/p\rangle$ as a function of the hadron $p$, $\eta$ and $\phi$.
The $\langle E/p\rangle$ variation with pseudorapidity is associated with geometrical effects. Approximatelly constant $\langle E/p\rangle$ values are observed in the more uniform region in the long barrel ($|\eta|<0.8$) while the larger variations appear in the gap/crack regions and extended barrels of the detector. The effect also translates in the measured $E/p$ standard deviation, which is around 0.28 for $|\eta| < 1.0$ and increases to 0.55 for larger absolute values of pseudorapidity. The statistical uncertainties of the determinations are smaller than 0.1\%.

The ratio of the mean $E/p$ values obtained analysing experimental and simulated events in three $\eta$ regions ($|\eta| < 0.7$, $0.7\le|\eta|\le 1.0$ and $1.0 < |\eta| \le  1.7$) are reported in Table~\ref{tab:eop_bin}. The table also reports the ratios of the standard deviations of the $E/p$ distributions, $\sigma(E/p)$, obtained using experimental and simulated data. For each of three pseudorapidity regions, the values refer to isolated hadrons with momenta $2.0~\GeV \le p < 3.0~\GeV$, $3.0~\GeV \le p < 4.0~\GeV$, $4.0~\GeV \le p < 5.0~\GeV$, $5.0~\GeV \le p < 7.0~\GeV$ and $p \ge 7~\GeV$ respectively.

\begin{table}[t]
\caption{Ratios of the mean values and of the standard deviations of the $E/p$ distributions obtained from experimental and simulated data.
The results were obtained by analysing isolated hadrons with different pseudorapidity and momentum values.
Statistical and systematic uncertainties are combined in quadrature.}
\centering
\begin{tabular}{l c c c}
\toprule
$\langle E/p\rangle^{\text{Data}}/\langle E/p\rangle^{\text{MC}}$ & $|\eta|<0.7$ & $0.7 \le |\eta| \le 1.0$ & $1.0 < |\eta| \le 1.7$\\
\midrule
2.0~\GeV $\le p < 3.0$~\GeV & 0.98 $\pm$ 0.01\phantom{0} & 0.97 $\pm$ 0.01\phantom{0} & 1.03 $\pm$ 0.05\phantom{0}\\
3.0~\GeV $\le p < 4.0$~\GeV & 0.98 $\pm$ 0.01\phantom{0} & 0.97 $\pm$ 0.01\phantom{0} & 1.00 $\pm$ 0.03\phantom{0}\\
4.0~\GeV $\le p < 5.0$~\GeV & 0.97 $\pm$ 0.02\phantom{0} & 0.98 $\pm$ 0.02\phantom{0} & 0.99$\pm$ 0.02\phantom{0}\\
5.0~\GeV $\le p < 7.0$~\GeV & 0.98 $\pm$ 0.01\phantom{0} & 0.96 $\pm$ 0.02\phantom{0} & 0.97 $\pm$ 0.02\phantom{0}\\
$ p \ge 7.0$~\GeV & 1.01 $\pm$ 0.03\phantom{0} & 1.02 $\pm$ 0.05\phantom{0} & 0.98 $\pm$ 0.02\phantom{0}\\

\toprule
$\sigma(E/p)^{\text{Data}}/\sigma(E/p)^{\text{MC}}$ & $|\eta|<0.7$ & $0.7 \le |\eta| \le 1.0$ & $1.0 < |\eta| \le 1.7$\\
\midrule
2.0~\GeV $\le p < 3.0$~\GeV & 1.00 $\pm$ 0.03\phantom{0} & 1.04 $\pm$ 0.02\phantom{0} & 1.08 $\pm$ 0.05\phantom{0}\\
3.0~\GeV $\le p < 4.0$~\GeV & 1.02 $\pm$ 0.01\phantom{0} & 1.00 $\pm$ 0.03\phantom{0} & 1.04 $\pm$ 0.05\phantom{0}\\
4.0~\GeV $\le p < 5.0$~\GeV & 1.00 $\pm$ 0.02\phantom{0} & 1.02 $\pm$ 0.02\phantom{0} & 1.05$\pm$ 0.04\phantom{0}\\
5.0~\GeV $\le p < 7.0$~\GeV & 1.02 $\pm$ 0.03\phantom{0} & 1.06 $\pm$ 0.03\phantom{0} & 1.03 $\pm$ 0.02\phantom{0}\\
$ p \ge 7.0$~\GeV           & 1.00 $\pm$ 0.04\phantom{0} & 0.94 $\pm$ 0.06\phantom{0} & 1.03 $\pm$ 0.03\phantom{0}\\

\end{tabular}
\label{tab:eop_bin}
\end{table}

\begin{figure}[p]
\centering
\subfloat[]{\includegraphics[width=0.5\linewidth]{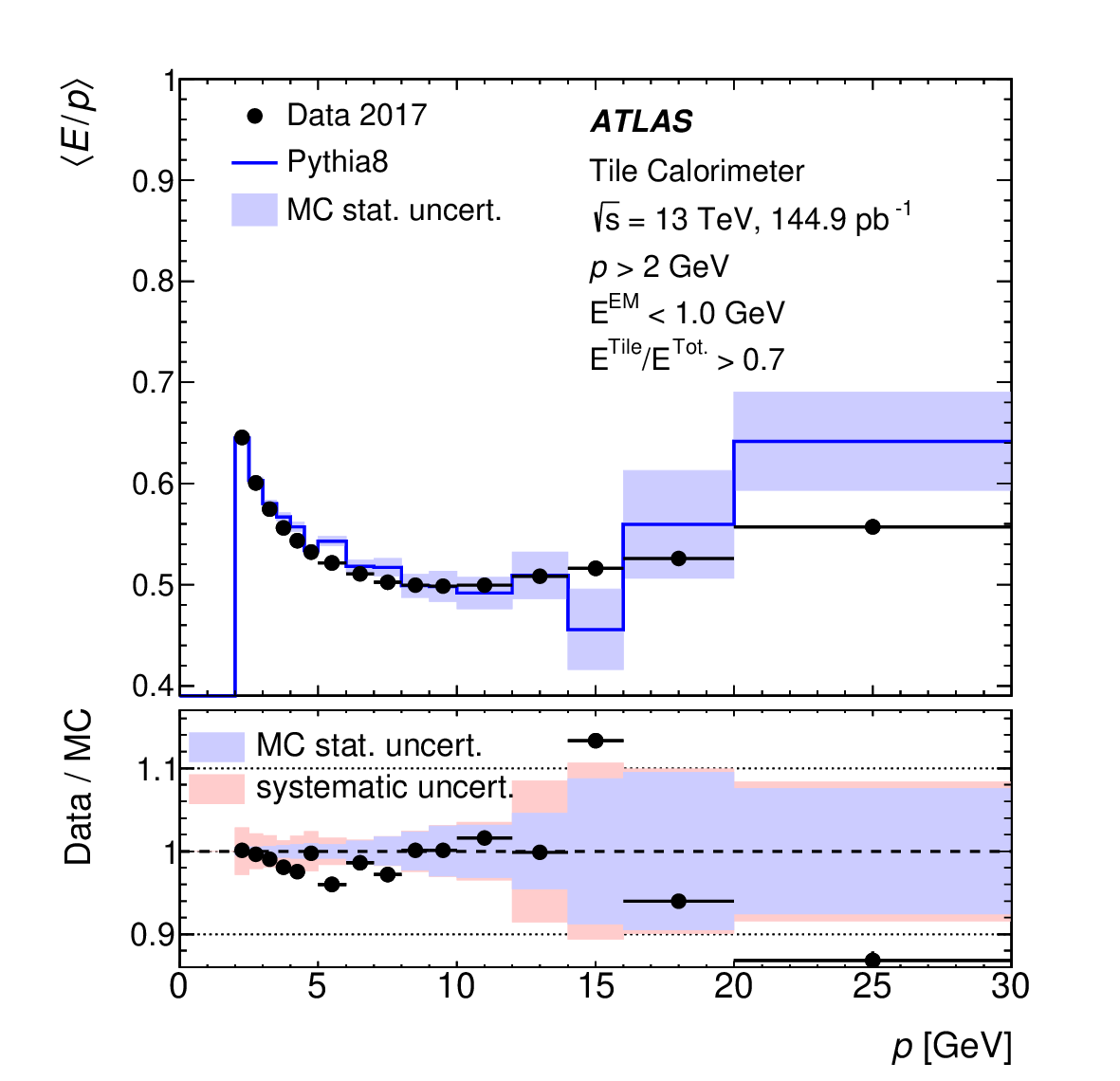}\label{fig:avgeop_vs_eta_p_phi_a}}
\subfloat[]{\includegraphics[width=0.5\linewidth]{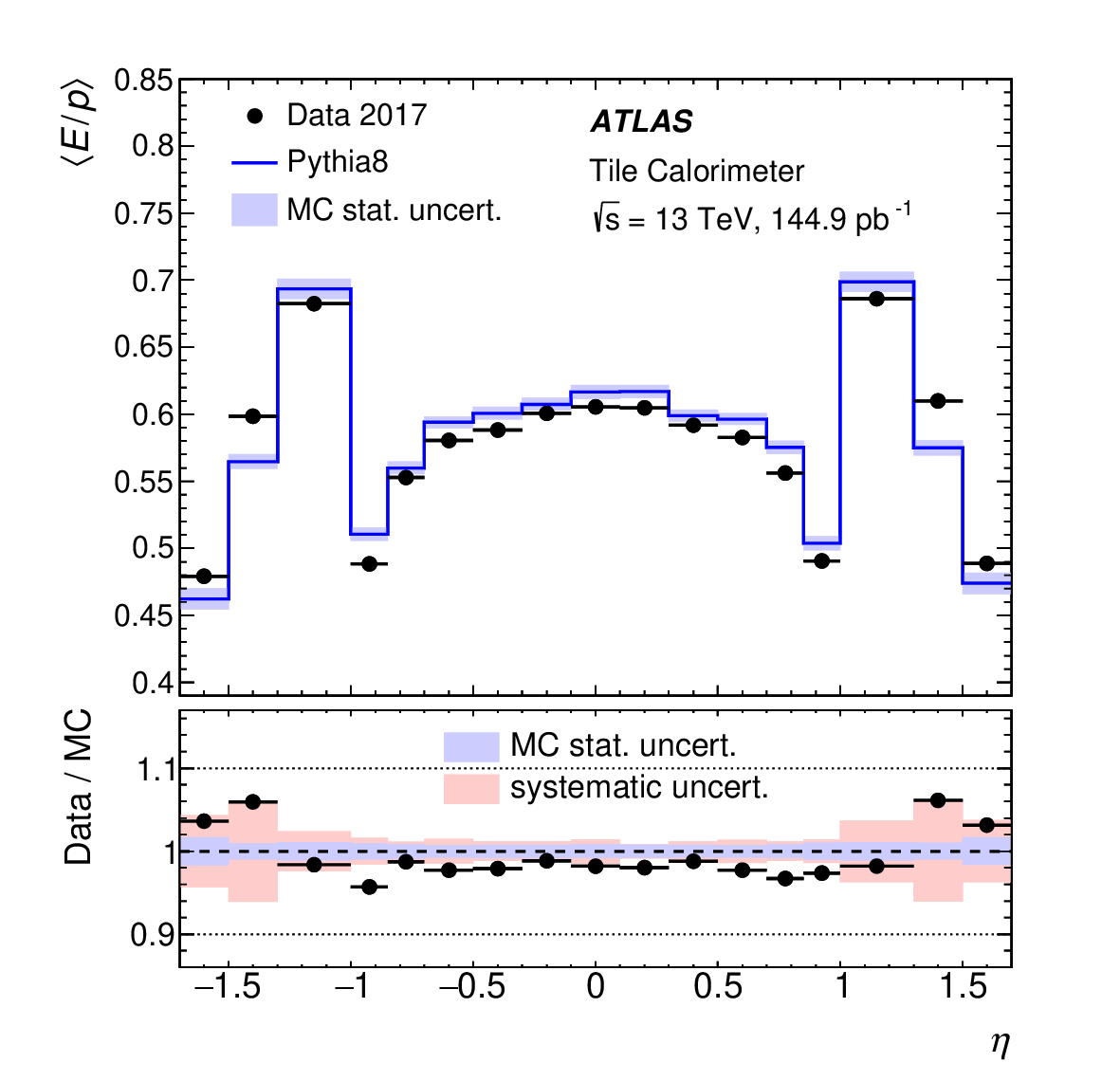}\label{fig:avgeop_vs_eta_p_phi_b}}\\
\subfloat[]{\includegraphics[width=0.5\linewidth]{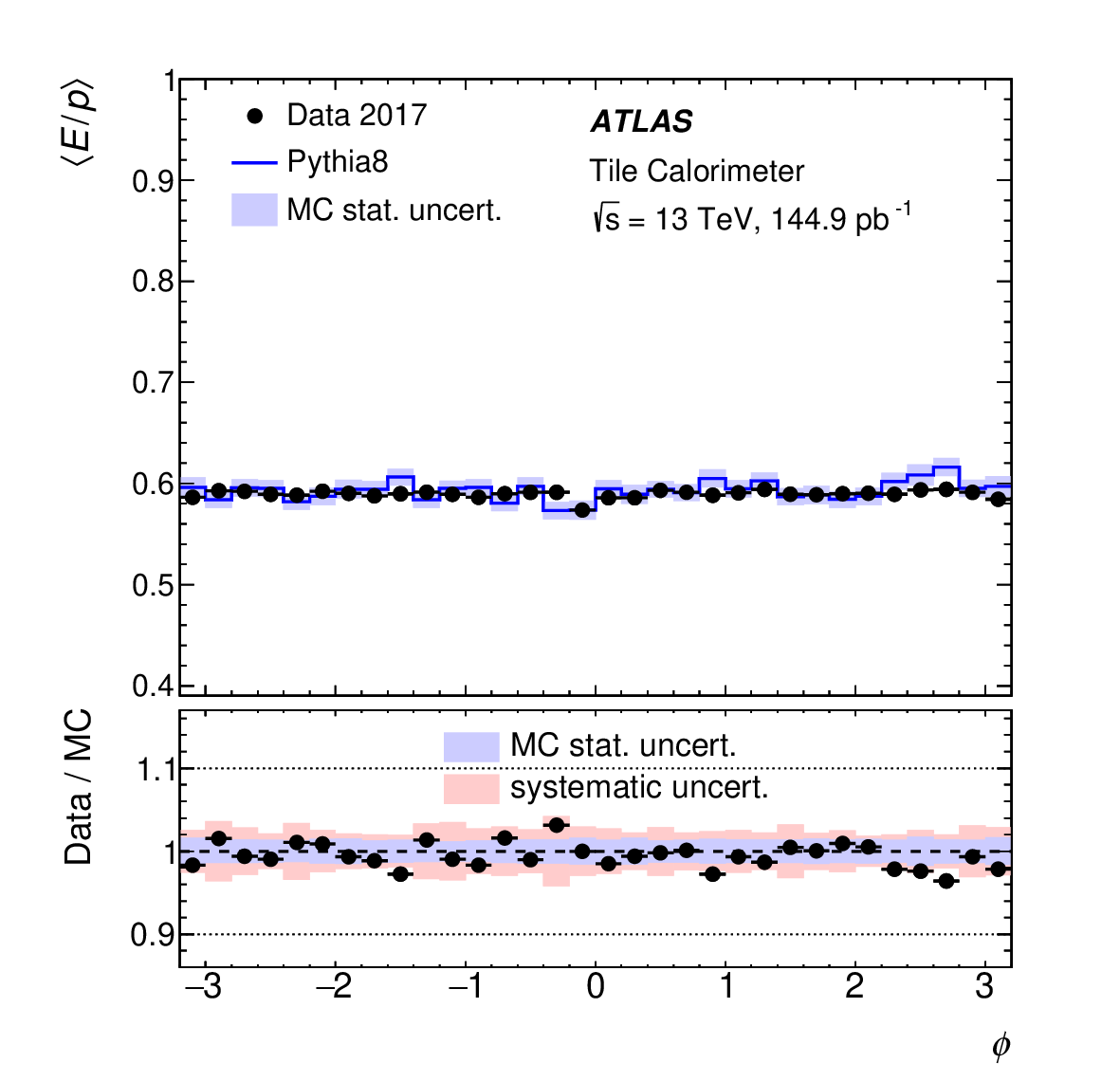}\label{fig:avgeop_vs_eta_p_phi_c}}
\caption{The average ratio of energy to momentum ($\langle E/p\rangle$) measurements as a function of (a) $p$, (b) $\eta$ and (c) $\phi$. The ratios of experimental to simulated data are plotted in the lower panels. MC statistical and experimental systematic uncertainties are shown.}
\label{fig:avgeop_vs_eta_p_phi}
\end{figure}

Two sources of systematic uncertainty are identified. The first one is from the contamination of neutral particles. The effects on the ratio of the experimental and simulation results are estimated by varying the selection criteria for $E_\mathrm{LAr}$ and $E_\mathrm{Tile}/(E_\mathrm{LAr}+E_\mathrm{Tile})$. To this end, different cuts on the reconstructed number of clusters are also applied. A systematic uncertainty of about 1\%, almost independent of $\eta$, is found. It corresponds to the maximum difference of the results obtained by applying the different cuts. In particular for large values of  $|\eta|$, another source of systematic uncertainty comes from the description in the simulation of the dead material in front of TileCal. Effects of few percent are observed by varying the energy deposited in the gap/crack scintillators in the dead material region. Statistical and systematic uncertainties are combined in quadrature in Table~\ref{tab:eop_bin}.
The comparison between the results obtained by analysing experimental and simulated data shows a good reconstruction of the energy at the EM scale of low momentum hadrons. The uncertainty on the mean value (standard deviation) of the energy distributions is better than 5\% (6\%).

\subsection{Timing performance with collision data}
\label{subsec:timing_performance_collision_data}

Correct channel timing obtained by analysing jets as discussed in Section~\ref{subsubsec:time_calibration} is essential for energy reconstruction, particle identification and searches for hypothetical long-lived particles. The performance of time calibration is presented in this section.

\FloatBarrier
\subsubsection{Mean time and time resolution}
\label{sec:mean_time_resolution}

\begin{figure}[t]
\centering
\includegraphics[width=0.5\linewidth]{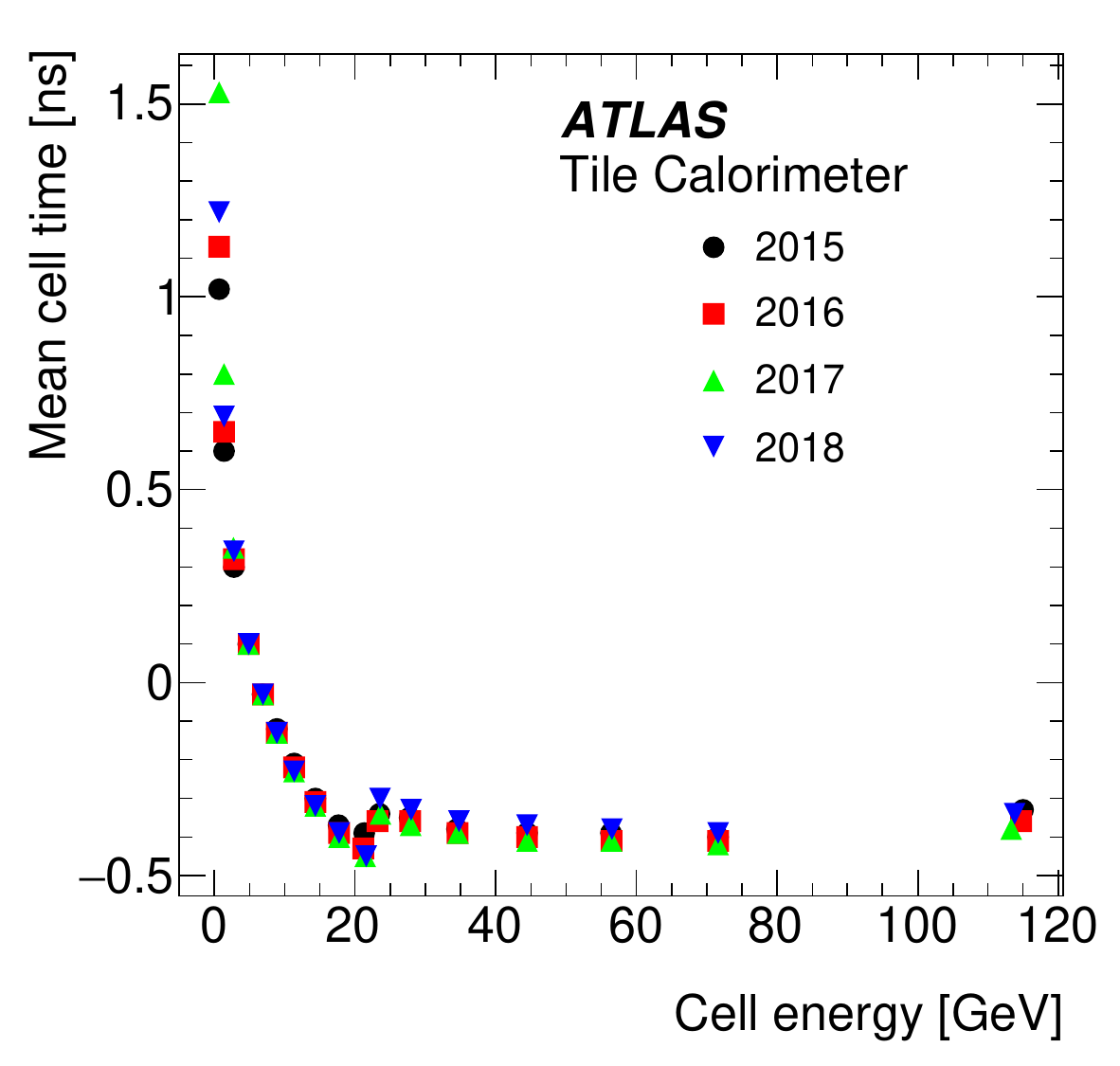}
\caption{The mean reconstructed cell time as a function of the cell energy. Results obtained in Run~2 (2015, 2016, 2017 and 2018), are shown. Statistical uncertainties are smaller than the size of the markers. The discontinuity close to 22~\GeV\ corresponds to
the ADC high-/low-gain transition.}
\label{fig:timing_performance_mean_run2}
\end{figure}

Figure~\ref{fig:timing_performance_mean_run2} compares the mean
of the cell time~(average of the reconstructed
times of channels associated with that cell) as a function of the cell energy measurements of all TileCal cells in each of the four years of Run~2, 2015, 2016, 2017 and 2018. For each energy bin, the mean cell time corresponds to the mean value, $\mu$, of a Gaussian function fit to the cell time distribution in the range of $\mu \pm 2\sigma$ ($\sigma$ is the standard deviation of the Gaussian function). To do this, an iterative procedure is applied. The mean cell time decreases with the cell energy, reaching the plateau at energies above 40~\GeV.
The time behaviour remains
essentially unchanged during the whole
Run~2.  The differences appear only in
the two lowest energy bins, below 4~\GeV, dominated by \pileup conditions. At higher energies, the year-to-year differences are within 0.05~ns in agreement with previous results~\cite{TD_proceedings}.

\begin{figure}[t]
\centering
\subfloat[]{\includegraphics[width=0.5\linewidth]{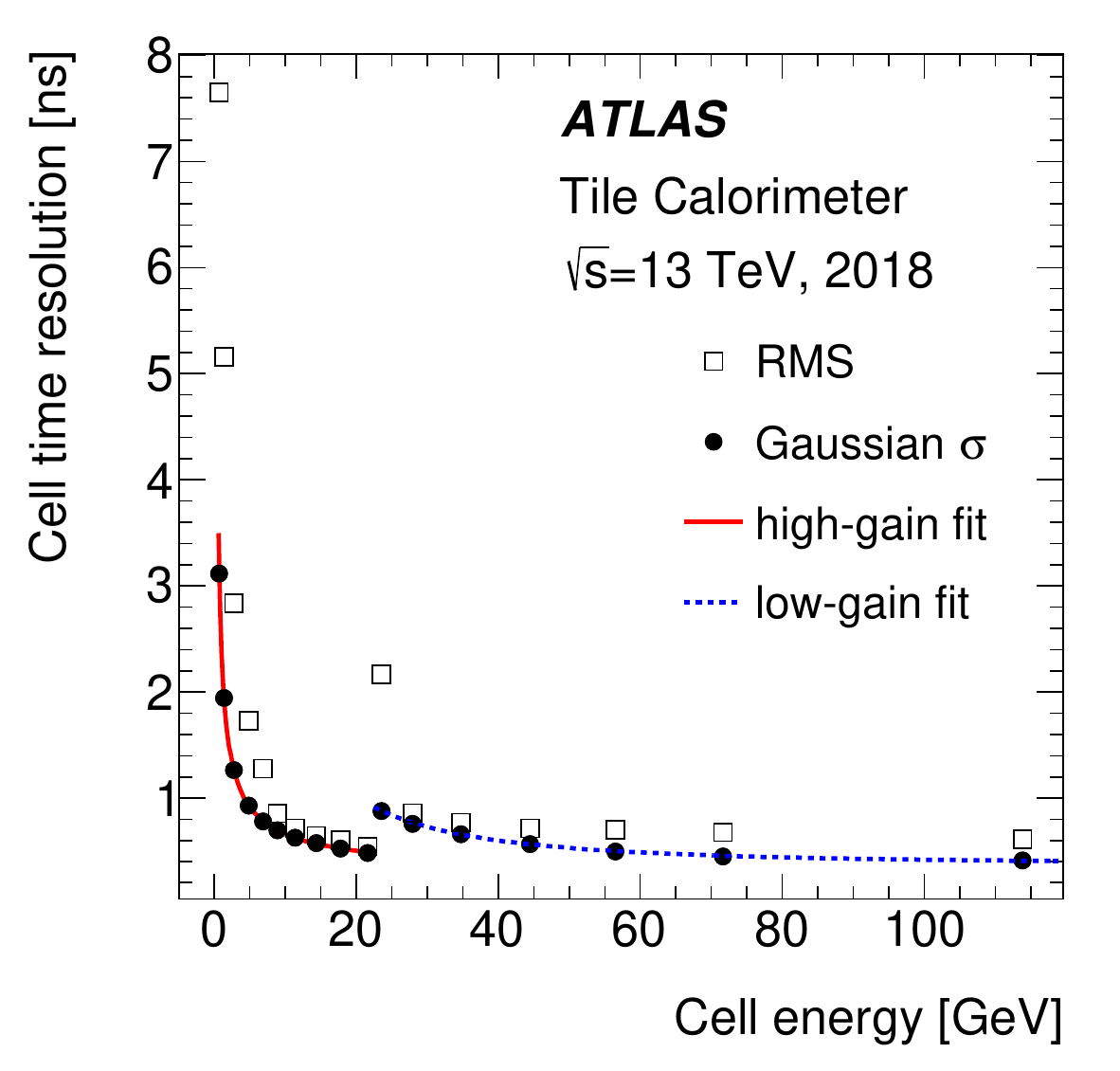}\label{fig:timing_performance_resolution_a}}
\subfloat[]{\includegraphics[width=0.5\linewidth]{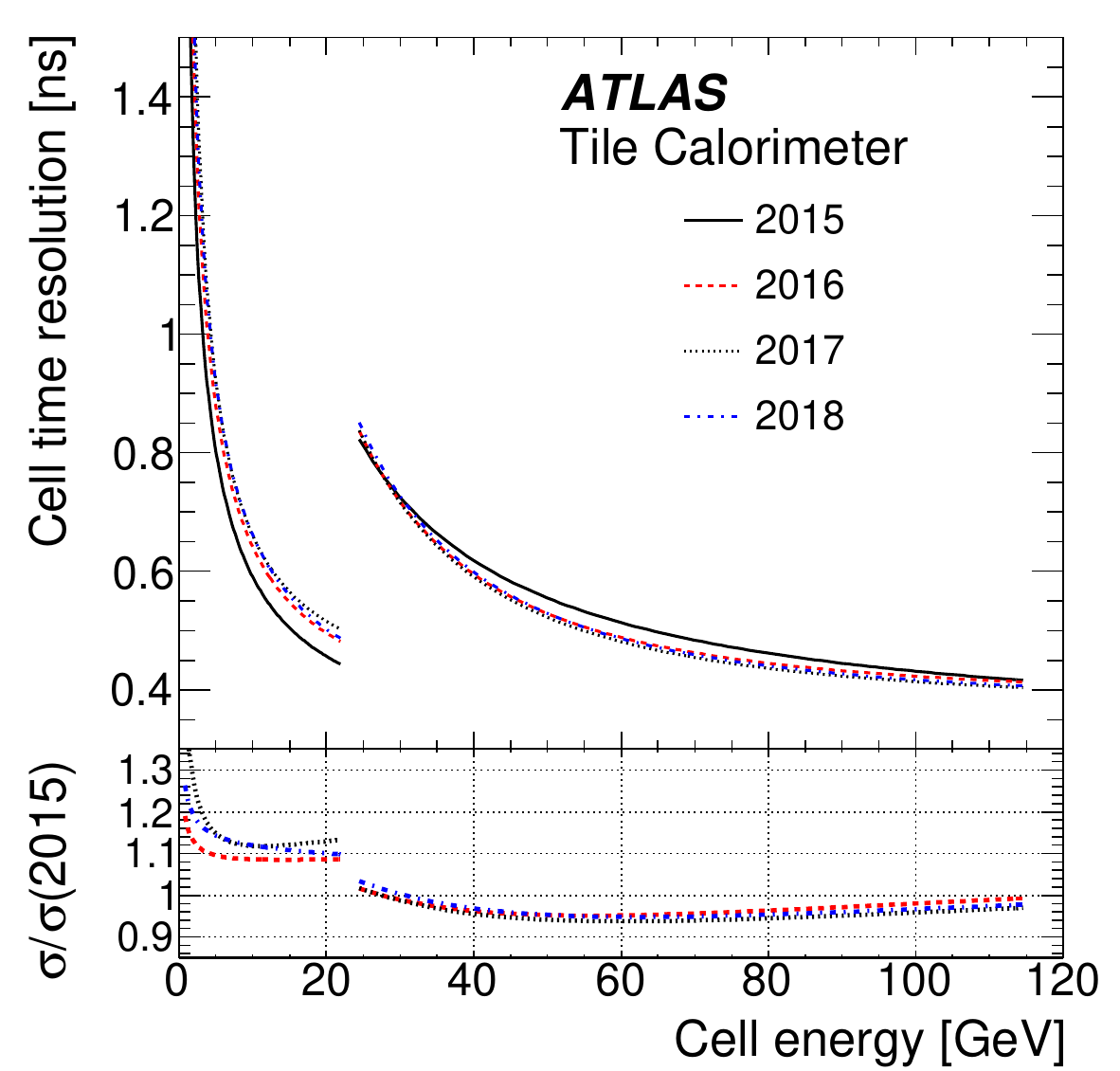}\label{fig:timing_performance_resolution_b}}
\caption{(a) The time resolution (full circles) and  RMS of the time distribution (open squares) as a function of the cell energy. The results were obtained by analysing 2018 data. Statistical uncertainties are smaller than the size of the markers. The fits of the
function in Eq.~(\ref{eq:resolution_fit_function}) are superimposed on high- and low-gain resolution data as indicated by solid and dashed curves, respectively.
(b) A comparison of the time resolution fit functions obtained by analysing 2015, 2016, 2017, 2018 data samples~\cite{TD_proceedings} as a function of the cell energy. The 2017 high-gain fit is affected by a worse time resolution in the lower energy bin that leads to slightly different fit parameters $ p_0 $ and $ p_1 $ (see Eq.~(\ref{eq:resolution_fit_function})) compared to the results of the other years. The effect translates into an increase of the 2017 and 2015 resolution ratio near the high-gain end point. The lower panel shows the fitted time resolution obtained by analysing 2016, 2017 and 2018 data relative to the values obtained analysing 2015 data.}
\end{figure}

Time resolution corresponds to the $\sigma$ of the Gaussian function fit to the cell time distribution discussed above.
The values of time resolution and of the RMS of the cell time distribution in each cell energy bin obtained using 2018 data are shown in Figure~\ref{fig:timing_performance_resolution_a}.
The time resolution is relatively stable between individual runs.
The RMS values, being more affected by the tails of the time distribution driven by the \pileup conditions in the given run, are more unstable. The fit functions

\begin{equation}
\label{eq:resolution_fit_function}
\sigma=\sqrt{p_0^2+(p_1/\sqrt{E})^2+(p_2/E)^2}
\end{equation}

are superimposed on HG and LG resolution data in the figure. The fit functions obtained for the four different data-taking years of Run~2 are
compared in Figure~\ref{fig:timing_performance_resolution_b}. Since the calibration procedure for the HG setting remained unchanged during the whole Run~2, the differences in time resolution for cell energies smaller than 22~\GeV\ are mainly caused by different \pileup conditions. The best resolution is observed in 2015. The difference relative to other years is 10\% for cell energies around 20~\GeV\ in HG.

The time calibration for LG was the same during the years 2016--2018. The time resolution is slightly worse in 2015 due to an older calibration procedure. It relied on the HG time calibration taking into account a constant offset due to faster signal propagation at LG instead of direct determination of the LG time calibration constants.
The updated procedure, as described in Section~\ref{subsubsec:time_calibration}, improves the time resolution for LG by up to 5\% despite higher \pileup in 2016--2018 compared to 2015.

\subsubsection{Effect of \pileup on the time resolution}
\label{sec:resolution_pileup}
The dependence of time resolution on \pileup was investigated by analysing the data on a run-to-run basis.
The mean number of reconstructed primary vertices per bunch crossing within a luminosity  block, $N_{\text{PV}}$, is used as a proxy for the \pileup conditions in each run. Results are shown in Figure~\ref{fig:timing_performance_resolution_pileup}. Low \pileup runs show slightly better time resolution than runs with higher \pileup. This feature is fully compatible with the differences in the time resolution reported in Figure~\ref{fig:timing_performance_resolution_b}.

\begin{figure}[t]
\centering
\includegraphics[width=0.5\linewidth]{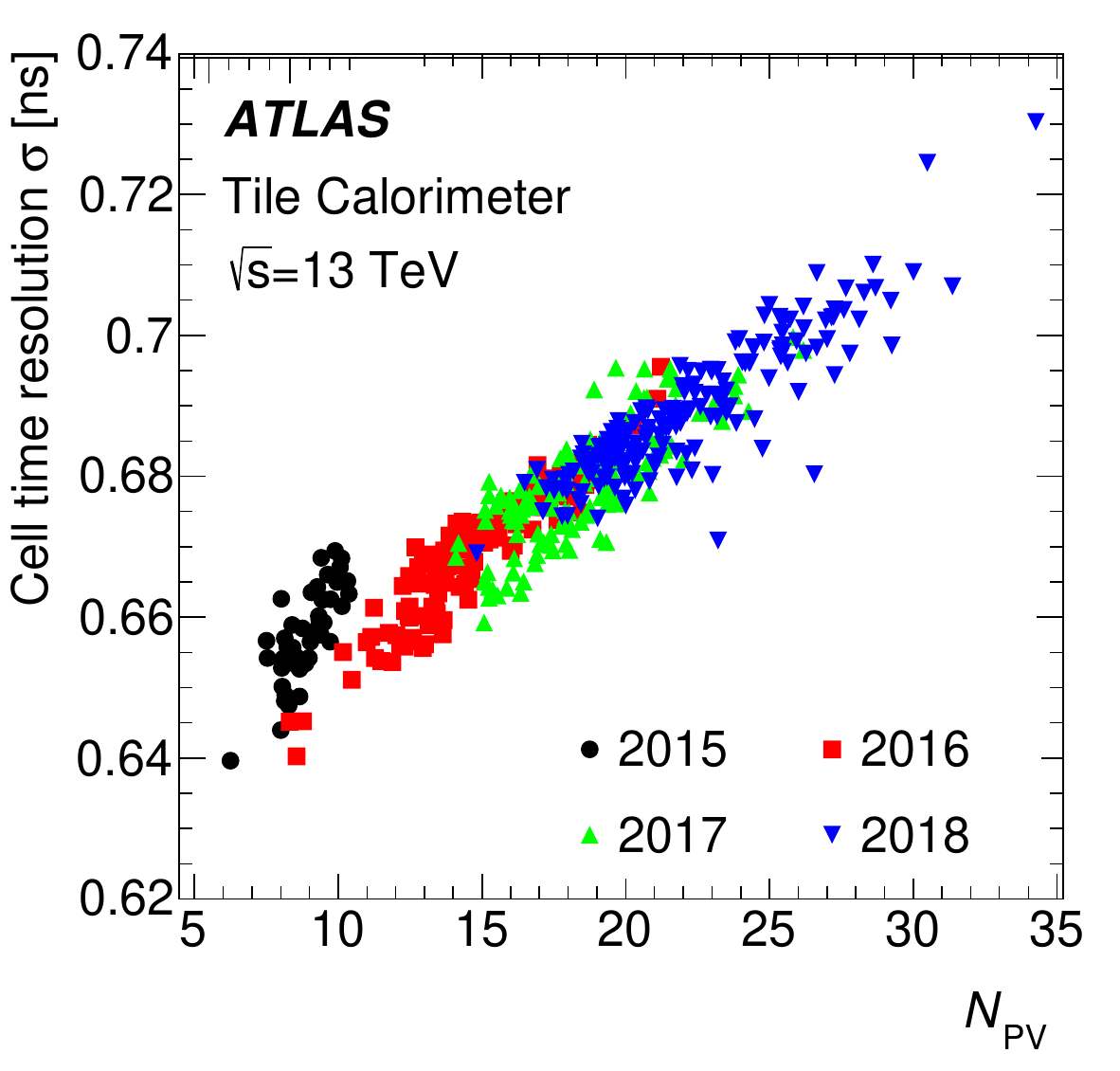}
\caption{The cell time resolution ($\sigma$) determined in
the energy range $\SI{8}{\GeV} < E_{\mathrm{cell}} < \SI{10}{\GeV}$ as a function of the mean number of reconstructed primary vertices per bunch crossing within a luminosity block, $N_{\text{PV}}$, in the run. Each point corresponds to one run and is obtained using all the cells of the detector. In the analysis, only runs with 25~ns bunch spacing and the number of bunches in the LHC larger than 1000 are used.
The worse resolution in 2015 is compatible with the larger run-to-run differences provided the phase changes are not strictly coupled to individual run boundaries, as discussed in Section~\ref{sec:run_to_run}.
}
\label{fig:timing_performance_resolution_pileup}
\end{figure}

\subsubsection{Run-to-run differences}
\label{sec:run_to_run}
The differences in the mean reconstructed time were also investigated
on run-to-run basis. These differences are due to a phase between the LHC clock signal and the bunch-crossings in ATLAS that can vary slightly in time. Large run-to-run differences can negatively influence the time resolution integrated over several runs. While the run-to-run RMS was rather large in 2015 (up to 0.1~ns), it significantly improved since the beginning of 2016 and stayed at the level of 0.03~ns or better. This value is negligible compared to the time resolution constant term $p_0$ shown in Figure~\ref{fig:timing_performance_resolution_a} and the run-to-run differences have no impact on the time resolution even if integrating over many runs.

ATLAS monitors this phase with the beam pick-up based
timing system (BPTX)~\cite{OHM2010558}.\footnote{The BPTX stations are beam position monitors installed in the LHC tunnel \SI{175}{m} away from
the ATLAS interaction point.} The improvement in the run-to-run RMS
observed in the TileCal is in line with the automated correction
for large phases based on BPTX measurements that was adopted in ATLAS in 2016.

\subsection{Electronic and \pileup noise}
\label{subsec:signal_noise}

The total noise per cell is calculated taking into account two components, electronic noise and a contribution from \pileup interactions (\pileup noise). These two contributions are added in quadrature to estimate the total noise (see Section~\ref{subsubsec:pileup_noise}).
The total cell noise is used as input to the topological clustering algorithm~\cite{PERF-2014-07}. See also Section~\ref{subsec:energy_response_hadrons}.

\subsubsection{Electronic noise}
\label{subsubsec:electronic_noise}

The electronic noise is measured by fluctuations of the  channel pedestal $ped$~(Eq.~(\ref{eq:of})) and is largely independent of LHC beam conditions. The distributions are obtained using dedicated runs with no beam in the LHC\@. The reconstruction of pedestal data mirrors that of the data-taking period, using the non-iterative OF technique. The cell electronic noise is given by

\begin{equation}
\sigma_\text{electronic} = \sqrt{\sum^{N_c}_i{(\sigma^i_\text{electronic})^2}}
\label{eq:cell_electronic_noise}
\end{equation}

where $\sigma^i_\text{electronic}$ is the standard deviation (RMS) of the distribution of pedestal values of channel $i$ and $N_c$ is the number of channels in the cell. Most of the TileCal cells are read out by two channels.
Both are read out in HG and LG. Therefore, for each cell there are four combinations (low--low, low--high, high--low and high--high). The
high--high combination is typically used when the deposited energy in a cell is
below 15~\GeV\, while above that the low--low one is most commonly used. In the
region around 15~\GeV\ the combination of low--high/high--low is
used. Figure~\ref{fig:etadist-preHV} shows the $\phi$-averaged electronic noise (RMS)
for all cells as a function of $\eta$, for the high--high gain combination, measured in a single representative run taken in 2017.
Different calorimeter layers are presented.
The average cell noise for regular cells
is approximately 20.5~\MeV\ with a standard deviation of 2.5~\MeV.
However, cells located in the region $|\eta| \approx 1$ show noise values closer to 27.5~\MeV. These cells are formed by channels physically located near the LVPS\@.
Detailed discussions of the electronic noise is given in Ref.~\cite{TCAL-2017-01}.

\begin{figure}[t]
\centering
\includegraphics[width=0.7\linewidth]{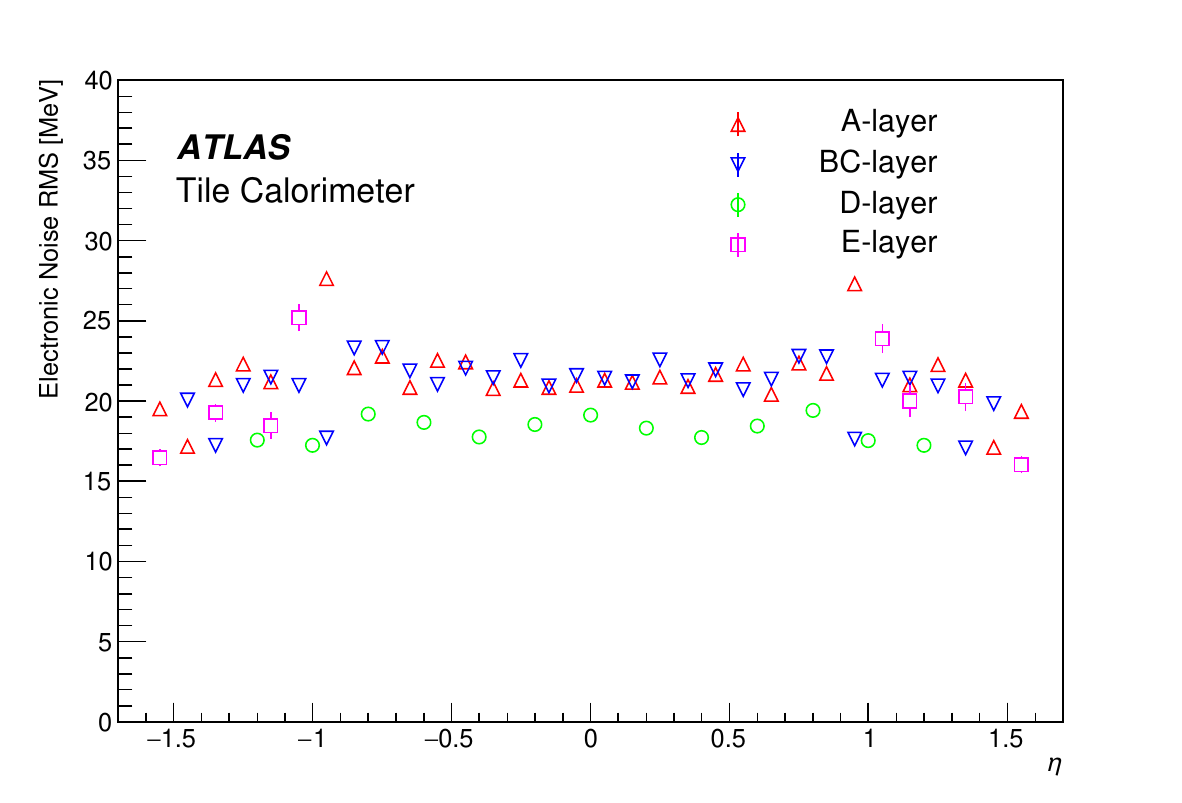}
\caption{The $\phi$-averaged electronic noise
as a function of $\eta$ of the cell, with both contributing read-out channels in high-gain mode. For each cell type the average value over all modules is taken. The statistical uncertainties are smaller than the marker size. Values are extracted using a single representative pedestal calibration run taken in 2017. The different cell types are shown separately for each layer.
The transition between the long and extended barrels can be seen in the range of $0.7 < |\eta| < 1.0$.
}
\label{fig:etadist-preHV}
\end{figure}

\subsubsection{\Pileup noise}
\label{subsubsec:pileup_noise}

The total noise is the sum of all sources of noise contributing to
the signals in TileCal cells. The electronic noise introduced in Section
\ref{subsubsec:electronic_noise} and the \pileup noise described in this section are the two main contributions. \Pileup noise has two
components that lead to the overlay of multiple signals, they are called in-time and out-of-time \pileup. In-time \pileup results in the overlay of signals from multiple proton--proton interactions within the triggered bunch crossing. On the other hand, out-of-time \pileup is the contribution of signals from neighbouring bunch crossings. It is present if the width of the electrical pulse, shown in Figure~\ref{fig:ps_lo_hi}, is larger than the bunch spacing. During Run~2, the TileCal operated in high \pileup conditions with a proton bunch spacing of 25~ns used in most runs. The mean number of interactions per bunch crossing within a luminosity block,
$\langle\mu\rangle$, ranged from about 8 to 70. At low $\langle\mu\rangle$, the total noise in TileCal cells is
dominated by the electronic noise component. However, for larger values of $\langle\mu\rangle$, the \pileup noise becomes the largest contribution resulting from both in-time and out-of-time \pileup.

\Pileup noise is studied in zero-bias data selected by a random trigger.
The trigger accepts events from collisions occurring a fixed number of LHC bunch crossings after an accepted high-energy L1 trigger whose rate scales with luminosity~\cite{TRIG-2019-04}.
This triggering provides a data sample that is not biased by any residual signal in the calorimeter system. The study is performed using events from several runs collected in 2016 with $14<\langle\mu\rangle<44$ providing a number of events comparable with available MC simulations.
Minimum-bias MC samples for \pileup noise studies were generated with \textsc{Pythia 8} using the NNPDF3.0NNLO parton distribution fuction set and the A3 minimum-bias tune~\cite{Ball:2014uwa,ATL-PHYS-PUB-2016-017} and with $0<\langle\mu\rangle<50$.

\begin{figure}[t]
\centering
\includegraphics[width=.7\textwidth]{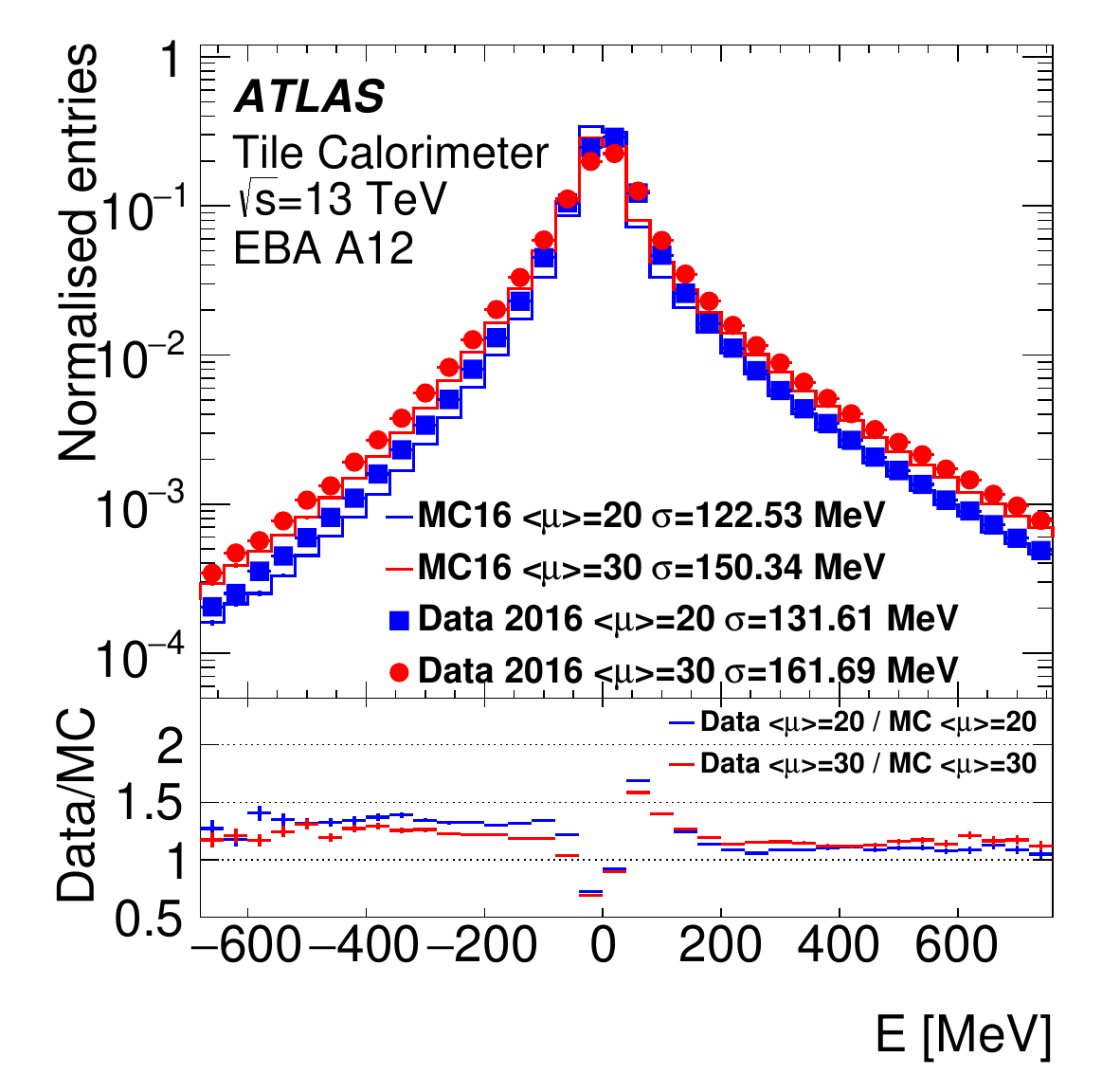}
\caption{Normalised energy distributions in cells A12 ($1.1<|\eta|<1.2$) observed in $pp$ collision data with 25~ns bunch spacing at $\sqrt{s}=13$~\TeV\ collected in 2016 in the zero-bias stream and in the \textsc{Pythia 8} MC simulation with the A3 minimum-bias tune.
An integration over all modules is performed.
The depicted distributions correspond to two different \pileup conditions with $\langle\mu\rangle=20$ (squares) and $\langle\mu\rangle=30$ (circles).
The ratio of the data to the MC simulation is shown in the lower panel.}
\label{fig:noise_energy}
\end{figure}

\Cref{fig:noise_energy} shows the energy distribution in
A12 cells located in the $1.1<|\eta|<1.2$ region of the detector.
Those cells receive the highest radiation dose in layer A (see Figure~\ref{fig:Tile_TID_prelim}).
The energy distributions are centred around zero in absence of high-energy signals. Increasing \pileup widens the energy distribution both in data and MC simulation. Reasonable agreement between data and the simulation is found above approximately 200~\MeV. However, below this energy, the simulated energy distribution is narrower than in data. The total noise measured as the standard deviation (RMS) of the total energy distribution in A12 cells is lower by about 7\% in MC simulation compared to data.

\begin{figure}[t]
\centering
\includegraphics[width=.7\textwidth]{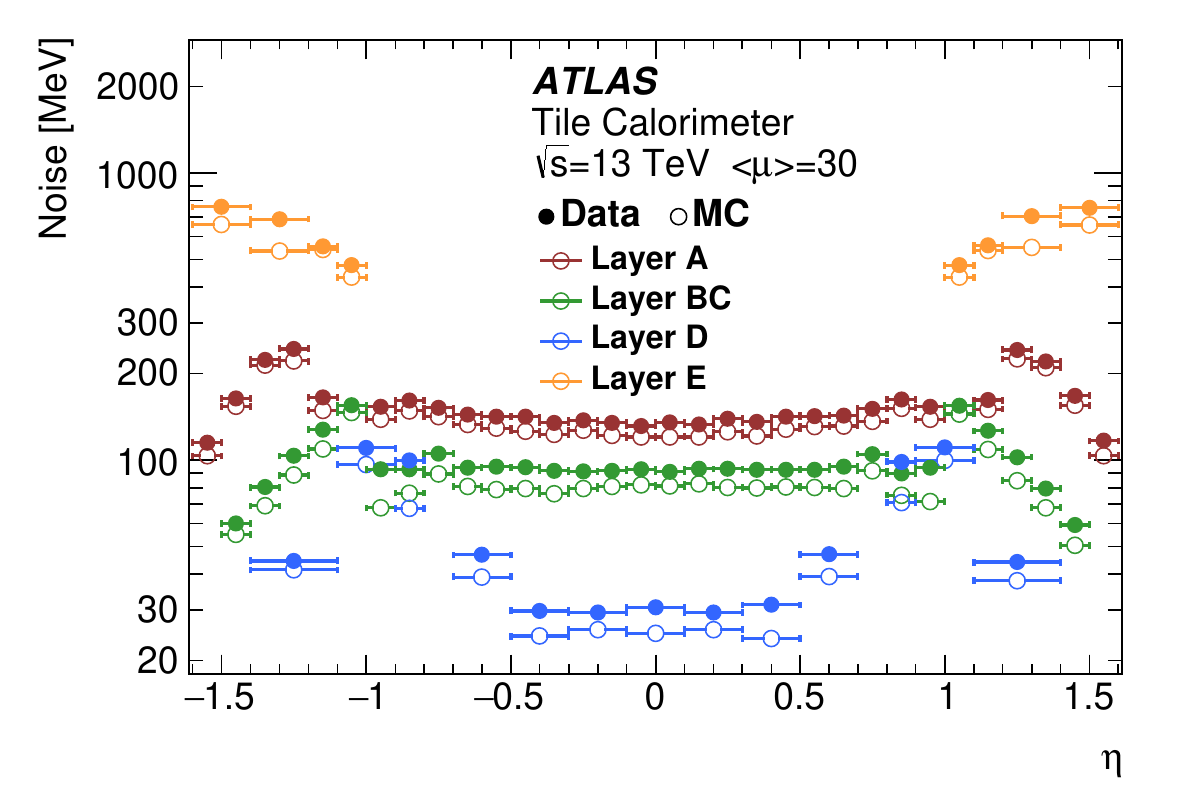}
\caption{The total noise in TileCal cells, as a function of $\eta$, observed in $pp$ collision data with 25~ns bunch spacing at $\sqrt{s}=13$~\TeV\ collected in 2016 in the zero-bias stream with an average number of interactions $\langle\mu\rangle=30$ per bunch crossing and in the \textsc{Pythia 8} MC simulation with the A3 minimum-bias tune.
The noise is estimated as the standard deviation (RMS) of the measured cell energy distribution.
The data (MC simulation) are plotted with closed (open) markers. The cells of different layers
are shown with different colours.}
\label{fig:noise_eta}
\end{figure}

\Cref{fig:noise_eta} shows the total noise (electronic noise and the contribution from \pileup)
in different TileCal layers, for events with $\langle\mu\rangle=30$, as the function of $\eta$ obtained using experimental and simulated data.
The electronic noise component is small and almost flat for standard cells, as seen in Figure~\ref{fig:etadist-preHV}.
Due to the distribution of upstream material and the distance of cells from the interaction point, the \pileup seen in the TileCal is not expected to be uniform. The total noise is approximately constant with $\eta$ in the LB, while the variations in the EB are due to larger amounts of various absorbers traversed by soft particles.
The largest effect of \pileup is seen in the cells of layers~A and E that are the closest to the collision point and experience the highest particle flux (see Figure~\ref{fig:Tile_TID_prelim}).
The \pileup noise diminishes for cells in the outermost layers BC/B and D.
The simulation models the $\eta$ dependence of the noise. However, for
all $\eta$ values, the simulation predicts 20\% lower noise than the
data as shown in Figure~\ref{fig:noise_eta}.
The largest part of the difference comes from the negative energy tail seen in Figure~\ref{fig:noise_energy}, which is a feature of the energy reconstruction caused by out-of-time \pileup.
The total noise measured in MC is used as input to the topological clustering algorithm for data and MC.

\begin{figure}[t]
\centering
\subfloat[]{\includegraphics[width=0.495\linewidth]{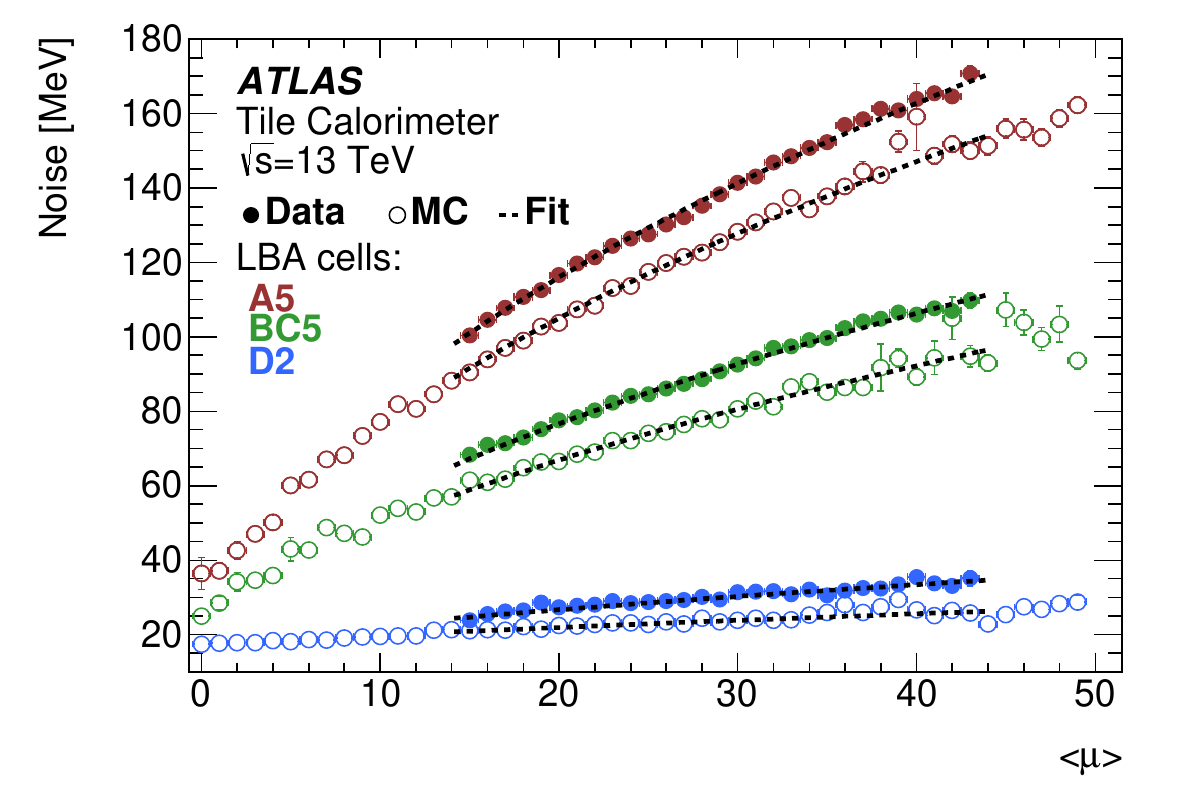}\label{fig:noise_vs_mu_lba}}
\subfloat[]{\includegraphics[width=0.495\linewidth]{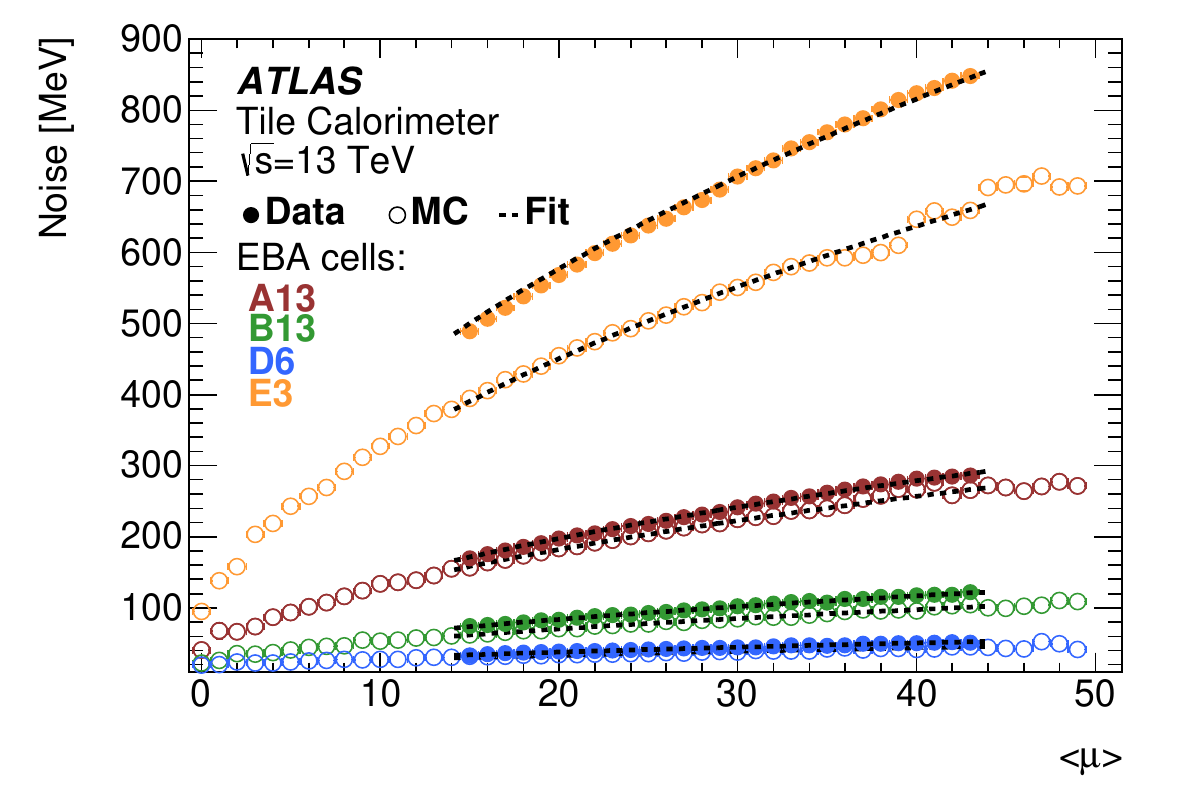}\label{fig:noise_vs_mu_eba}}
\caption{The total noise in TileCal cells, as a function of the average number of interactions per bunch crossing $\langle\mu\rangle$, observed in $pp$ collision data with 25~ns bunch spacing at $\sqrt{s}=13$~\TeV\ collected in 2016 in the zero-bias stream with an average number of $\langle\mu\rangle$ equal to 30, and in \textsc{Pythia 8} MC simulation with the A3 minimum-bias tune.
The noise is estimated as the standard deviation of the energy distribution per cell.
The data (MC simulation) are plotted with closed (open) markers.
The noise is shown for cells located in the region (a) $0.4<|\eta|<0.5$ in the LB and (b) $1.2<|\eta|<1.3$ in the EB.
Due to statistical limitations, the total noise measured in data is shown only for $\langle\mu\rangle>15$. The fit functions from Eq.~(\ref{eq:total_noise_fit}) are overlayed on the experimental and simulated data points (dashed curves) in the figure.
}
\label{fig:noise_vs_mu}
\end{figure}

The \pileup noise is measured in all TileCal cells using the dependence of the total noise on $\langle\mu\rangle$. As an example the results of the experimental and simulated data for the cells A5, B5, D2, A13, B13, D6 and E3 are shown in \Cref{fig:noise_vs_mu}. The results for cells of the same type located in Side C ($\eta<0$) and  Side A ($\eta>0$) are averaged assuming similar \pileup in both sides of the TileCal.
The total noise, $\sigma_\text{total}$, distribution is fitted using the function:

\begin{equation}
\sigma_\text{total} = \sqrt{\sigma_\text{electronics}^2 + \langle\mu\rangle \sigma_\text{\pileup}^2}
\label{eq:total_noise_fit}
\end{equation}

where $\sigma_\text{electronics}$ and $\sigma_\text{\pileup}$ are the electronic and \pileup noise respectively.
The statistical errors in the measured total noise are obtained using pseudo experiments from MC simulation~\cite{ATL-PHYS-PUB-2021-011}.  The use of statistical errors in the fit increases its reliability.
Electronic noise $\sigma_\text{electronics}$ that corresponds to $\langle\mu\rangle=0$,
is measured in pedestal calibration data as discussed in~\Cref{subsubsec:electronic_noise}.
\Pileup noise is derived from a fit using $\sigma_\text{\pileup}$ as the free parameter. The fit functions are overlayed on the experimental and simulated data points in the figure. The experimental (simulated) determinations of $\sigma_\text{\pileup}$ are 25.5 (23.0)~\MeV\  for cell A5, 16.5 (14.2)~\MeV\ for the cell BC5, 4.5 (2.95)~\MeV\ for the cell D2, 43.9 (40.5)~\MeV\ for cell A13, 18.2 (15.1)~\MeV\ for cell B13, 7.54 (6.41)~\MeV\ for cell D6 and 129.0 (100.6)~\MeV\ for cell E3.

The determined \pileup noise is
uploaded to the conditions database, for use by the physics object reconstruction algorithms
to calculate total noise for any $\langle\mu\rangle$ condition. This is used online by the HLT and offline by the algorithm building topological clusters used to reconstruct jets~\cite{PERF-2014-07}.

\subsection{Performance of the tile muon trigger system}
\label{subsubsec:performance_of_the_tilemuon_trigger}

\begin{figure}[t]
\centering
\includegraphics[width=0.6\textwidth]{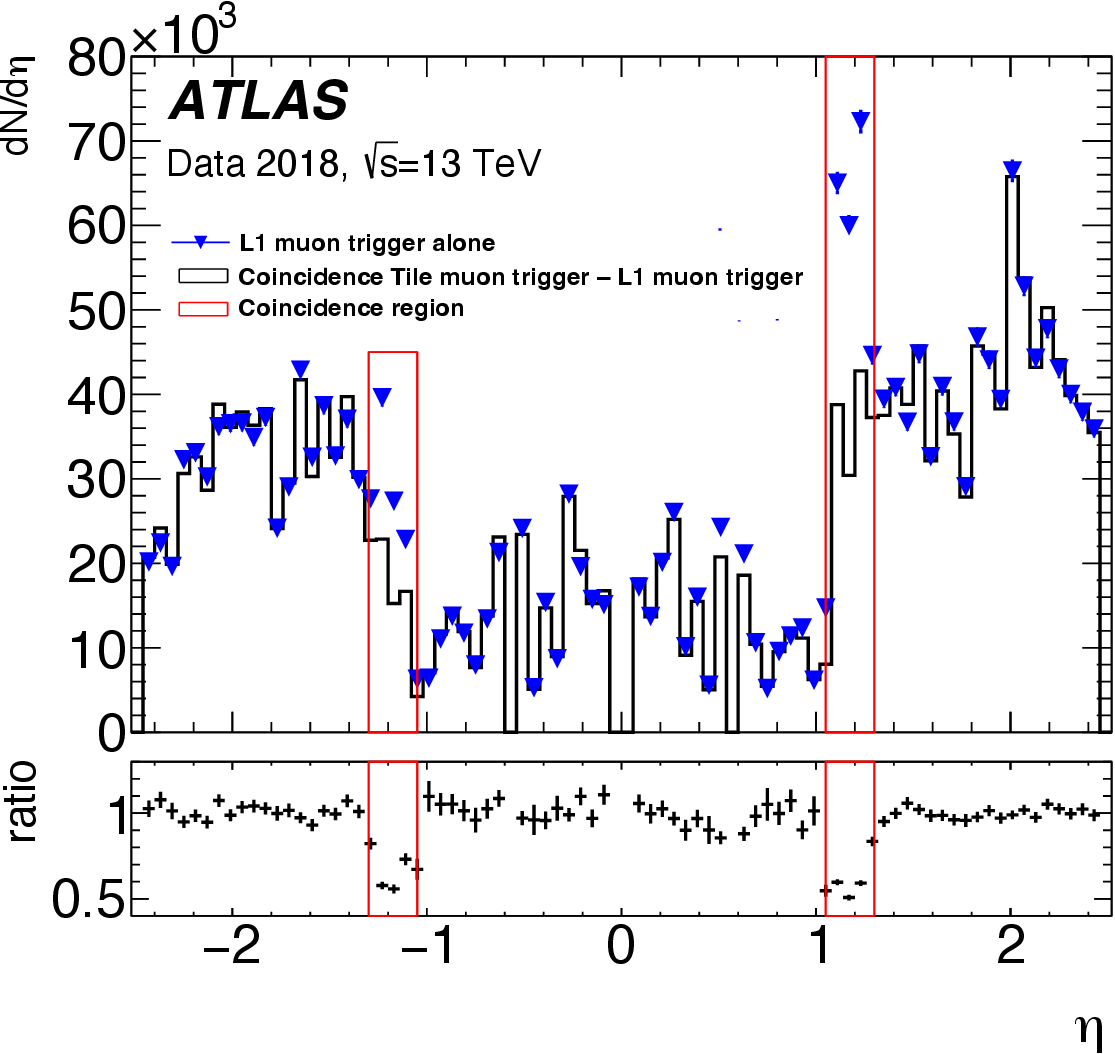}
\caption{The $\eta$ distribution of particles with transverse momentum larger than 20~\GeV\ measured in events selected using the L1 muon trigger alone (triangles) and the tile muon system coincidence (line). The ratios of the bin contents are shown in the lower panel. The coincidence regions, $1.0<|\eta|<1.3$, are indicated by the vertical lines. The $\eta$ of the tracks is reconstructed using online information. The asymmetry of the distribution is due to the different acceptance of the muon spectrometer within the toroidal magnetic field.
}
\label{eta_distribution}
\end{figure}

The tile muon boards, described in Section~\ref{subsec:tile_muon_trigger_system}, were installed and integrated into the ATLAS data acquisition system during the LHC technical stop in 2015. The system has been operating since the beginning of 2018. This section presents results of the tile muon system performance studies using data collected in 2018.

The $\eta$ distribution obtained using the tile muon trigger can be compared with the one obtained using the L1 muon trigger in Figure~\ref{eta_distribution}.
The data show that the tile muon trigger coincidence reduces the acceptance rate in the region $1.0<|\eta|<1.3$ by about 50\%. The total muon trigger rate is reduced by about 6\%. Studies made using $Z\to\mu\mu$ events show that the tile muon trigger inefficiency is less than 2.5\%, compatible with the expected geometrical inefficiency due to the thin gaps between TileCal modules~\cite{BPeralva}.

\subsection{Performance of the minimum-bias trigger scintillators}
\label{subsec:efficiency_mbts}

The events used to study the response of the TileCal to isolated hadrons (Section~\ref{subsec:energy_response_hadrons}) were collected in 2017, more than two years after the start of Run 2. As already discussed they were selected with the MBTS system (Section~\ref{subsec:mbts_intro}) requiring a signal over threshold in at least one MBTS counter in either of the two sides of the detector.

The MBTS trigger efficiency is measured as a function of the multiplicity of tracks reconstructed in the inner detector~\cite{STDM-2015-02}. Figure~\ref{fig:pp_mult_13TeV_pt500_a} shows the results obtained using data collected in 2015 at the beginning of Run~2. The trigger efficiency of the events with only one charged particle with $\pt >500$~\MeV\ emitted in the range of $|\eta|<2.5$ is almost 99\%. It reaches 100\% for events with more than 10 particles. At the same time, only about 20\% of events triggered by the MBTS do not have reconstructed tracks, which means that the fraction of false triggers is even lower, since this category also contains real $pp$ interactions where no particles are produced in the acceptance of the inner detector. The measurements were performed after a long shutdown period in which new scintillating counters were installed and the performance of the detector was not yet affected by the radiation effects discussed in Section~\ref{sec:detector_ageing}. In 2015, the PMT HV was set around 700~V, and the counter threshold value was about 100~mV. In 2017, after more than two years of operation, the MBTS counter response degraded and the data used to study the response of the TileCal to isolated hadrons (see Section~\ref{subsec:energy_response_hadrons}) were collected with the PMTs HV increased to 750--800~V and the threshold value reduced down to around 50 mV. As shown in Figure~\ref{fig:pp_mult_13TeV_pt500_b} the trigger efficiency drops from 99\% to 80\% for events with the lowest multiplicity. It is about 95\% for events with 10 charged particles and does not reach 100\% even for events with 40 charged particles.

\begin{figure}[t]
\centering
\subfloat[]{\includegraphics[width=0.5\linewidth]{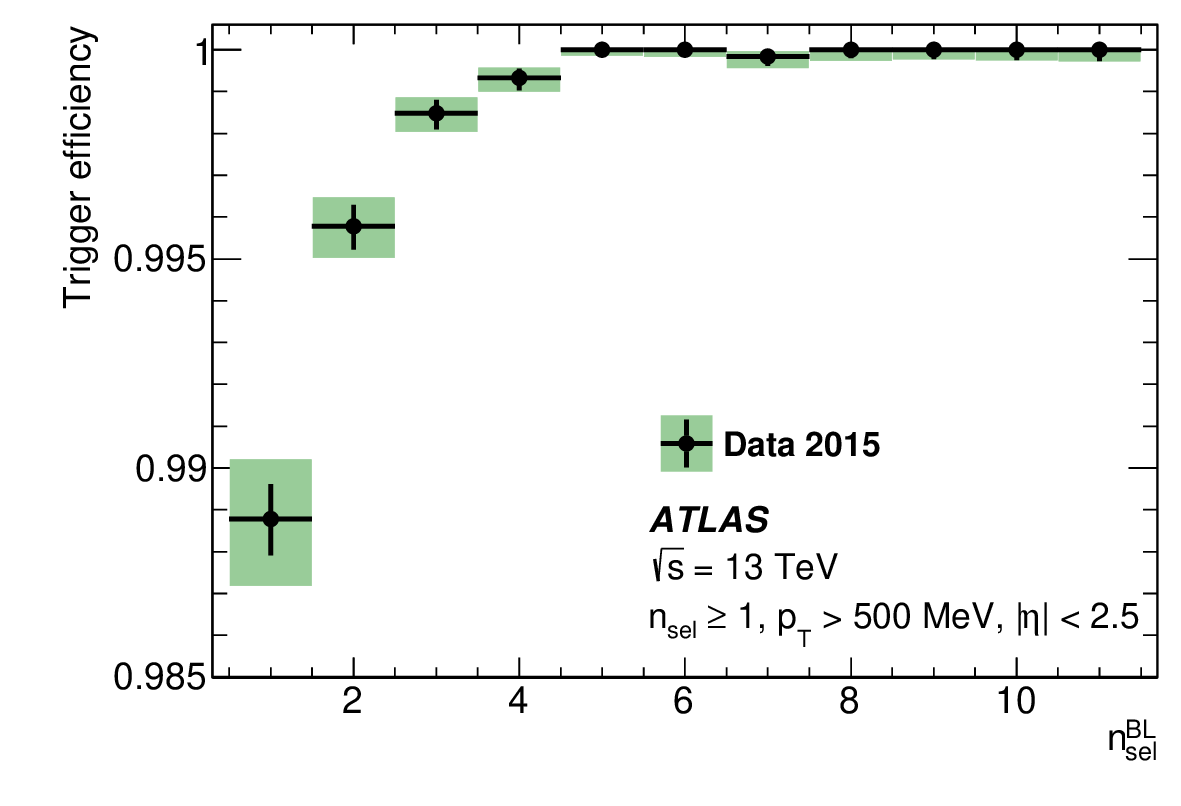}\label{fig:pp_mult_13TeV_pt500_a}}
\subfloat[]{\includegraphics[width=0.5\linewidth]{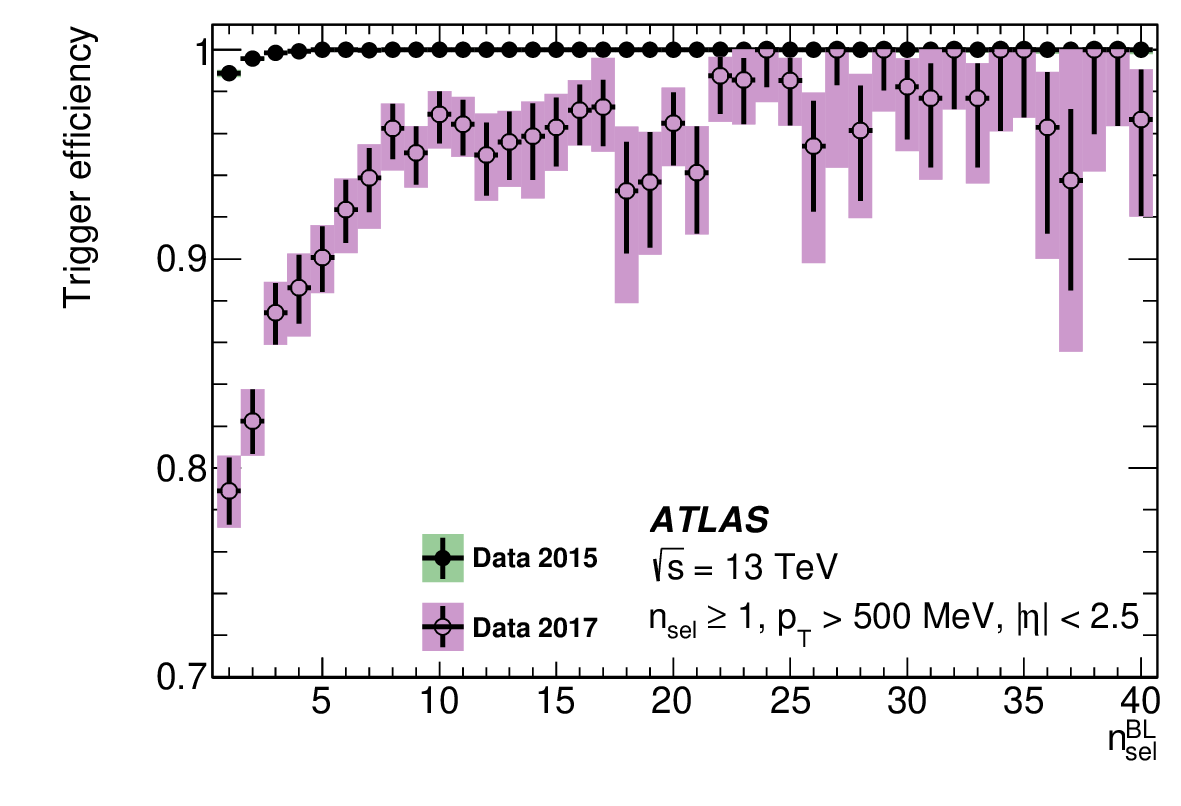}\label{fig:pp_mult_13TeV_pt500_b}}
\caption{(a) The efficiency of the MBTS trigger during $pp$ collisions at $\sqrt{s}=13~\TeV$ as a function of the multiplicity of charged particles compatible with the beam line, $n^{\mathrm{BL}}_{\mathrm{sel}}$, with $\pt>500~\MeV$ and $|\eta| < 2.5$ reconstructed in the inner detector. The data were collected in 2015.
The statistical uncertainties are shown as vertical bars, the sum in quadrature of statistical and systematic uncertainties is shown as the height of the shaded rectangles~\cite{STDM-2015-02}. Systematic uncertainties dominate. (b) The efficiency of the MBTS trigger during $pp$ collisions at $\sqrt{s}$=13~\TeV\ as a
function of the multiplicity of charged particles compatible with the beam line, $n^{\mathrm{BL}}_{\mathrm{sel}}$, with $\pt>500~\MeV$ and $|\eta| < 2.5$ reconstructed in the inner detector. Results from the analysis of data collected in 2017 are compared with those from 2015 presented in (a). The statistical uncertainties are shown as
vertical bars, the sum in quadrature of statistical and systematic uncertainties of the 2017 data are shown by the shaded rectangles. For 2017 data, statistical uncertainties dominate,
while the uncertainties for 2015 data in (b) are smaller than the size of the symbols because of the much larger range of the vertical axis than in (a).}
\end{figure}

\FloatBarrier
\subsection{Summary of performance studies}
\label{subsec:summary_performance_studies}

Muons produced by $W \to \mu\nu$ decays in $pp$ collisions were used to check the calibration of the TileCal. The non-uniformity of the cell response, due to variations in the optical components, electronics of the read-out systems and potential cell miscalibration, is determined. A value of 2.4\% is found by analysing the events of the three periods 2015--2016, 2017 and 2018. The analyses show no clear evidence of problematic cells. The radial layer intercalibration of the TileCal was also evaluated. For all layers the results are consistent with a correct setting of the scale used to reconstruct the jet energy within 2\% in all three periods. Good stability of the detector response over time was observed. The studies show 95\% confidence intervals for the variations of the cell response between 2015--2016 and 2017 and between 2017 and 2018 of [$-1.3$\%, 3.3\%] and [$-2.4$\%, 2.5\%], respectively.

The measurements of the $E/p$ ratio using isolated hadrons show good uniformity of the response across the azimuth angle $\phi$. An agreement between experimental and simulated results is observed, confirming the goodness of the calibration of the cell energy at the EM scale. The uncertainty on the determinations of the mean value (standard deviation) is better than 5\% (6\%).

The time calibration was extensively checked and monitored during Run~2. Using jet measurements, the resolution of the time determination in the TileCal cells is found to be better than 1~ns for energies deposited in the cells larger than 5~\GeV. The huge reduction of the timing jumps, the most critical issue in Run~1, allowed the investigation of and the correction for other timing-related problems as specified in Section~\ref{subsubsec:time_calibration}. In the HG regime, despite significantly larger \pileup, a slightly better time resolution is obtained in Run~2. The time resolution for LG benefits from an improved calibration procedure.

Noise measurements were compared with the simulated expectations. In the case of the A12 cells that receive the highest radiation dose in the regular TileCal cells, the noise is lower by about 7\% in MC simulation compared to data.

In 2015, at the beginning of Run 2, the MBTS trigger efficiency was measured to be larger than 98.5\%.
In 2017, the responses of the radiation-damaged detectors were found to be 0.5\% (inner counters) and 10\% (outer counters) of the corresponding values measured in 2015.
Despite this large response deterioration, in 2017 the MBTS trigger efficiency was larger than 80\%.

The tile muon trigger system uses the TileCal information combined with the TGC muon chambers to reduce the background muon trigger rate due to low momentum protons emerging from the endcap toroid and the beam shielding. The total muon trigger rate reduces by about 6\%. The $Z\to \mu\mu$ events show that the tile muon trigger efficiency is about 97.5\%, compatible with the expected geometrical inefficiency due to thin gaps between TileCal modules~\cite{TRIG-2018-01}.


\FloatBarrier

\section{Conclusion}
\label{sec:conclusion}
%

This paper presents a description of the ATLAS tile calorimeter signal reconstruction, calibration, data quality monitoring, and performance during the Run~2 data-taking period of the LHC (2015--2018).

A precision of about 1\% was demonstrated for the individual calorimeter calibration systems.
Information collected from the D layer of the EB and the MBTS used for trigger purposes have been presented. The investigation of ageing and radiation damage effects allows the expected light yield of the calorimeter to be extrapolated to the end of the data-taking period at the LHC (HL-LHC). The cells exposed to the most severe radiation conditions are expected to degrade to $75^{+9}_{-25}\%$ ($35^{+17}_{-27}\%$). The radiation damage to the MBTS is also assessed. Data quality activities lead the TileCal to contribute with an efficiency of 99.65\% to high-quality ATLAS data-taking and with less than 1.1\% of all cells deemed non-operational at the end of each data-taking year. Thorough monitoring of cell performance allows for quick and efficient repairs during yearly maintenance periods.

Performance is assessed with isolated muons, hadrons and jets. The proton--proton collisions at the LHC at a centre-of-mass energy of 13~\TeV\ are used in the analyses. The results obtained by analysing muon data are consistent with a correct setting of the EM scale used to reconstruct the jet energy within 2\% during all of the Run~2 period. The MC modelling of the single pion energy response is shown to agree with the data collected within the assigned uncertainties.
The timing performance determined from jet measurements is stable during Run~2. Compared to Run~1, the time resolution shows small increase due to \pileup effects, but stays well below the 0.7~ns above 10~\GeV\ in high-gain.

Large increases in noise, coming from increased \pileup conditions during Run~2, are extracted and modelled for each TileCal cell. Despite radiation damage, the MBTS allows the online selection of events with the highest efficiency and the lowest possible bias, as well as calorimeter performance studies. Coincidence of the D cells of the EB with the muon trigger system results in a 50\% reduction of the excess trigger rates induced by low energy background protons.

Overall, during the LHC Run 2, the TileCal performed in accordance with expectations.


%
\section*{Acknowledgements}

%
%

%
%

We thank CERN for the very successful operation of the LHC and its injectors, as well as the support staff at
CERN and at our institutions worldwide without whom ATLAS could not be operated efficiently.

The crucial computing support from all WLCG partners is acknowledged gratefully, in particular from CERN, the ATLAS Tier-1 facilities at TRIUMF/SFU (Canada), NDGF (Denmark, Norway, Sweden), CC-IN2P3 (France), KIT/GridKA (Germany), INFN-CNAF (Italy), NL-T1 (Netherlands), PIC (Spain), RAL (UK) and BNL (USA), the Tier-2 facilities worldwide and large non-WLCG resource providers. Major contributors of computing resources are listed in Ref.~\cite{ATL-SOFT-PUB-2023-001}.

We gratefully acknowledge the support of ANPCyT, Argentina; YerPhI, Armenia; ARC, Australia; BMWFW and FWF, Austria; ANAS, Azerbaijan; CNPq and FAPESP, Brazil; NSERC, NRC and CFI, Canada; CERN; ANID, Chile; CAS, MOST and NSFC, China; Minciencias, Colombia; MEYS CR, Czech Republic; DNRF and DNSRC, Denmark; IN2P3-CNRS and CEA-DRF/IRFU, France; SRNSFG, Georgia; BMBF, HGF and MPG, Germany; GSRI, Greece; RGC and Hong Kong SAR, China; ISF and Benoziyo Center, Israel; INFN, Italy; MEXT and JSPS, Japan; CNRST, Morocco; NWO, Netherlands; RCN, Norway; FCT, Portugal; MNE/IFA, Romania; MESTD, Serbia; MSSR, Slovakia; ARRS and MIZ\v{S}, Slovenia; DSI/NRF, South Africa; MICINN, Spain; SRC and Wallenberg Foundation, Sweden; SERI, SNSF and Cantons of Bern and Geneva, Switzerland; MOST, Taipei; TENMAK, T\"urkiye; STFC, United Kingdom; DOE and NSF, United States of America.

Individual groups and members have received support from BCKDF, CANARIE, CRC and DRAC, Canada; CERN-CZ, PRIMUS 21/SCI/017 and UNCE SCI/013, Czech Republic; COST, ERC, ERDF, Horizon 2020, ICSC-NextGenerationEU and Marie Sk{\l}odowska-Curie Actions, European Union; Investissements d'Avenir Labex, Investissements d'Avenir Idex and ANR, France; DFG and AvH Foundation, Germany; Herakleitos, Thales and Aristeia programmes co-financed by EU-ESF and the Greek NSRF, Greece; BSF-NSF and MINERVA, Israel; Norwegian Financial Mechanism 2014-2021, Norway; NCN and NAWA, Poland; La Caixa Banking Foundation, CERCA Programme Generalitat de Catalunya and PROMETEO and GenT Programmes Generalitat Valenciana, Spain; G\"{o}ran Gustafssons Stiftelse, Sweden; The Royal Society and Leverhulme Trust, United Kingdom.

In addition, individual members wish to acknowledge support from Chile: Agencia Nacional de Investigaci\'on y Desarrollo (FONDECYT 1190886, FONDECYT 1210400, FONDECYT 1230812, FONDECYT 1230987); China: National Natural Science Foundation of China (NSFC - 12175119, NSFC 12275265, NSFC-12075060); Czech Republic: PRIMUS Research Programme (PRIMUS/21/SCI/017); EU: H2020 European Research Council (ERC - 101002463); European Union: European Research Council (ERC - 948254), Horizon 2020 Framework Programme (MUCCA - CHIST-ERA-19-XAI-00), European Union, Future Artificial Intelligence Research (FAIR-NextGenerationEU PE00000013), Italian Center for High Performance Computing, Big Data and Quantum Computing (ICSC, NextGenerationEU), Marie Sklodowska-Curie Actions (EU H2020 MSC IF GRANT NO 101033496); France: Agence Nationale de la Recherche (ANR-20-CE31-0013, ANR-21-CE31-0013, ANR-21-CE31-0022), Investissements d'Avenir Idex (ANR-11-LABX-0012), Investissements d'Avenir Labex (ANR-11-LABX-0012); Germany: Baden-Württemberg Stiftung (BW Stiftung-Postdoc Eliteprogramme), Deutsche Forschungsgemeinschaft (DFG - 469666862, DFG - CR 312/5-1); Italy: Istituto Nazionale di Fisica Nucleare (FELLINI G.A. n. 754496, ICSC, NextGenerationEU); Japan: Japan Society for the Promotion of Science (JSPS KAKENHI JP21H05085, JSPS KAKENHI JP22H01227, JSPS KAKENHI JP22H04944, JSPS KAKENHI JP22KK0227); Netherlands: Netherlands Organisation for Scientific Research (NWO Veni 2020 - VI.Veni.202.179); Norway: Research Council of Norway (RCN-314472); Poland: Polish National Agency for Academic Exchange (PPN/PPO/2020/1/00002/U/00001), Polish National Science Centre (NCN 2021/42/E/ST2/00350, NCN OPUS nr 2022/47/B/ST2/03059, NCN UMO-2019/34/E/ST2/00393, UMO-2020/37/B/ST2/01043, UMO-2021/40/C/ST2/00187); Slovenia: Slovenian Research Agency (ARIS grant J1-3010); Spain: BBVA Foundation (LEO22-1-603), Generalitat Valenciana (Artemisa, FEDER, IDIFEDER/2018/048), La Caixa Banking Foundation (LCF/BQ/PI20/11760025), Ministry of Science and Innovation (MCIN \& NextGenEU PCI2022-135018-2, MICIN \& FEDER PID2021-125273NB, RYC2019-028510-I, RYC2020-030254-I, RYC2021-031273-I, RYC2022-038164-I), PROMETEO and GenT Programmes Generalitat Valenciana (CIDEGENT/2019/023, CIDEGENT/2019/027); Sweden: Swedish Research Council (VR 2018-00482, VR 2022-03845, VR 2022-04683, VR grant 2021-03651), Knut and Alice Wallenberg Foundation (KAW 2017.0100, KAW 2018.0157, KAW 2018.0458, KAW 2019.0447); Switzerland: Swiss National Science Foundation (SNSF - PCEFP2\_194658); United Kingdom: Leverhulme Trust (Leverhulme Trust RPG-2020-004); United States of America: U.S. Department of Energy (ECA DE-AC02-76SF00515), Neubauer Family Foundation.

%
%


%
%
%
%
%
%
%

%
%
%
%
%
%

%
%

%
%
%
\printbibliography
\clearpage
\input{atlas_authlist}
%

%
%
%
%

%
%
%
%

\end{document}

%% file: atlas_authlist.tex
 
\begin{flushleft}
\hypersetup{urlcolor=black}
{\Large The ATLAS Collaboration}

\bigskip

\AtlasOrcid[0000-0002-6665-4934]{G.~Aad}$^\textrm{\scriptsize 102}$,
\AtlasOrcid[0000-0002-5888-2734]{B.~Abbott}$^\textrm{\scriptsize 120}$,
\AtlasOrcid{J.~Abdallah}$^\textrm{\scriptsize 8}$,
\AtlasOrcid[0000-0002-1002-1652]{K.~Abeling}$^\textrm{\scriptsize 55}$,
\AtlasOrcid[0000-0001-5763-2760]{N.J.~Abicht}$^\textrm{\scriptsize 49}$,
\AtlasOrcid[0000-0002-8496-9294]{S.H.~Abidi}$^\textrm{\scriptsize 29}$,
\AtlasOrcid[0000-0002-9987-2292]{A.~Aboulhorma}$^\textrm{\scriptsize 35e}$,
\AtlasOrcid[0000-0001-5329-6640]{H.~Abramowicz}$^\textrm{\scriptsize 151}$,
\AtlasOrcid[0000-0002-1599-2896]{H.~Abreu}$^\textrm{\scriptsize 150}$,
\AtlasOrcid[0000-0003-0403-3697]{Y.~Abulaiti}$^\textrm{\scriptsize 117}$
% ...
\bigskip
\\

% ...
$^{8}$Department of Physics, University of Texas at Arlington, Arlington TX; United States of America.\\
% ...
$^{29}$Physics Department, Brookhaven National Laboratory, Upton NY; United States of America.\\
% ...
$^{35}$$^{(a)}$Facult\'e des Sciences Ain Chock, R\'eseau Universitaire de Physique des Hautes Energies - Universit\'e Hassan II, Casablanca;$^{(b)}$Facult\'{e} des Sciences, Universit\'{e} Ibn-Tofail, K\'{e}nitra;$^{(c)}$Facult\'e des Sciences Semlalia, Universit\'e Cadi Ayyad, LPHEA-Marrakech;$^{(d)}$LPMR, Facult\'e des Sciences, Universit\'e Mohamed Premier, Oujda;$^{(e)}$Facult\'e des sciences, Universit\'e Mohammed V, Rabat;$^{(f)}$Institute of Applied Physics, Mohammed VI Polytechnic University, Ben Guerir; Morocco.\\
% ...
$^{49}$Fakult\"{a}t Physik , Technische Universit{\"a}t Dortmund, Dortmund; Germany.\\
% ...
$^{55}$II. Physikalisches Institut, Georg-August-Universit\"{a}t G\"ottingen, G\"ottingen; Germany.\\
% ...
$^{102}$CPPM, Aix-Marseille Universit\'e, CNRS/IN2P3, Marseille; France.\\
% ...
$^{117}$Department of Physics, New York University, New York NY; United States of America.\\
% ...
$^{120}$Homer L. Dodge Department of Physics and Astronomy, University of Oklahoma, Norman OK; United States of America.\\
% ...
$^{150}$Department of Physics, Technion, Israel Institute of Technology, Haifa; Israel.\\
$^{151}$Raymond and Beverly Sackler School of Physics and Astronomy, Tel Aviv University, Tel Aviv; Israel.\\
% ...
$^{a}$ Also Affiliated with an institute covered by a cooperation agreement with CERN.\\
$^{b}$ Also at An-Najah National University, Nablus; Palestine.\\
$^{c}$ Also at Borough of Manhattan Community College, City University of New York, New York NY; United States of America.\\
$^{d}$ Also at Center for High Energy Physics, Peking University; China.\\
$^{e}$ Also at Center for Interdisciplinary Research and Innovation (CIRI-AUTH), Thessaloniki; Greece.\\
$^{f}$ Also at Centro Studi e Ricerche Enrico Fermi; Italy.\\
% ...

\end{flushleft}
